\documentclass[traditabstract]{aa}  

\usepackage{amsmath}
\usepackage{graphicx}
\usepackage{natbib}
\usepackage{morefloats}

\bibpunct{(}{)}{;}{a}{}{,} 
\usepackage{txfonts}

\begin{document}

 \title{The ``toothbrush-relic'': evidence for a coherent linear 2-Mpc scale shock wave in a massive merging galaxy cluster?} 

\titlerunning{Complex diffuse radio emission in 1RXS~J0603.3+4214}

   \author{R.~J. van Weeren\inst{1,2,3}\thanks{Einstein Fellow,  \email{rvanweeren@cfa.harvard.edu}}
         \and H.~J.~A. R\"ottgering\inst{1}
         \and H.~T.~Intema \inst{4}
         \and L.~Rudnick \inst{5}
         \and M.~Br\"uggen \inst{6,7}
         \and M.~Hoeft \inst{8}
         \and J.~B.~R. Oonk \inst{2}
          }
          
   \institute{Leiden Observatory, Leiden University,
              P.O. Box 9513, NL-2300 RA Leiden, The Netherlands
                 \and Netherlands Institute for Radio Astronomy (ASTRON), Postbus 2, 7990 AA Dwingeloo, The Netherlands 
                 \and Harvard-Smithsonian Center for Astrophysics, 60 Garden Street, Cambridge, MA 02138, USA
                 \and National Radio Astronomy Observatory, 520 Edgemont Road, Charlottesville, VA 22903-2475, USA
                 \and Minnesota Institute for Astrophysics, University of Minnesota, 116 Church St. S.E., Minneapolis, MN 55455, USA
                 \and Jacobs University Bremen, P.O. Box 750561, 28725 Bremen, Germany
                 \and Hamburger Sternwarte, Gojenbergsweg 112, 21029 Hamburg, Germany
                 \and Th\"uringer Landessternwarte Tautenburg, Sternwarte 5, 07778, Tautenburg, Germany
                 }


 
\abstract
   {Some merging galaxy clusters host diffuse extended radio emission, so-called radio halos and relics, unrelated to individual galaxies.  The origin of these halos and relics is still debated, although there is compelling evidence now that they are related to cluster merger events. Here we present detailed Westerbork Synthesis Radio Telescope (WSRT) and Giant Metrewave Radio Telescope (GMRT) radio observations between 147~MHz and 4.9~GHz of a new radio-selected  galaxy cluster 1RXS~J0603.3+4214, for which we find a redshift of 0.225. The cluster is detected as an extended X-ray source in the ROSAT All Sky Survey with an X-ray luminosity of $L_{\rm{X,~0.1-2.4~keV}} \sim 1 \times 10^{45}$~erg~s$^{-1}$. The cluster hosts a large bright 1.9~Mpc radio relic, an elongated $\sim2$~Mpc radio halo, and two fainter smaller radio relics. The large radio relic has a peculiar linear morphology. For this relic we observe a clear spectral index gradient from the front of the relic towards the back, in the direction towards the cluster center. Parts of this relic are highly polarized with a polarization fraction of up to 60\%. We performed Rotation Measure (RM) Synthesis between 1.2 and 1.7~GHz. The results suggest that for the west part of the large relic some of the Faraday rotation is caused by ICM and not only due to galactic foregrounds. We also carried out a detailed spectral analysis of this radio relic and created radio color-color diagrams. We find (i) an injection spectral index of $-0.6$ to $-0.7$, (ii) steepening spectral index and increasing spectral curvature in the post-shock region, and (iii) an overall power-law spectrum between 74~MHz and 4.9~GHz with $\alpha=-1.10 \pm 0.02$.  Mixing of emission in the beam from regions with different spectral ages is probably the dominant factor that determines the shape of the radio spectra. Changes in the magnetic field, total electron content, or adiabatic gains/losses do not play a major role. A model in which particles are (re)accelerated in a first order Fermi process at the front of the relic provides the best match to the observed spectra. We speculate that in the post-shock region particles are re-accelerated by merger induced turbulence to form the radio halo as the relic and halo are connected. The orientation of the bright relic and halo indicate a north-south merger event, but the peculiar linear shape and the presence of another relic, perpendicular to the bright relic, suggest a more complex merger event. Deep X-ray observations will be needed to determine the merger scenario. 
      }
  
   \maketitle

\section{Introduction}
Radio halos and relics are diffuse sources found in galaxy clusters that show signs of an ongoing merger event \citep[see][for an overview]{2005AdSpR..36..729F, 2008SSRv..134...93F, 2011SSRv..tmp..138B}. These sources reveal the presence of relativistic  particles and cluster-wide   magnetic fields  \citep[e.g.,][]{1977ApJ...212....1J} within the intracluster medium (ICM). In the hierarchical model of structure formation galaxy cluster form through a sequence of mergers with smaller substructures. These merger events create shocks and turbulence in the ICM and could amplify magnetic fields. Several models have been put forward which link the presence of these diffuse radio sources to  cluster merger events \citep[e.g.,][]{1998A&A...332..395E,2001MNRAS.320..365B,2001ApJ...557..560P, 2005MNRAS.357.1313C,2010arXiv1011.0729K, 2011A&A...527A..99E}.

\emph{Radio halos} are smooth extended radio sources that roughly follow the X-ray emission from the ICM. They are typically unpolarized and have an extent of about a Mpc. {There are two main classes of models proposed to explain the origin of radio halos: (1) relativistic electrons in the ICM are re-accelerated in-situ through interaction with turbulence generated in the ICM by cluster-cluster mergers \citep[][]{2001MNRAS.320..365B, 2001ApJ...557..560P}, or (2) the energetic electrons are secondary products of proton-proton collisions \citep[e.g.,][]{1980ApJ...239L..93D, 1999APh....12..169B, 2000A&A...362..151D, 2001ApJ...562..233M, 2010ApJ...722..737K, 2011A&A...527A..99E}.  In addition, models using a combination of the two mechanisms have been considered \citep{2005MNRAS.363.1173B,2008SSRv..134..311D,  2011MNRAS.410..127B}. }

\emph{Radio relics} are usually divided up into three different classes  \citep[see][for an overview]{2004rcfg.proc..335K}. (1) {\it Radio Gischt} are large extended sources mostly found in the outskirts of galaxy clusters. They are proposed to trace shock waves \citep{1998A&A...332..395E, 2001ApJ...562..233M}, in which particles are accelerated by the diffusive shock acceleration mechanism \citep[DSA; e.g.,][]{1977DoSSR.234R1306K, 1977ICRC...11..132A, 1978MNRAS.182..147B, 1978MNRAS.182..443B, 1978ApJ...221L..29B, 1983RPPh...46..973D, 1987PhR...154....1B, 1991SSRv...58..259J, 2001RPPh...64..429M} in a first-order Fermi process. However, the efficiency with which collisionless shocks can accelerate particles is unknown and may not be enough to produce the observed radio brightness of relics. A closely linked scenario is that of  shock re-acceleration of pre-accelerated electrons in the ICM, which is a more efficient mechanism for weak shocks \citep[e.g.,][]{2005ApJ...627..733M, 2008A&A...486..347G, 2011ApJ...734...18K, 2012ApJ...756...97K}.
(2) {\it AGN relics} trace old radio plasma from previous episodes of AGN activity. (3) {\it Radio phoenices} are AGN relics that have been compressed adiabatically by merger shock waves boosting the radio emission \citep{2001A&A...366...26E, 2002MNRAS.331.1011E}. The radio spectra of these sources should be steep ($\alpha \lesssim -1.5$, {$F_{\nu} \propto \nu^{\alpha}$, where $\alpha$ is the spectral index) and curved  due to synchrotron and Inverse Compton (IC) losses.

A different model for radio relics (radio gischt) and halos has been proposed by \cite{2010arXiv1011.0729K}. It is based on a secondary cosmic ray electron model, where the time evolution of the magnetic fields and cosmic ray distribution are taken into account to explain both halos and gischt.  For the outer edges of relics it is predicted that the spectral index should be $\alpha \simeq -1$. In addition, the magnetic fields ($B$)  are expected to be strong, with $B \gtrsim B_{\rm{CMB}} > 3$~$\mu$Gauss, and $B_{\rm{CMB}}$ the equivalent magnetic field strength of the cosmic microwave background.

\begin{table*}
\begin{center}
\caption{GMRT observations}
\begin{tabular}{llllll}
\hline
\hline
&  147~MHz   & 241~MHz (dual)  & 325~MHz &610~MHz (dual), 610~MHz & 1280~MHz \\
\hline
Observation date &    Apr 29, 2010 &Nov 20, 2009&  Apr 25, 2010 & Nov 20, 2009 \& 30 Apr, 2010 & Nov 12, 2009 \\
Usable bandwidth &  15~MHz & 6~MHz &  30~MHz & 30~MHz & 30 MHz \\
Channel width &31.25~kHz &62.5~kHz & 62.5~kHz & 62.5~kHz&  62.5~kHz \\
Polarization & RR+LL & LL &  RR+LL& RR, RR+LL  & RR+LL \\
Integration time &  4 sec & 8 sec & 8 sec & 8 sec & 16 sec  \\
Total on-source time &  5.5 hr &3.5 hr & 6.5 hr & 3.5 + 6.0 hr & 4.0 hr \\
Beam size                                        &  $26\arcsec \times  22\arcsec$ &$16\arcsec \times 12\arcsec$ &  $12\arcsec \times  8.7\arcsec$ & $5.1\arcsec \times 4.1\arcsec$ & $5.8\arcsec \times 2.0\arcsec$  \\
Rms noise ($\sigma_{\rm{rms}}$) & 0.92~mJy~beam$^{-1}$ & 300~$\mu$Jy~beam$^{-1}$ &  79~$\mu$Jy~beam$^{-1}$ &  26~$\mu$Jy~beam$^{-1}$ & 32~$\mu$Jy~beam$^{-1}$ \\
Briggs weighting  & 0.5 & 0.5 & 0.0 & 0.0 & 0.5\\
\hline
\end{tabular}
\label{tab:rx42gmrtobservations}
\end{center}
\end{table*}

Although there is now substantial evidence that radio relics and halos are related to galaxy cluster mergers \citep[e.g.,][]{2010ApJ...721L..82C}, the detailed physics are still not understood. Radio spectra are a crucial way to separate between the physical models for the origin of radio relics and halos. Obtaining high-quality spectra is however difficult as these sources are extended and have a low surface brightness. In particular, spectral curvature provides important information about the underlying physical processes. However, curvature in the synchrotron spectrum is only observable if the observed frequency range is wide enough. In addition, obtaining images at matched resolutions over a wide-frequency range is often not possible.

We  carried out an observing campaign to search for new diffuse radio sources in clusters \citep{2011A&A...528A..38V, 2011A&A...533A..35V}.  By inspecting the WENSS \citep{1997A&AS..124..259R} and NVSS \citep{1998AJ....115.1693C} surveys we came across the complex radio source B3~0559+422B. The radio source roughly coincided with an extended X-ray source in the ROSAT All Sky Survey. In addition, an overdensity of galaxies following the X-ray emission, was visible in 2MASS images. The diffuse radio source, extended X-ray emission and galaxy overdensity strongly suggested the presence of a previously unidentified  galaxy cluster located at moderate redshift. In this paper we present detailed radio observations of this new galaxy cluster 1RXS~J0603.3+4214 and investigate the spectral and polarimetric properties of the diffuse emission.

The layout of this paper is as follows. In Sect.~\ref{sec:rx42obs-reduction} we give an overview of the observations and the data reduction. The WSRT and GMRT images are presented in  Sect.~\ref{sec:rx42results}. The radio spectra and polarization data are analyzed in Sects~\ref{sec:rx42spectralindex} to \ref{sec:rx42pola}. We end with a discussion and conclusions in Sects.~\ref{sec:rx42discussion} and \ref{sec:rx42conclusion}.

Throughout this paper we assume a $\Lambda$CDM 
cosmology with $H_{0} = 71$~km~s$^{-1}$~Mpc$^{-1}$, $\Omega_{m} = 0.3$, and $\Omega_{\Lambda} = 0.7$. All images are in the J2000 coordinate system.

\section{Observations \& Data Reduction}
\label{sec:rx42obs-reduction}

\subsection{GMRT observations}
GMRT observations were taken using the GMRT software backend \citep[GSB; ][]{2010ExA....28...25R}. An overview of the observations is given in Table~\ref{tab:rx42gmrtobservations}.

The data were reduced with the  NRAO Astronomical Image Processing System (AIPS) package. The data were visually inspected for the presence of {radio frequency interference (RFI)} which was subsequently removed. For the 147 and 241 MHz data, RFI was fitted and subtracted using the technique described by \cite{2009ApJ...696..885A}, implemented in  Obit \citep{2008PASP..120..439C}. Standard bandpass and gain calibration were carried out, followed by several rounds of phase self-calibration 
and two final rounds of amplitude and phase self-calibration. The fluxes for the calibrator sources were set by the \cite{perleyandtaylor} extension to the \cite{1977A&A....61...99B} scale. Images were made using ``briggs'' weighting \citep{briggs_phd}, see Table~\ref{tab:rx42gmrtobservations}.  
were cleaned down to $2$ times the rms noise level ($2\sigma_{\mathrm{rms}}$) within the clean boxes and corrected for the primary beam attenuation\footnote{http://gmrt.ncra.tifr.res.in/gmrt\_hpage/Users/doc/manual/
 
 UsersManual/node27.html}. For more details about the data reduction see  \cite{2011A&A...527A.114V}. The 147~MHz data were further calibrated for ionospheric phase distortions, as these can become quite severe at this frequency, with the SPAM package \citep{2009A&A...501.1185I}. At 325~MHz we removed several sources using the ``peeling''-method \citep[e.g.,][]{2004SPIE.5489..817N}.
{We assume a 5\% uncertainty in the calibration of the absolute flux-scale \citep{2004ApJ...612..974C}. }

\subsection{WSRT observations}
\label{sec:rx42wsrt}
\begin{table*}
\begin{center}
\caption{WSRT observations}
\begin{tabular}{llllll}
\hline
\hline
& 25 cm, 1221~MHz & 21 cm, 1382~MHz& 18 cm, 1714~MHz& 13 cm, 2272~MHz & 6 cm, 4.9~GHz\\
\hline
Bandwidth            & $8\times20$~MHz & $8\times20$~MHz  & $8\times20$~MHz  & $8\times20$~MHz &$8\times20$~MHz \\
Number of channels per IF                     & 64 &  64  &  64 & 64  &64 \\
Polarization	                 & XX, YY, XY, XY & XX, YY, XY, XY& XX, YY, XY, XY&RR, LL, RL, LR&  XX, YY, XY, XY \\
Observation dates	      & Aug 28, 2010 &Sep 10, 2010& Sep 3, 2010 & Sep 9, 2010 &Sep 2 \&6, 2010\\ 
Total on-source time     & 12~hr&12~hr&12~hr&12~hr&24hr\\
Beam size                      &$29.0\arcsec \times 19.0\arcsec$ &$27.1\arcsec \times 16.4\arcsec$ &$23.6\arcsec \times 15.6\arcsec$ &$16.1\arcsec\times9.8\arcsec$ & $7.0\arcsec \times 4.7\arcsec$ \\
Rms noise ($\sigma_{\rm{rms}}$) & 39~$\mu$Jy~beam$^{-1}$ &  27~$\mu$Jy~beam$^{-1}$ &  25~$\mu$Jy~beam$^{-1}$  &40~$\mu$Jy~beam$^{-1}$&41~$\mu$Jy~beam$^
{-1}$ \\
Briggs weighting  & 0.0 & 0.5 & 1.0 & 0.5 & 0.5\\
\hline
\hline
\end{tabular}
\label{tab:rx42wsrtobservations}
\end{center}
\end{table*}
WSRT observations of 1RXS~J0603.3+4214 were taken in the L-band, 13cm and 6cm bands,  
see Table \ref{tab:rx42wsrtobservations} for details. The data were calibrated with the CASA\footnote{http://casa.nrao.edu/} package. We first removed time ranges affected by shadowing and RFI. The data were then calibrated for the bandpass response and subsequent gain solutions were transferred to the target source. Channel dependent leakage terms were found using an unpolarized calibrator source and the polarization angles were determined from 3C286.  The fluxes for the calibrator sources were set by the \cite{perleyandtaylor} extension to the \cite{1977A&A....61...99B} scale.  The data were exported into AIPS for several rounds of phase self-calibration, followed by two rounds of amplitude and phase self-calibration, see also \cite{2011A&A...527A.114V}. Images were cleaned with manually placed clean boxes and corrected for the primary beam attenuation. A deep image was created by combining the images from individual IF's from the 18, 21, and 25 cm bands. The images were convolved to the same resolution and combined with a spectral index scaling of $-1$. The resolution of this combined image is $29\arcsec~\times~19\arcsec$. {We assume a 5\% uncertainty in the calibration of the absolute flux-scale.} 

We used the WSRT L-band observations to perform RM-Synthesis (see  Sect.~\ref{sec:rx42rmsyn}). For this, we created full polarization maps of every two neighboring frequency channels (i.e.,  the image bandwidth is 62.5~kHz). We inspected all these images and removed maps that had high noise levels, were affected by RFI or had other artifacts. In the end, 280 channels maps between 1170 and 1786~MHz were retained for RM-synthesis.

\subsection{WHT spectroscopy \& imaging}
Optical images of 1RXS~J0603.3+4214  were taken with the PFIP camera on the 4.2m  William Herschel Telescope (WHT) between 15 and 19 April, 2009 with V, R and I filters. The seeing varied between 1.0\arcsec~and 1.5\arcsec and the total integration time  was  1500~s per filter. The data were reduced with IRAF \citep{1986SPIE..627..733T, 1993ASPC...52..173T} and the \emph{mscred} package \citep{1998ASPC..145...53V}. Images were flat-fielded and corrected for the bias offset. R and I band images were also fringe corrected. The individual exposures were averaged, with pixels being rejected above $3.0\sigma_{\mathrm{rms}}$ to remove cosmic rays and other artifacts. The optical images were strongly affected by the bright star BD+42~1474 (V$_{\rm{mag}}$= 8.62), see Fig.~\ref{fig:rx42optical}. 

To determine the redshift of the cluster, WHT ISIS spectra of galaxies located on the optical images were taken on 10 and 11 February, 2011. For the blue arm we used the R300B grating and for the red arm the R316R grating. The slit-width was 1.5\arcsec. Flat-fielding, bias correction, and  wavelength calibration were performed in IDL\footnote{http://www.ittvis.com}. The total exposure time per galaxy was  1500~s in both the blue and red arms. In the end, the spectra for the blue and red arm were merged into single spectrum for each galaxy.

\section{Results: redshift, X-rays, and radio continuum maps}
\label{sec:rx42results}

\subsection{Redshift of  1RXS~J0603.3+4214}
\label{sec:rx42spectrum}

The brightest cluster members are visible in the 2MASS survey images \citep{2006AJ....131.1163S} and some of them are also listed in the 2MASS Extended sources catalog \citep{2003yCat.7233....0S}. The cluster is not listed in NED or SIMBAD and probably remained unidentified because it is located relatively close to the galactic plane at a galactic latitude $b= 9.4\degr$ and nearby the bright star BD+42~1474.

No published redshifts are available for any of the galaxies in the cluster. Galaxies for which WHT ISIS spectra were obtained are listed in Table~\ref{tab:rx42redshift} and the spectra are shown in Fig.~\ref{fig:rx42spectra}. The spectra are typical for passive red elliptical galaxies mostly found in clusters. We find that the five galaxies are located at $0.220\leq z \leq 0.228$. Taking the average value we adopt $z=0.225 \pm 0.04$ for the cluster, with the uncertainty in the redshift given by the standard deviation.

\begin{table}
\begin{center}
\caption{Redshifts}
\begin{tabular}{lll}
\hline
\hline
Galaxy & 2MASS K mag & $z$ \\
\hline
2MASX J06031667+4214416 & 13.204 & 0.227\\  
2MASX J06030757+4216215 & 13.610 & 0.222\\  
2MASS J06032605+4214050 & 15.075 & 0.228\\  
2MASS J06030644+4215241 & 15.201 & 0.227\\  
2MASX J06032432+4209306 & 13.858 & 0.220\\  
\hline
\hline
\end{tabular}
\label{tab:rx42redshift}
\end{center}
\end{table}

\begin{figure*}
\begin{center}
\includegraphics[angle =90, trim =0cm 0cm 0cm 0cm,width=0.99\textwidth]{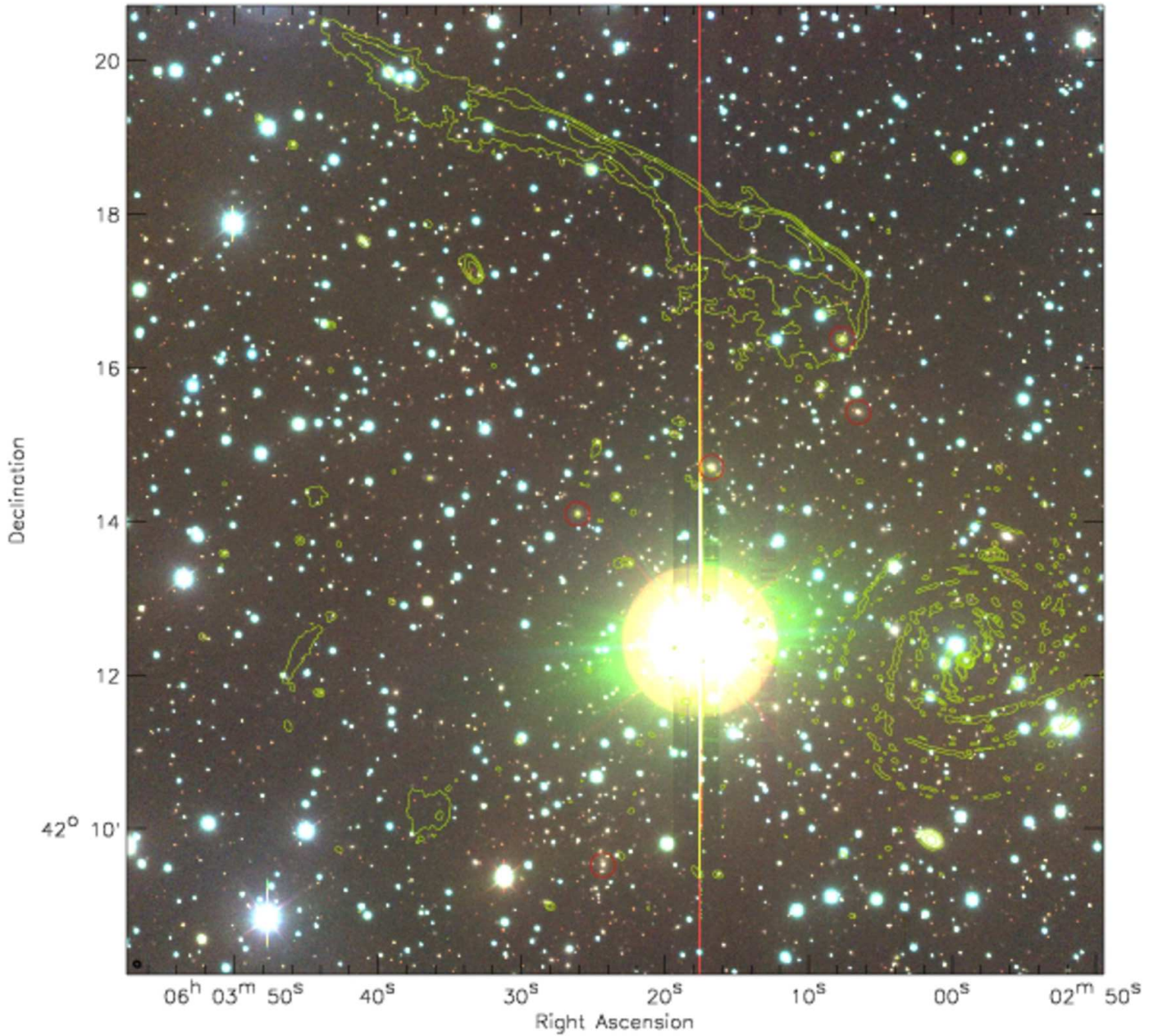}
\end{center}
\caption{WHT V, R, I color image. Yellow contours are from the GMRT 610~MHz image (Fig.~\ref{fig:rx42gmrt610}) and drawn at levels of ${[1, 2, 4, 8, \ldots]} \times 0.15$~mJy~beam$^{-1}$. Galaxies for which spectra were obtained are marked with red circles, see also Table~\ref{tab:rx42redshift} and Fig.~\ref{fig:rx42spectra}. The spiral pattern (yellow contours) at the bottom right is due to residual calibration errors around the source B3~0559+422A.}
\label{fig:rx42optical}
\end{figure*}

\begin{figure}
\begin{center}
\includegraphics[angle =90, trim =0cm 0cm 0cm 0cm,width=0.49\textwidth]{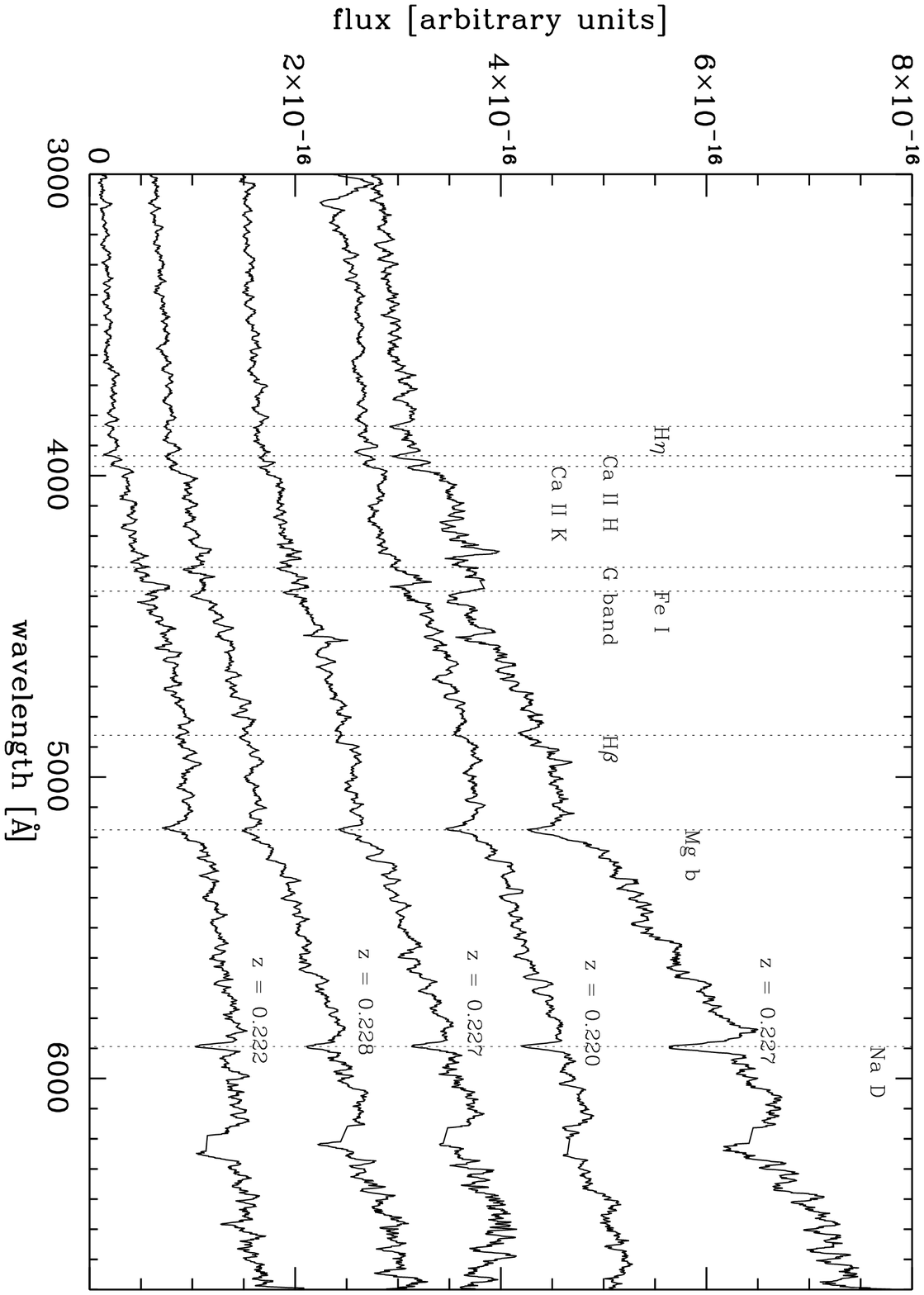}
\end{center}
\caption{Rest-frame WHT ISIS spectra for the galaxies listed in Table~\ref{tab:rx42redshift}. The order (from top to bottom) is the same as in Table~\ref{tab:rx42redshift}. Various absorption features are indicated.}
\label{fig:rx42spectra}
\end{figure}

\subsection{X-ray emission from the ICM}
\label{sec:rx42xray}
1RXS J0603.1+4214 is seen the ROSAT All-Sky Survey as an extended source and listed as  1RXS~J060313.4+421231, 1RXS~J060322.3+421305, and 1RXS~J060314.8+421439 \citep{1999A&A...349..389V, 2000IAUC.7432R...1V}, see  Fig.~\ref{fig:rx42wsrtlband} (left panel). 
Using the redshift and ROSAT count rate (0.21 PSPC cts~s$^{-1}$) we find an X-ray 
luminosity ($L_{\rm{X,~0.1-2.4~keV}}$) of $\sim 1 \times 10^{45}$~erg~s$^{-1}$.  
The X-ray emission is extended in the north-south direction. Additional emission extends to the east and west of the main X-ray peak. The high X-ray luminosity and extended emission are consistent with a massive cluster undergoing a major merger event.

\begin{figure*}
\begin{center}
\includegraphics[angle =90, trim =0cm 0cm 0cm 0cm,width=0.49\textwidth]{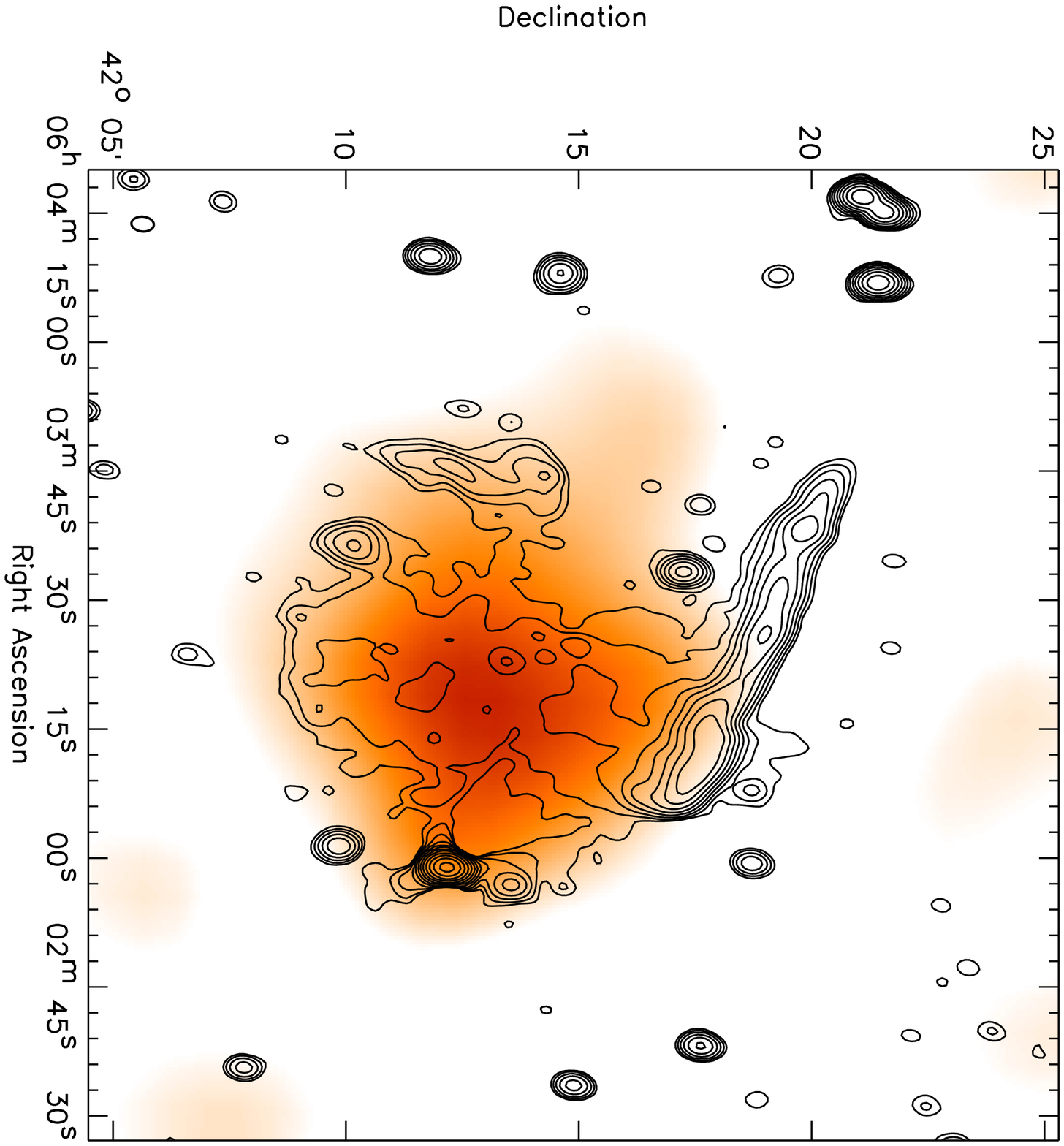}
\includegraphics[angle =90, trim =0cm 0cm 0cm 0cm,width=0.49\textwidth]{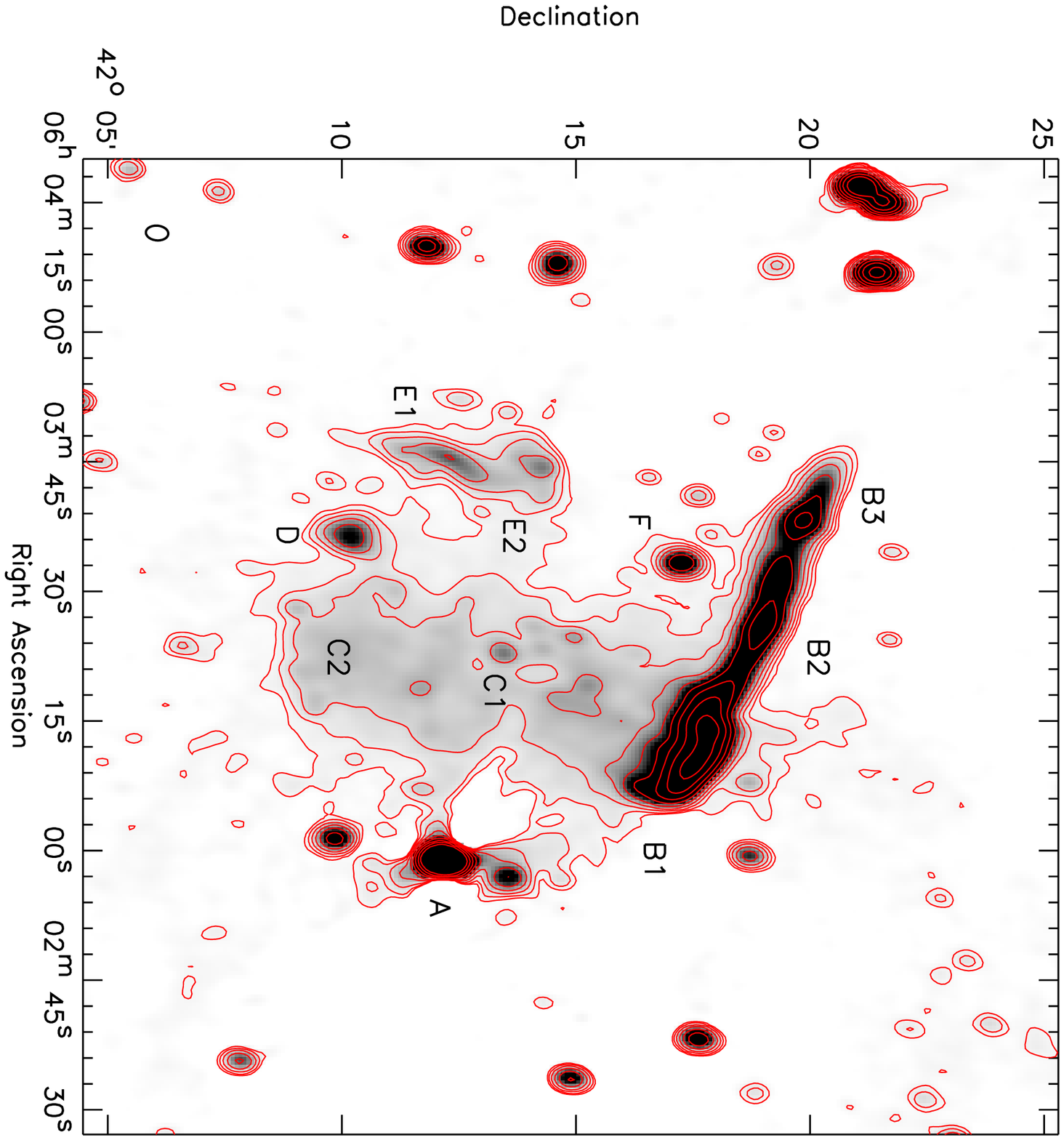}
\end{center}
\caption{Left: X-ray emission from ROSAT. The  image from the ROSAT All Sky Survey was smoothed with a 200\arcsec~FWHM Gaussian and is shown in orange colors. Solid contours are from the WSRT L-band image and drawn at levels of ${[1, 2, 4, 8, \ldots]} \times 0.15$~mJy~beam$^{-1}$. Right: WSRT L-band image (1160--1780~MHz) with sources labelled. Contour levels are drawn at $[1, 2, 4, 8, \ldots] \times 80$~$\mu$Jy~beam$^{-1}$. The beam size is $29\arcsec \times 19\arcsec$ and indicated in the bottom left corner of the image.}
\label{fig:rx42wsrtlband}
\end{figure*}

\subsection{Radio continuum maps}
\label{sec:rx42continuum}
The WSRT L-band map is displayed in Fig.~\ref{fig:rx42wsrtlband}. We labeled some of the most prominent sources in this image. The GMRT 147, 241, 325, 610, 1280 MHz, and WSRT 21 and 13~cm images are shown in Figs.~\ref{fig:rx42gmrt610} to \ref{fig:rx42gmrt241} (for the 4.9~GHz image see Fig.~\ref{fig:rx42pol}). The properties of the diffuse source in the cluster are summarized in Table~\ref{tab:rx42diffuse}. 

Source A (B3~0559+422A) is a compact flat spectrum source with a flux of 0.29 Jy at 1.4 GHz \citep{2007MNRAS.376..371J,  2002ApJS..141...13B,  1999A&AS..135..273M, 1992MNRAS.254..655P}.  The radio source is associated with a star-like object in the WHT images and listed as a quasar by \cite{2009A&A...505..385A}, but no redshift is reported. Source F is a compact source (but resolved in the 1280 and 610~MHz images)  and has an optical counterpart in the WHT images. 

The most prominent source in the GMRT and WSRT images is B3 0559+422B (source B). The source is also visible in the 74 MHz VLSS survey \citep{2007AJ....134.1245C} and listed as \object{VLSS~J0603.2+4217}. In the L-band image, the source has a largest angular size of 8.7 \arcmin, which corresponds to a physical size of almost 1900~kpc. The source consists of a bright western part (B1) and a fainter linear extension to the northeast (B2, B3). The radio emission brightens and fades  two times along this extension, while the width of the source also varies.
In the high-resolution 1280, 610, and 325 MHz images, the source displays a complex filamentary morphology. Some ``streams'' of emission extent from the northern part of B1 to the south. The northern boundary of the source is sharp, while the emission fades more slowly at the southern part. We classify source B as a radio relic because of the lack of optical counterparts, its polarization and spectral properties (see Sects. \ref{sec:rx42spectralindex} to \ref{sec:rx42pola}), and the large physical elongated size and peripheral location in the cluster. 

Extending from B1 to the south, there is a patch of low surface brightness emission (C: subdivided into C1and C2). A hint of emission is visible north of B1 in the L-band and 325~MHz images, possibly associated with C1. The surface brightness decreases more rapidly at the south side of C (C2) than at the east and west sides. If we consider the faint emission north of B1 to be part of C then the source has a largest angular size of about 10\arcmin~(i.e., a physical extent of 2.1~Mpc). Only counting the emission south of B1 the source extends about 8\arcmin. 
 We classify source C as the giant elongated radio halo because it roughly follows the X-ray emission from the ICM and has different spectral index properties than source B.  The radio power of $6.8 \times 10^{24}$~W~Hz$^{-1}$ falls on the $L_{\rm{X}}$--$P_{\rm{1.4GHz}}$ correlation for giant radio halos \citep[e.g.,][]{2000ApJ...544..686L,2006MNRAS.369.1577C}. However, at the most southern part of C2 the surface brightness increases slightly. This is best seen in the 147 and 241~MHz images (Fig.~\ref{fig:rx42gmrt241}). The southern end of C2 could therefore be the ``counter'' radio relic of B1 (see Sect.~\ref{sec:rx42discussion}). Although, we can also not exclude the possibility that it is part of the radio halo.
 
Another diffuse elongated source is located to the east of C1. The source consist of two parts, E1 and E2. We classify source E as a radio relic because we could not find an  optical counterpart associated with it and it is located at the eastern boundary of the X-ray emission. The extent of E is 4\arcmin which corresponds to 860~kpc.  Source D is another diffuse source without any optical counterpart in the WHT images, which we also classify as a radio relic. It has a largest angular size of 1.0\arcmin~in the GMRT 610 MHz image and displays hints of a complex morphology (Fig.~\ref{fig:rx42610cut}), very different from a typical radio galaxy.

\begin{table}
\begin{center}
\caption{Diffuse radio sources in 1RXS~J0603.3+4214}
\begin{tabular}{lllll}
\hline
\hline
Source & $S_{\rm{1382 MHz}}$ & $P_{\rm{1.4 GHz}}$ &LLS$^{a}$ & $\alpha$\\
              &      mJy                                   &    $10^{24}$~W~Hz$^{-1}$  &kpc    \\
\hline
B&$319.5\pm20.8$ &60& 1870 & $-1.10 \pm 0.02$ \\
C&$35.9\pm 2.6$ & 6.8& 1700--2100  & $-1.15 \pm 0.06$\\
D&  $5.38\pm 0.31$& 1.0&215  & $-1.10 \pm 0.05$\\
E& $9.48\pm0.97$ & 1.8 & 860 & $-1.0 \pm 0.2$\\
\hline
\hline
\end{tabular}
\label{tab:rx42diffuse}
\end{center}
$^{a}$ largest linear size\\
\end{table}

\begin{figure*}
\begin{center}
\includegraphics[angle =90, trim =0cm 0cm 0cm 0cm,width=1.0\textwidth]{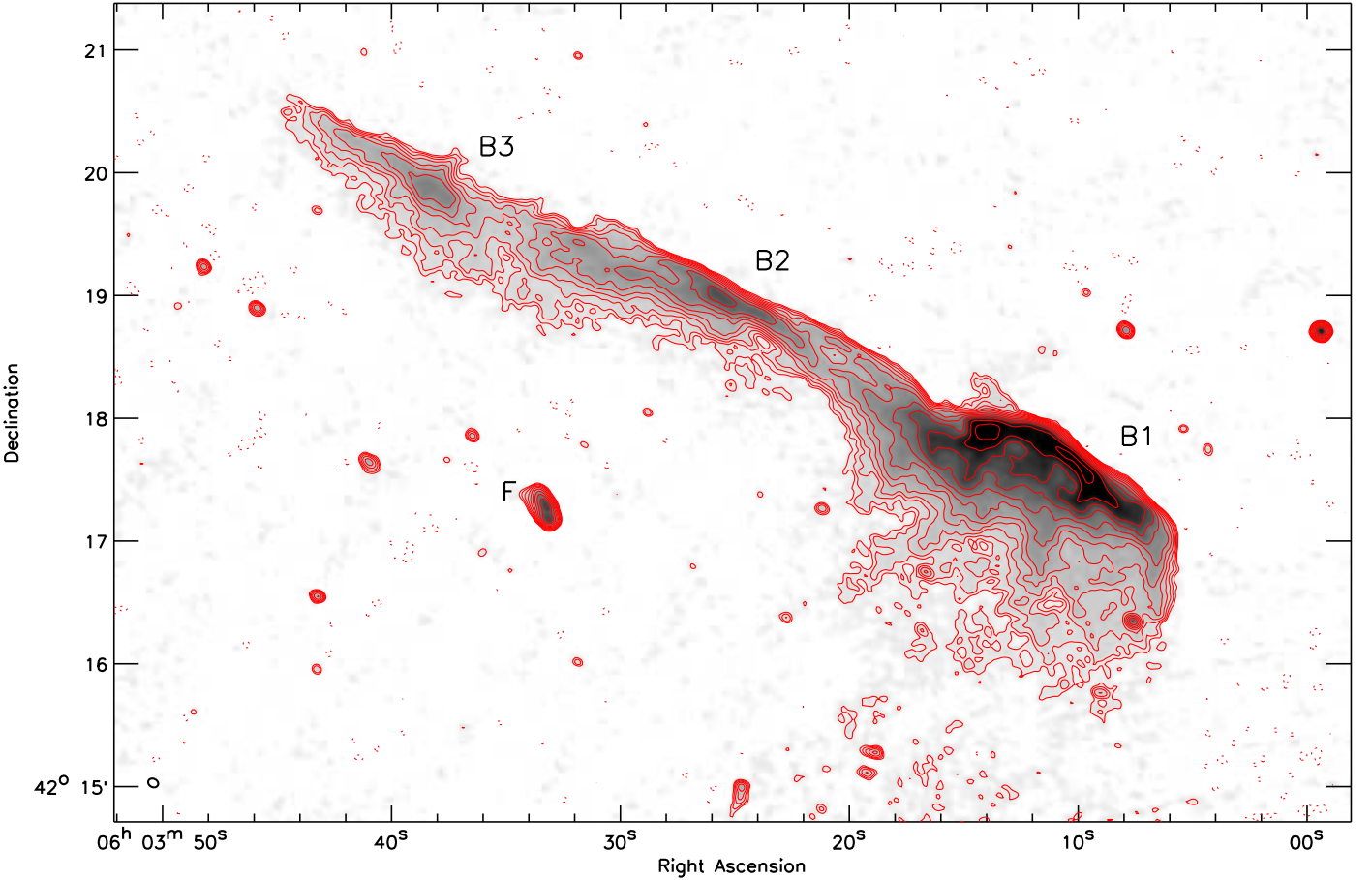}
\end{center}
\caption{GMRT 610 MHz image. Contour levels are drawn at $\sqrt{[1, 2, 4, 8, \ldots]} \times 4\sigma_{\mathrm{rms}}$.  Negative  $-3\sigma_{\mathrm{rms}}$ contours are shown by the dotted lines. The beam size is $5.1\arcsec \times 4.1\arcsec$ and indicated in the bottom left corner of the image.}
\label{fig:rx42gmrt610}
\end{figure*}

\begin{figure}
\begin{center}
\includegraphics[angle =90, trim =0cm 0cm 0cm 0cm,width=0.5\textwidth]{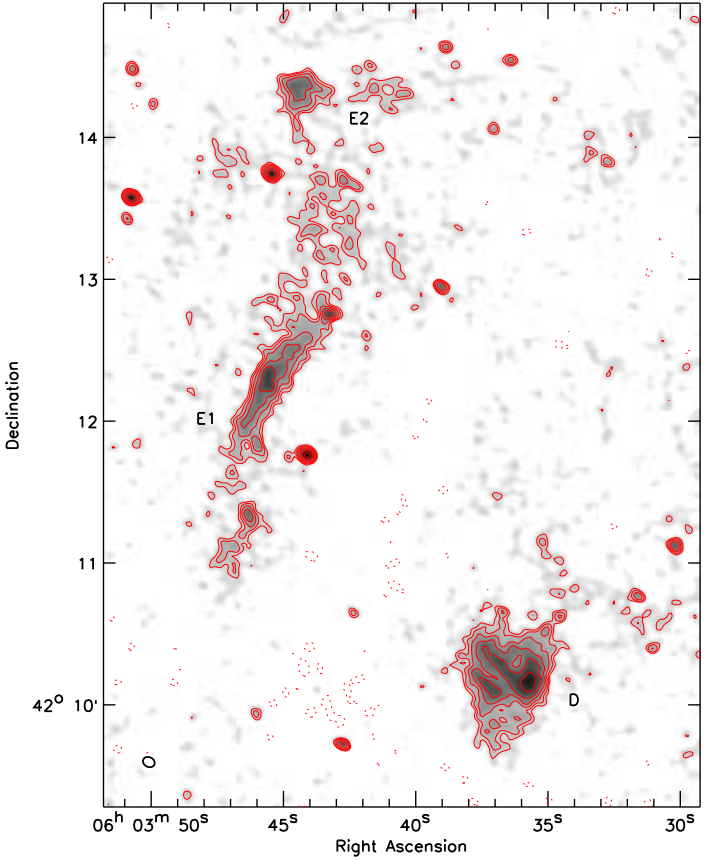}
\end{center}
\caption{GMRT 610 MHz cutout around sources D and E. Contours are drawn as in Fig.~\ref{fig:rx42gmrt610}. }
\label{fig:rx42610cut}
\end{figure}

\begin{figure*}
\begin{center}
\includegraphics[angle =90, trim =0cm 0cm 0cm 0cm,width=0.49\textwidth]{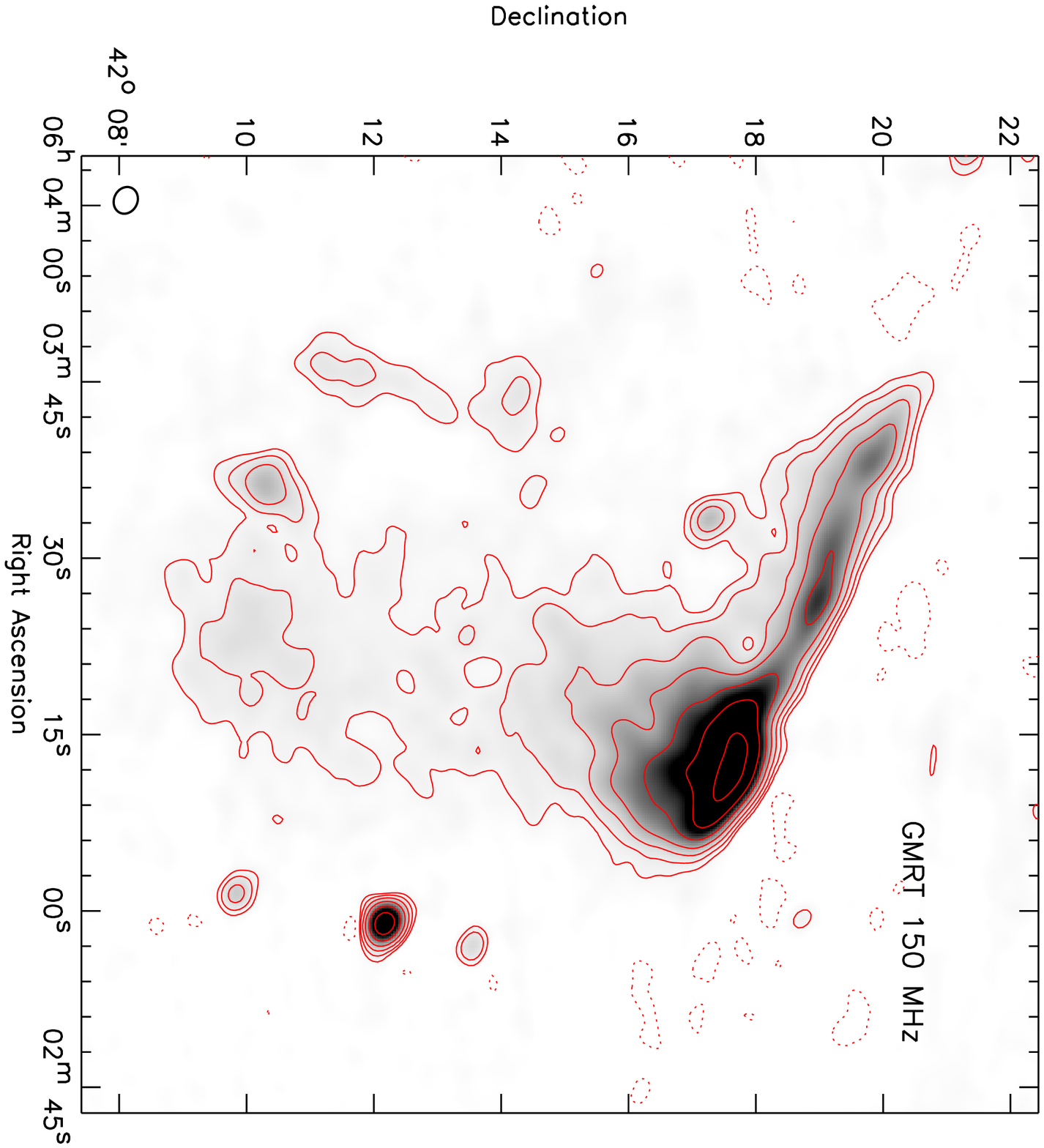}
\includegraphics[angle =90, trim =0cm 0cm 0cm 0cm,width=0.49\textwidth]{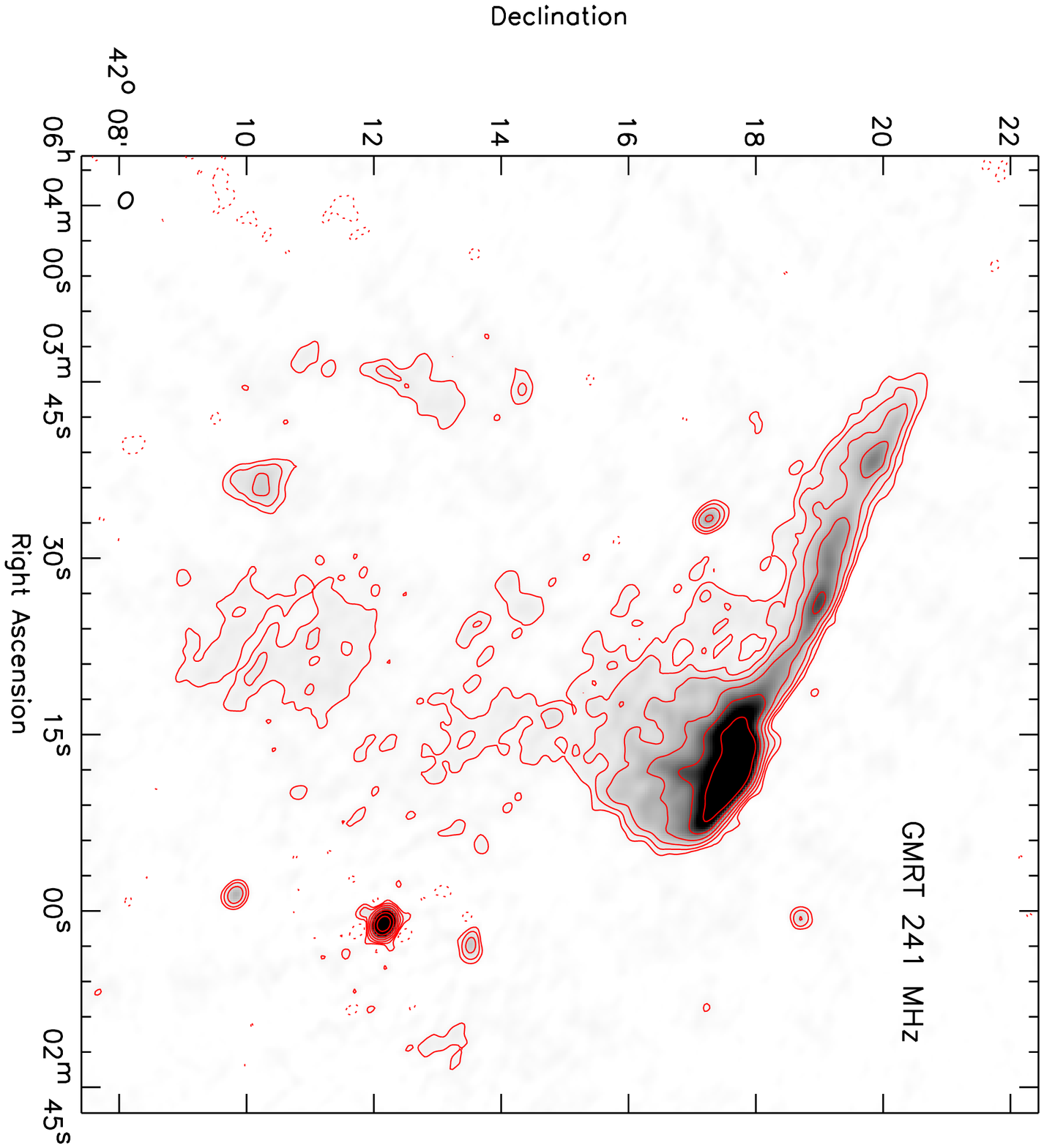}
\includegraphics[angle =90, trim =0cm 0cm 0cm 0cm,width=0.49\textwidth]{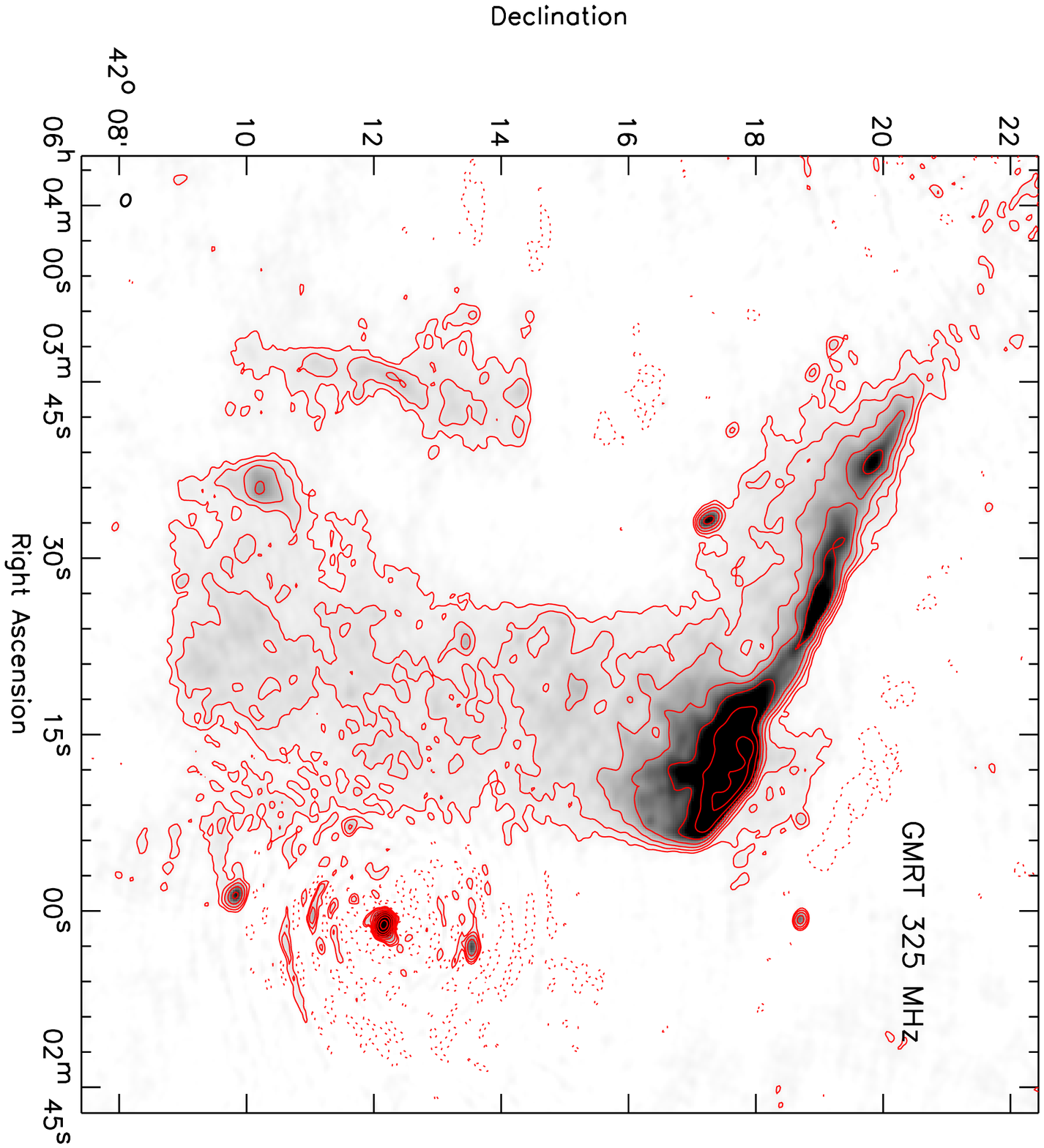}
\includegraphics[angle =90, trim =0cm 0cm 0cm 0cm,width=0.49\textwidth]{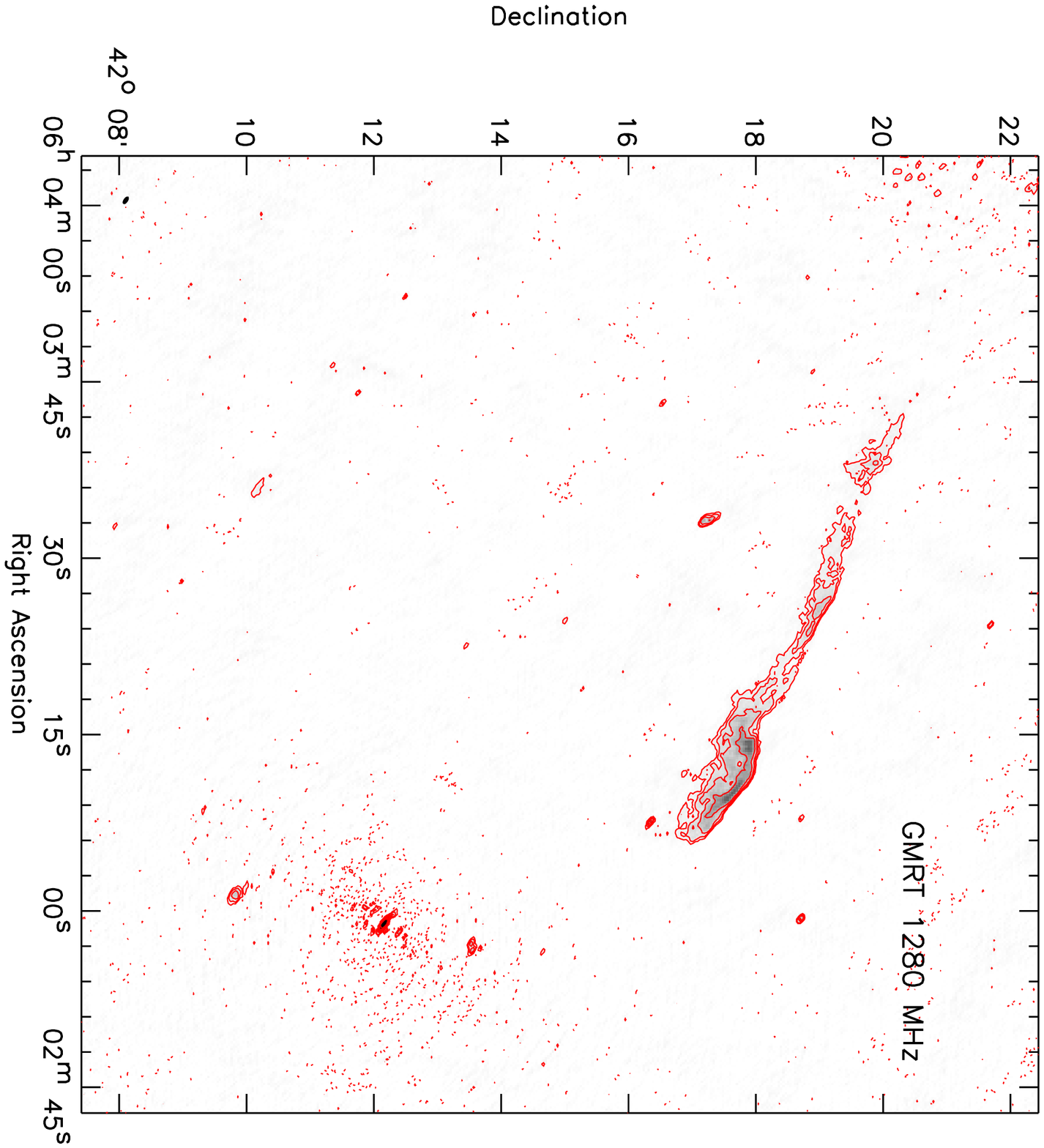}
\includegraphics[angle =90, trim =0cm 0cm 0cm 0cm,width=0.49\textwidth]{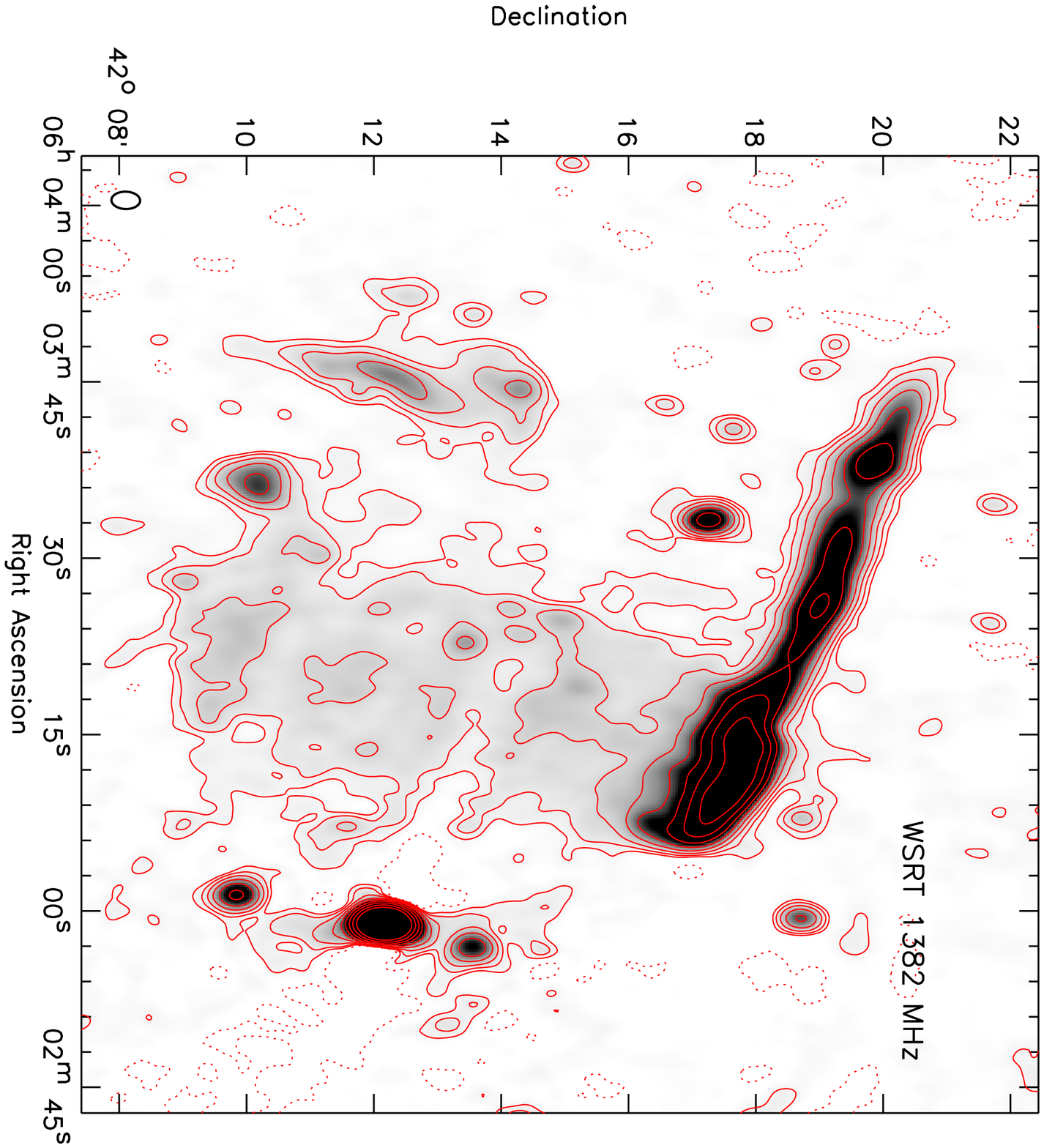}
\includegraphics[angle =90, trim =0cm 0cm 0cm 0cm,width=0.49\textwidth]{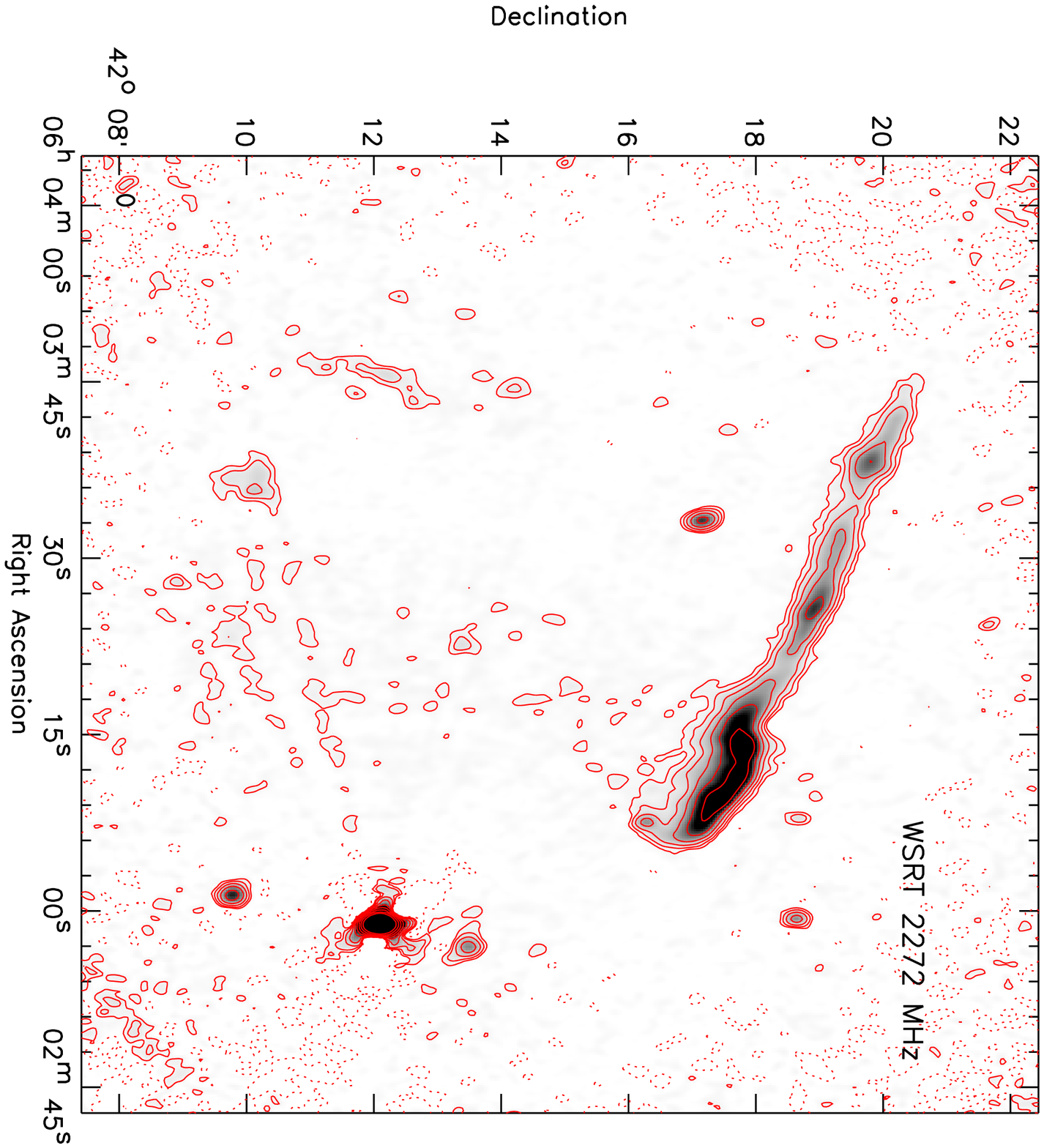}
\end{center}
\caption{{GMRT 150, 241, 325, 1280~MHz and WSRT 1382 and 2272~MHz images. Contour levels are drawn at $[1, 2, 4, 8, \ldots] \times 4\sigma_{\mathrm{rms}}$.  Negative  $-3\sigma_{\mathrm{rms}}$ contours are shown by the dotted lines. The beam size is shown in the bottom left corner of the images.}}
\label{fig:rx42gmrt241}
\end{figure*}


\section{Radio spectra}
\label{sec:rx42spectralindex}
 \subsection{Spectral index maps}
We created radio spectral index maps using the GMRT and WSRT data.
We  first made radio maps at roughly the same resolution applying suitable tapers in the uv-plane and using uniform weighting. We discarded data below 0.25~k$\lambda$ to select only common inner uvranges. The maps were then convolved to the same resolution. A high-resolution ($7.9\arcsec~\times~6.2\arcsec$) spectral index map between 610 and 325~MHz is shown in Fig. \ref{fig:rx42spix610-1280}. Pixels with values below  $5\sigma_{\mathrm{rms}}$ were blanked.
A medium-resolution ($20\arcsec~\times~18\arcsec$) spectral index map between 2272 and 147~MHz was also made by fitting a second order polynomial in $\log{(S)}-\log{(\nu)}$ space through the flux measurements at 2272, 1714, 1382, 1221, 610, 325, 241, and 147~MHz, see Fig.~\ref{fig:rx42spix_poly} (left panel). Pixels with values below  $1.5\sigma_{\mathrm{rms}}$ were blanked. To map the spectral index across the low surface brightness halo emission in the cluster, we also made a low-resolution spectral index map. We convolved the eight maps with a  $35\arcsec$ FWHM Gaussian (giving a resolution of approximately 40\arcsec) and fitted a power-law spectral index to minimize the number of fitted parameters. Pixels with a spectral spectral index error $> 0.7$ were blanked (Fig.~\ref{fig:rx42spix_poly}, right panel). {The errors in the spectral index maps are computed on the basis of the $\sigma_{\mathrm{rms}}$ values for the individual maps at the various frequencies and the reported flux calibration uncertainties in Sect.~\ref{sec:rx42obs-reduction}. When maps at more than two frequencies are used, it is assumed that the adopted fitting function provides a good physical description of the spectral shape. Figures displaying the corresponding errors in the spectral maps are show in Appendix~\ref{sec:spixerror}.}

The spectral index across relic B displays a clear north-south gradient, with a spectral index of about $-0.6$ to  $-0.75$ on the north side of the relic, steepening to $\sim-1.9$
at the south side of B2 and B3 and to $\lesssim-2.5$ for B1. The spectral index gradient is visible over the entire length of the relic. The high-resolution 610--325~MHz spectral index map reveals the same general trends for relic B as in the medium-resolution map. However, the SNR on the spectral index is somewhat lower, mainly because of the smaller frequency span. We note that the spectral index at the front of B2 and B3 has a steeper spectral index (about $-0.9$ to $-1.1$) in the region where the surface brightness drops (i.e., at the intersection B1--B2 and B2--B3).

The spectral index across Relic E varies mostly between $-1.0$ to $-1.2$. There are no  systematic trends visible across the relic, except that the north part of E3 has a somewhat flatter spectral index, see Fig.~\ref{fig:rx42spix_poly}.  The spectral index for relic D steepens from $-1.0$ to $-1.3$ from south to north. The spectral index across the radio halo (C) is difficult to determine as the SNR is low, but it roughly varies between  $-1.5$ and $-0.8$ with an average of $-1.1$. There is a hint of spectral flattening for the center of the halo, while to the north and south of it the spectral index is steeper on average.

\subsection{Integrated radio spectra}
We determined the integrated radio spectra of the components, taking the same maps which were used to create the medium-resolution ($20\arcsec~\times~18\arcsec$) spectral index map to minimize the effects of different uv-coverage. In addition, we added flux measurements at 74~MHz and 4.9~GHz for relic B from the VLSS survey and the WSRT observations. The radio spectrum is shown in Fig.~\ref{fig:rx42fluxB} and is well fitted by a single power-law with $\alpha = -1.10 \pm 0.02$ between 74 and 4900~MHz. For relics D and E we find integrated spectral indices of $-1.10\pm0.05$ and $-1.0 \pm 0.2$. For source C we find $\alpha = -1.15 \pm 0.06$. The integrated radio spectra do not show clear evidence for spectral breaks or turnovers, although the radio spectrum for E is poorly determined.


\begin{figure*}
\begin{center}
\includegraphics[angle =90, trim =0cm 0cm 0cm 0cm,width=1.0\textwidth]{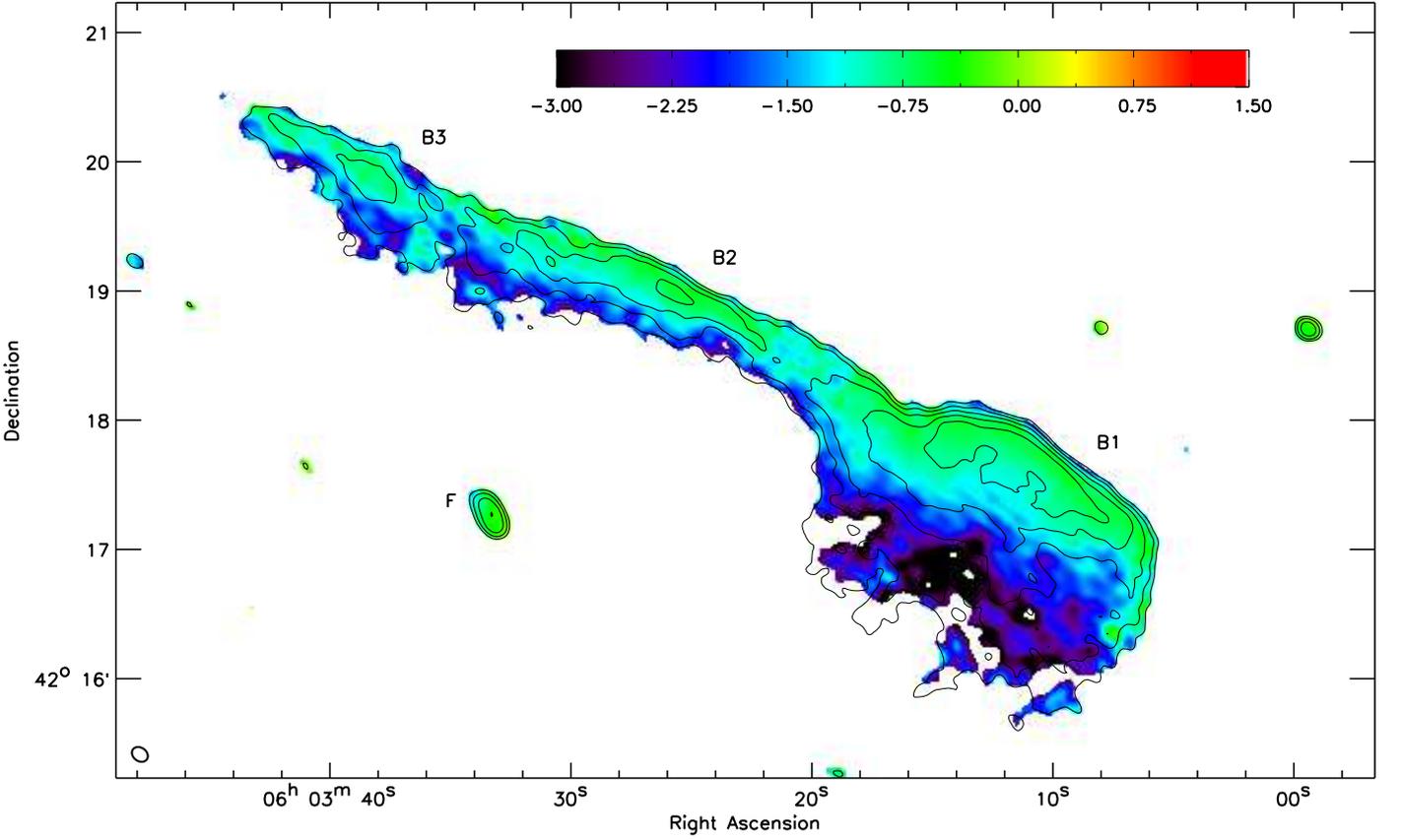} 
\end{center}
\caption{GMRT 610--325 MHz spectral index map. Contour levels from the GMRT 325~MHz image are drawn at $[1, 4, 16, 64, \ldots] \times 6\sigma_{\mathrm{rms}}$ and the spectral index map has a resolution of $7.9\arcsec \times 6.2\arcsec$. Pixels  below $5\sigma_{\mathrm{rms}}$ are blanked.}
\label{fig:rx42spix610-1280} 
\end{figure*}

\begin{figure*}
\begin{center}
\includegraphics[angle =90, trim =0cm 0cm 0cm 0cm,width=0.49\textwidth]{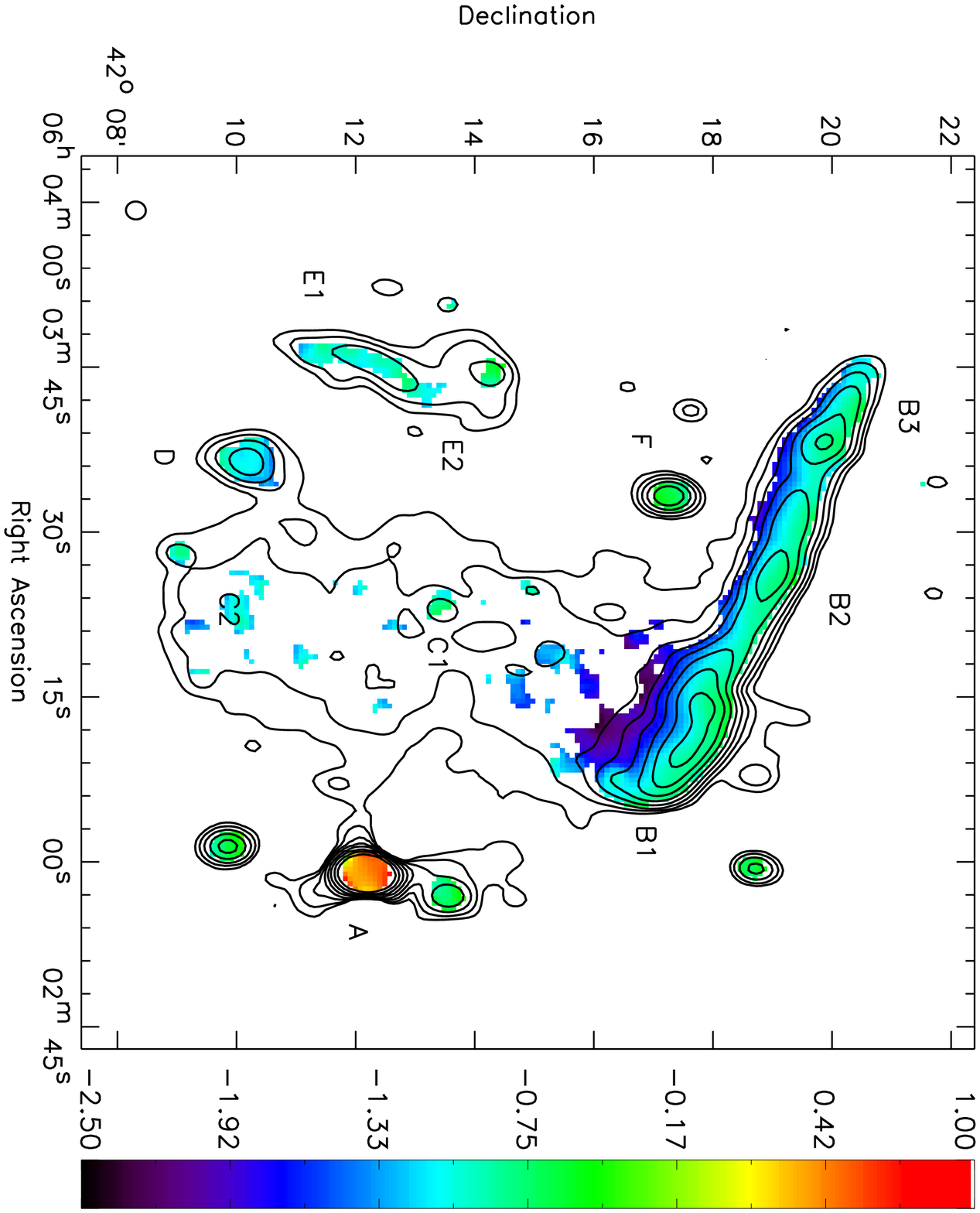}
\includegraphics[angle =90, trim =0cm 0cm 0cm 0cm,width=0.49\textwidth]{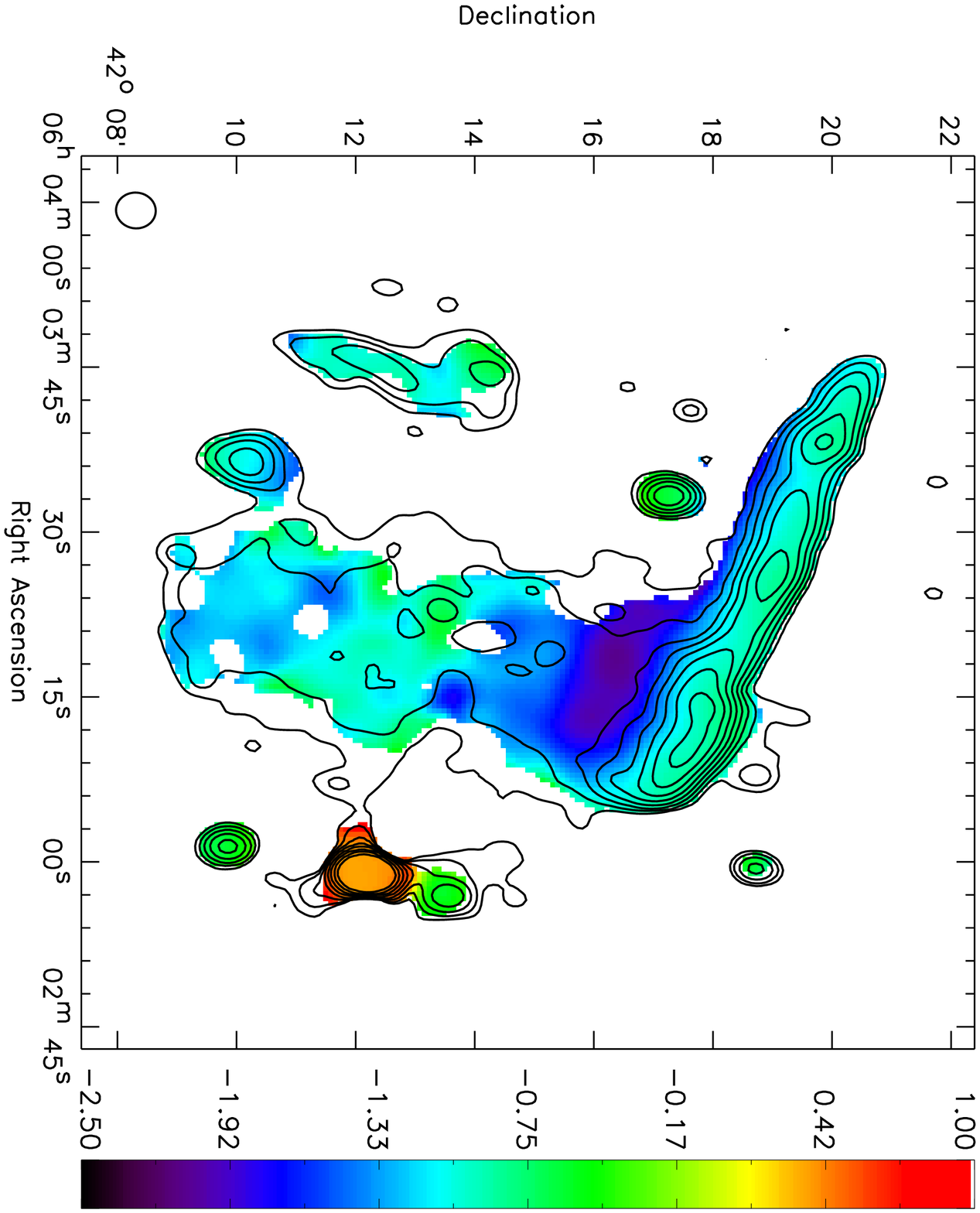}
\end{center}
\caption{Left: Fitted spectral index map between 2272 and 147~MHz. The spectral index map was made by fitting a second order polynomial in $\log{(S)}-\log{(\nu)}$  for each pixel in the maps at 2272, 1714, 1382, 1221, 610, 325, 241 and 147~MHz. Contour levels are drawn at $[1, 2, 4, 8, \ldots] \times 0.175$~mJy~beam$^{-1}$ and are from the L-band image in Fig.~\ref{fig:rx42wsrtlband}. The spectral index map has a resolution of $20\arcsec \times 18\arcsec$ and pixels  below $1.5\sigma_{\mathrm{rms}}$ were blanked. Right: Fitted spectral index (2272 and 147~MHz) for each pixel in the maps at 2272, 1714, 1382, 1221, 610, 325, 241 and 147~MHz. The individual maps were convolved with Gaussians of $35\arcsec$ FWHM and pixels with a spectral index error $> 0.7$ were blanked, contours are drawn as in the left panel.}
\label{fig:rx42spix_poly}
\end{figure*}

\begin{figure}
\begin{center}
\includegraphics[angle =90, trim =0cm 0cm 0cm 0cm,width=0.49\textwidth]{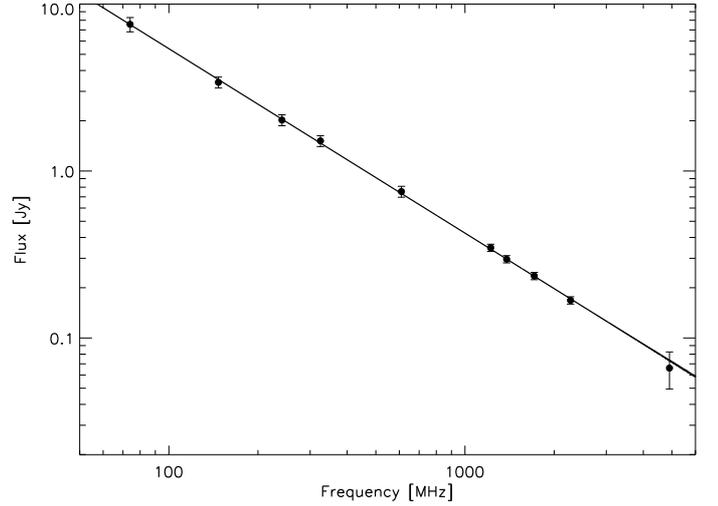}
\end{center}
\caption{Integrated radio spectrum for radio relic B.}
\label{fig:rx42fluxB}
\end{figure}

\begin{figure*}
\begin{center}
\includegraphics[angle =90, trim =0cm 0cm 0cm 0cm,width=0.33\textwidth]{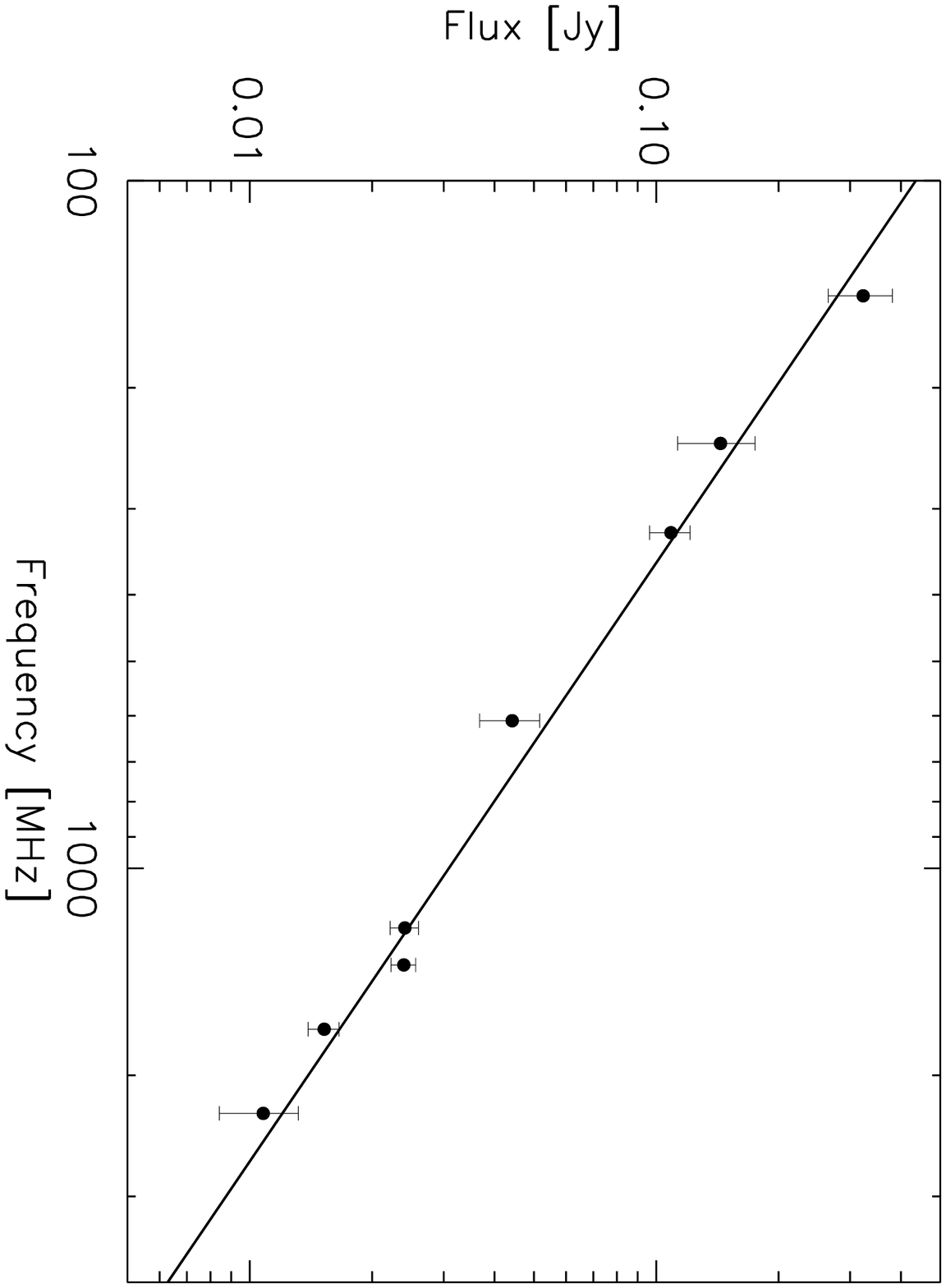}
\includegraphics[angle =90, trim =0cm 0cm 0cm 0cm,width=0.33\textwidth]{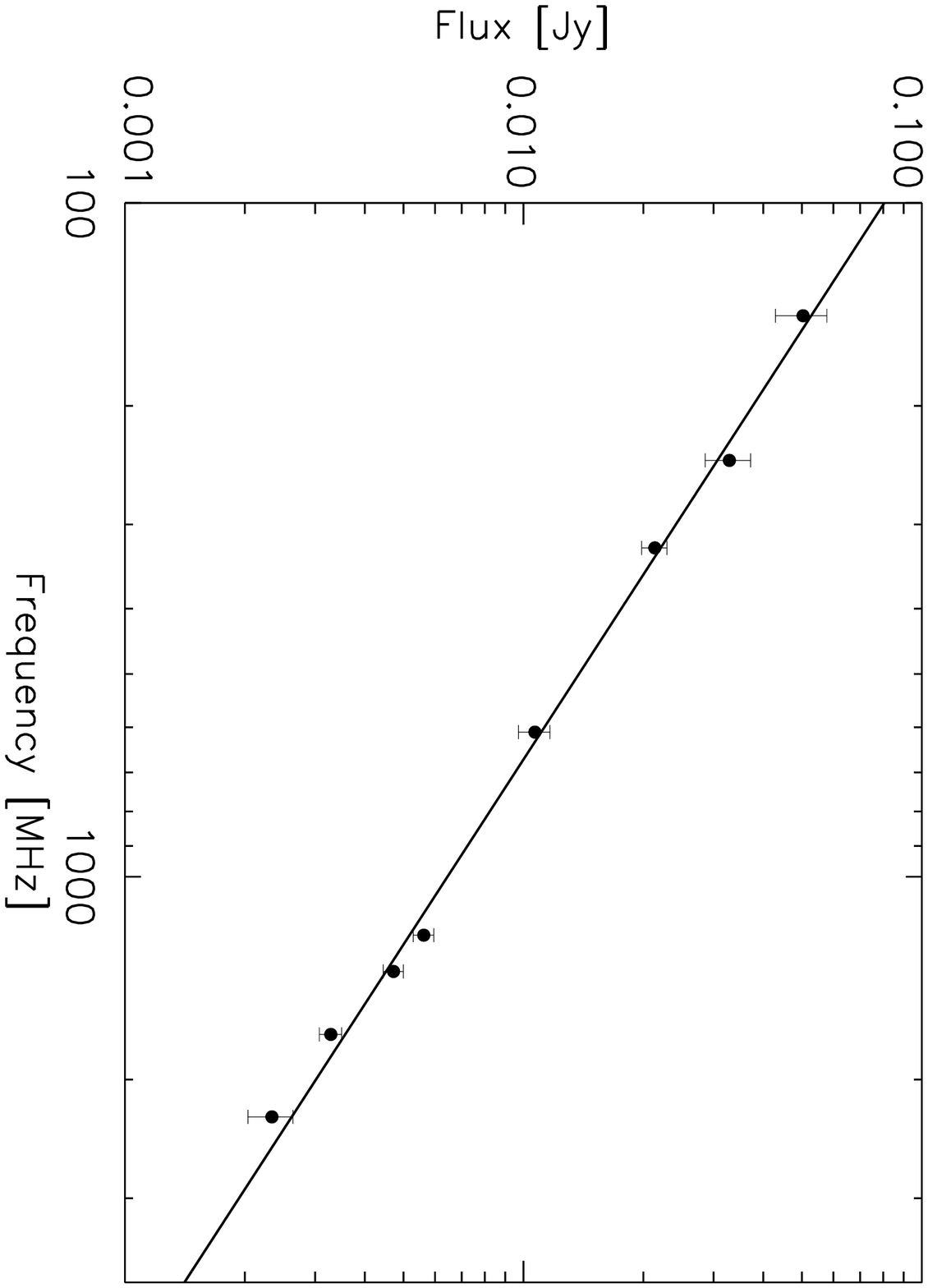}
\includegraphics[angle =90, trim =0cm 0cm 0cm 0cm,width=0.33\textwidth]{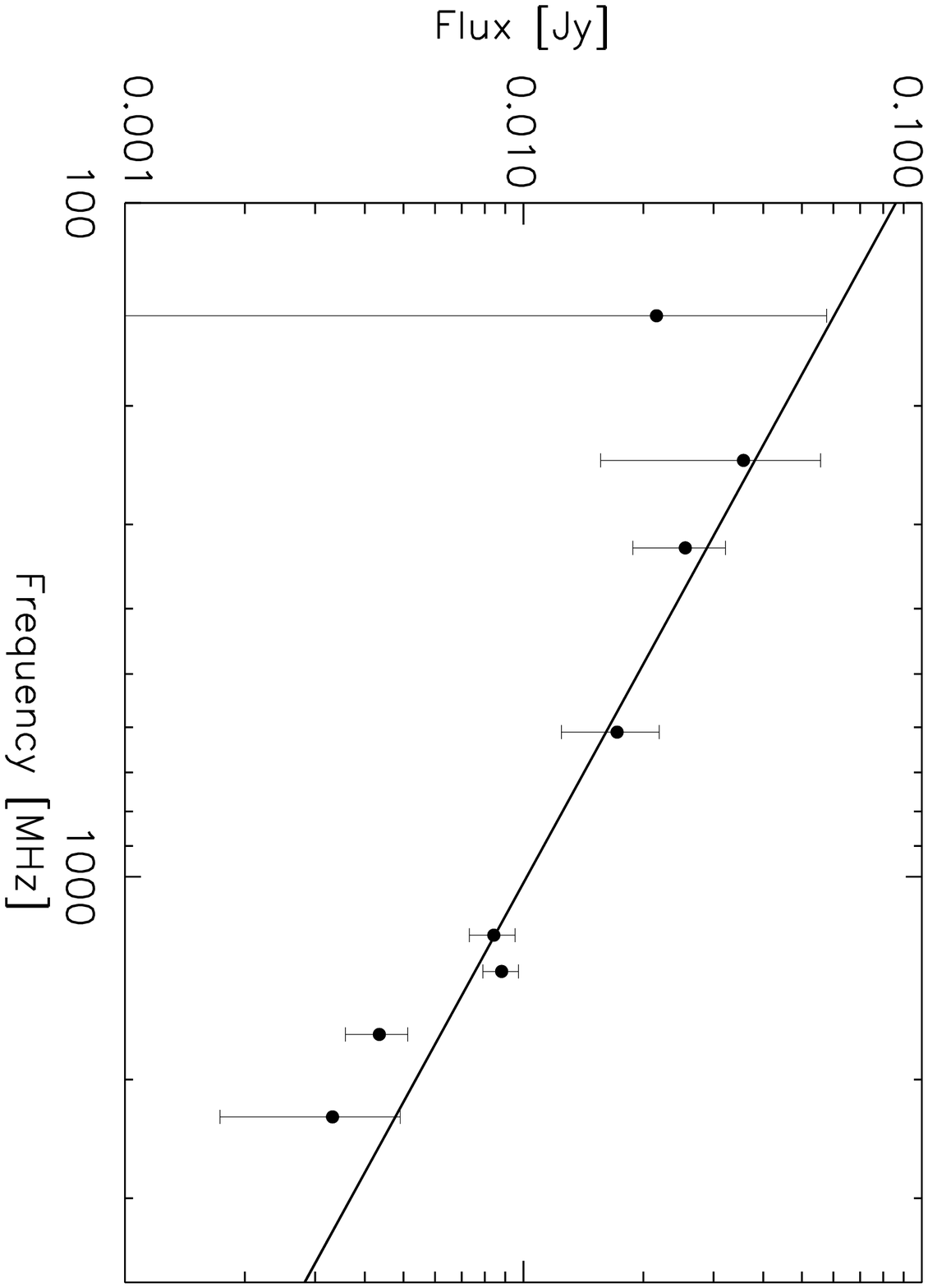}

\end{center}
\caption{Integrated radio spectra for sources C (left), D (middle), and E (right). The lines are fitted straight power-laws with indices given in Table~\ref{tab:rx42diffuse}.}
\label{fig:rx42fluxrest}
\end{figure*}

\section{Radio color-color diagrams}

We investigated the detailed spectral shape of the relic emission. We first divided the relic into two parts, B1 and B2+B3, which we analyzed separately. These parts were further subdivided into regions ($\alpha_{\rm{ref}}$) within a specific spectral index range of $0.1$ units, using the spectral index map from Fig.~\ref{fig:rx42spix_poly} (left panel).   The resulting spectra, including first order polynomial fits (in $\log{(I)}-\log{(\nu)}$ space) through the data points are shown in Fig.~\ref{fig:rx42flux_SB}. The fluxes were normalized by dividing by the number of pixels found in each region, i.e., we work in surface brightness units. 
Each spectra thus corresponds to a region from the spectral index map where  $\alpha_{\rm{ref}} - 0.05 < \alpha_{\rm{ref}}  < \alpha_{\rm{ref}} + 0.05$, with $\alpha_{\rm{ref}} = -0.65, -0.75, -0.85, $ etc. We limited $\alpha_{\rm{ref}}$ to $-1.85$ and $-1.65$ for B1 and B2+B3, respectively, to retain sufficient SNR at the highest frequency map (at 2272~MHz). We use these spectra as a starting point for creating the radio color-color diagrams.

\subsection{Spectral models}
\label{sec:rx42spectralmodels}
In the model put forward by \cite{1998A&A...332..395E}, relics trace shock waves in which particles are accelerated by the DSA mechanism. Without projection effects and mixing of emission, and all properties of the shock remaining constant, the spectra at the different locations should simply reflect the energy losses of the radiating particles. Assuming an edge-on planar shock-wave, the time since acceleration for particles at a given location behind the front of the relic is simply $l / v_{\rm{d}}$, with $l$ the distance from the front of the shock and $v_{\rm{d}}$ the shock downstream velocity. According to DSA-theory, the injection spectral index $\alpha_{\rm{inj}}$ is linked to the Mach number ($\mathcal{M}$) of the shock by  \citep[e.g.,][]{1987PhR...154....1B}
\begin{equation}
\alpha_{\rm{inj}} =  \frac{1}{2} - \frac{\mathcal{M}^2 +1} {\mathcal{M}^2 -1} \mbox{ .}
\label{eq:inj-mach}
\end{equation}
The integrated spectral index is steeper by about 0.5 units compared to $\alpha_{\rm{inj}}$ for a simple shock model where the electron cooling time is much shorter than the lifetime of the shock \citep{2002MNRAS.337..199M}. Most of the synchrotron radiation we observe comes from the ``critical'' frequency $\nu_{\rm{c}}$
\begin{equation}
\nu_{\rm{c}} = \frac{3\gamma^2 e B_\perp}{4\pi m_{\rm{e}}c } \mbox{ .}
\label{eq:rx42vc}
\end{equation}
Directly behind the front of the shock, the spectra should  have a power-law shape of the form $I(\nu) = I_0 \nu^{\alpha_{\rm{inj}}}$, under the usual assumption that one starts with a power-law distribution of relativistic electrons $N(E) = N_0 E^{-s}$, where $s = 1 - 2\alpha_{\rm{inj}}$.

Synchrotron spectra can be calculated using the standard formula \citep{1970ranp.book.....P}
\begin{equation}
I(\nu,t) \propto \int_{4\pi} \sin{(\theta)}  \int_{0}^{+\infty} F(x) N(E, \theta,t)  \,dEd\Omega  \mbox{ ,}   
\label{eq:Pacholczyk}
\end{equation} 
where $N(E,\theta, t)$ is the electron energy distribution, $\theta$ the pitch angle between the electron velocity and magnetic field vectors, $x = \nu / \nu_{\rm{c}}$, and $F(x) \equiv x \int_{x}^{+\infty} K_{5/3}(z) \, dz$, with $K_{5/3}(z)$ an irregular modified Bessel function. Synchrotron and IC (radiation) losses change the electron energy distribution over time at a rate
\begin{equation}
\frac{dE}{dt} = -  \xi E^2  = - (\xi_{\rm{syn}} + \xi_{\rm{IC}})E^2  \mbox{ .}
\end{equation}
The IC losses are given by $\xi_{\rm{IC}} = (2/3)c_{2} B^{2}_{\rm{CMB}}$ and the quantity $c_{2}=2e^4/3m_{\rm{e}}^4 c^7$ is a constant defined by \citep{1970ranp.book.....P}. Two main expressions for $\xi_{\rm{syn}}$ exist.

The Jaffe-Perola (JP) model takes into account that the pitch angles of the synchrotron emitting electrons are continuously isotropized on a timescale shorter than the radiative timescale  \citep{1973A&A....26..423J}
\begin{equation}
\xi_{\rm{syn}}  =  (2/3) c_2 B^2 \mbox{ .}
\end{equation}
For the Kardashev-Pacholczyk (KP) model, the pitch angle of the electrons remains in its original orientation with respect to the magnetic field 
\begin{equation}
\xi_{\rm{syn}}  =  c_2 B^2 \sin^2(\theta) \mbox{ .}
\end{equation}

The JP model is more realistic from a physical point of view, as an anisotropic pitch angle distribution will become more isotropic due to changes in the magnetic field strength between different regions and scattering by self-induced Alfv\'en waves \citep[e.g.,][]{1991ApJ...383..554C, 1993MNRAS.261...57T, 2001AJ....122.1172S}. The JP and the KP models assume a single impulsive injection that produces a power-law distribution of relativistic electrons. At $t=0$, $N(E)=N_0 E^{-s}$ and at some later time ($t$)
\begin{equation}
N(E,\theta, t) =   \begin{cases} N_0 E^{{-s_{\rm{inj}}}} \left(1 - \xi Et     \right)E^{{s_{\rm{inj}}} -2} & \text{if $E \leq  1/\xi t   $,} \\  0 & \text{if $E >  1/\xi t    $} \end{cases} \mbox{ .}
\end{equation}
With this equation for $N(E,\theta, t)$ and Eq.~\ref{eq:Pacholczyk} the synchrotron spectrum can be computed. The result is that the energy losses (both for the JP and KP models) cause the radio spectrum to steepen (see Fig.~\ref{fig:rx42specmodels}) above a break frequency ($\nu_{\rm{brk}}$), with \begin{equation}
\nu_{\rm{brk}}  \propto \frac{B}{\left(\left[B^2 + B^{2}_{\rm{CMB}}\right]t\right)^2   }    \mbox{ .}
\end{equation}
JP-spectra have an exponential high-frequency cutoff above $\nu_{\rm{brk}}$. KP-spectra steepen to a power-law with a slope $(4/3)\alpha_{\rm{inj}} -1$ above  $\nu_{\rm{brk}}$, because there are always some high energy electrons with low pitch angles to radiate at high frequencies in the KP case. 

In addition to the above mentioned JP and KP models, where all particles are injected in one single burst and then simply age, there is the continuous injection (CI) model \citep{1970ranp.book.....P}, i.e., a fresh supply of particles is injected continuously and the so-called KGJP or KGKP models \citep{1994A&A...285...27K}, where particles are injected for a fixed period of time after which the supply of newly injected electrons is switched off. This results in different expressions for $N(E,\theta, t)$, see eq.~6 from \cite{1994A&A...285...27K}. CI and KGJP/KP spectra are basically summations of individual JP and KP spectra with different amounts of spectral ageing. Examples of resulting spectral shapes used in this work are shown in Fig.~\ref{fig:rx42specmodels}.

\begin{figure*}
\begin{center}
\includegraphics[angle =90, trim =0cm 0cm 0cm 0cm,width=0.48\textwidth]{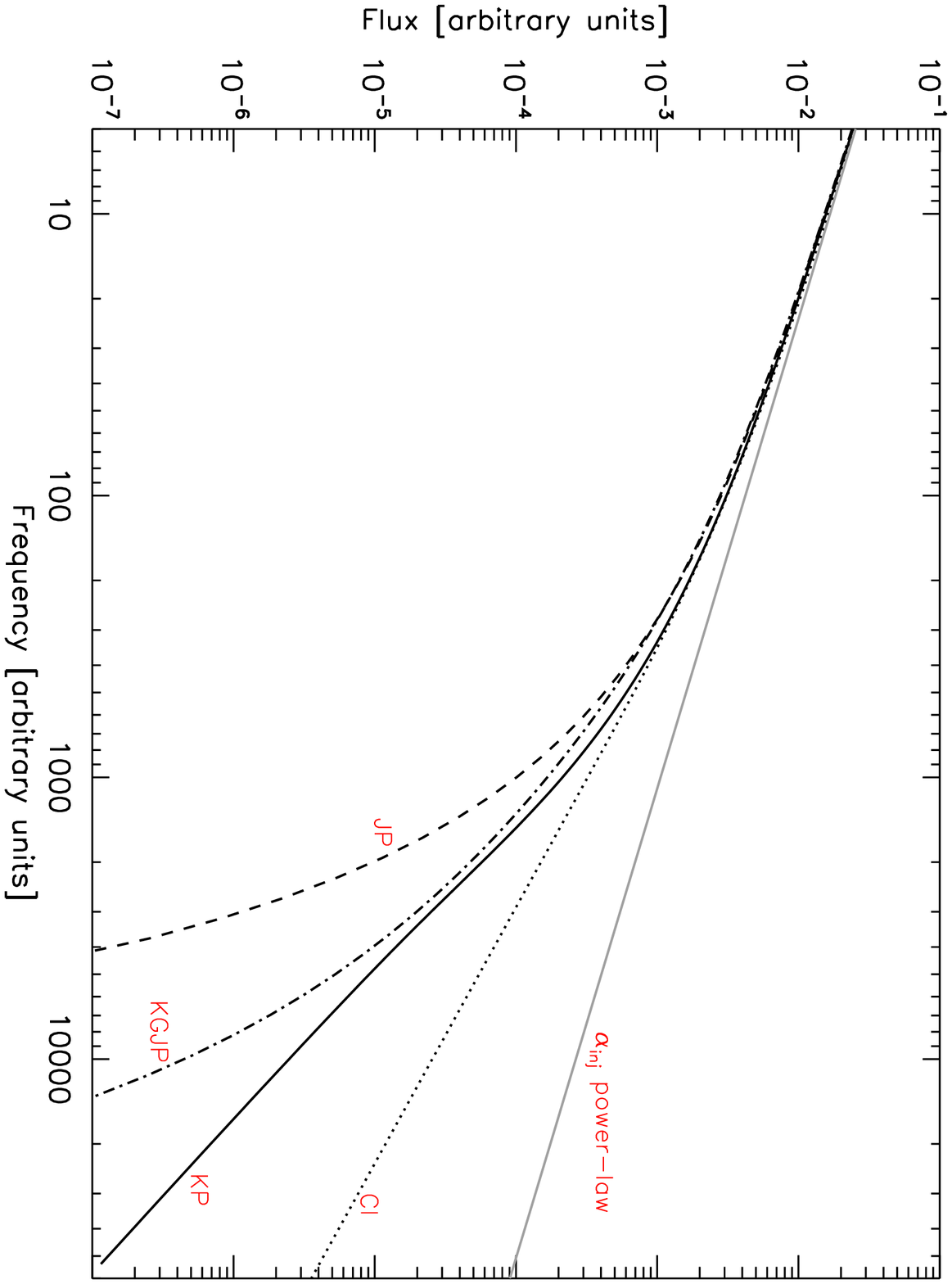}
\includegraphics[angle =90, trim =0cm 0cm 0cm 0cm,width=0.49\textwidth]{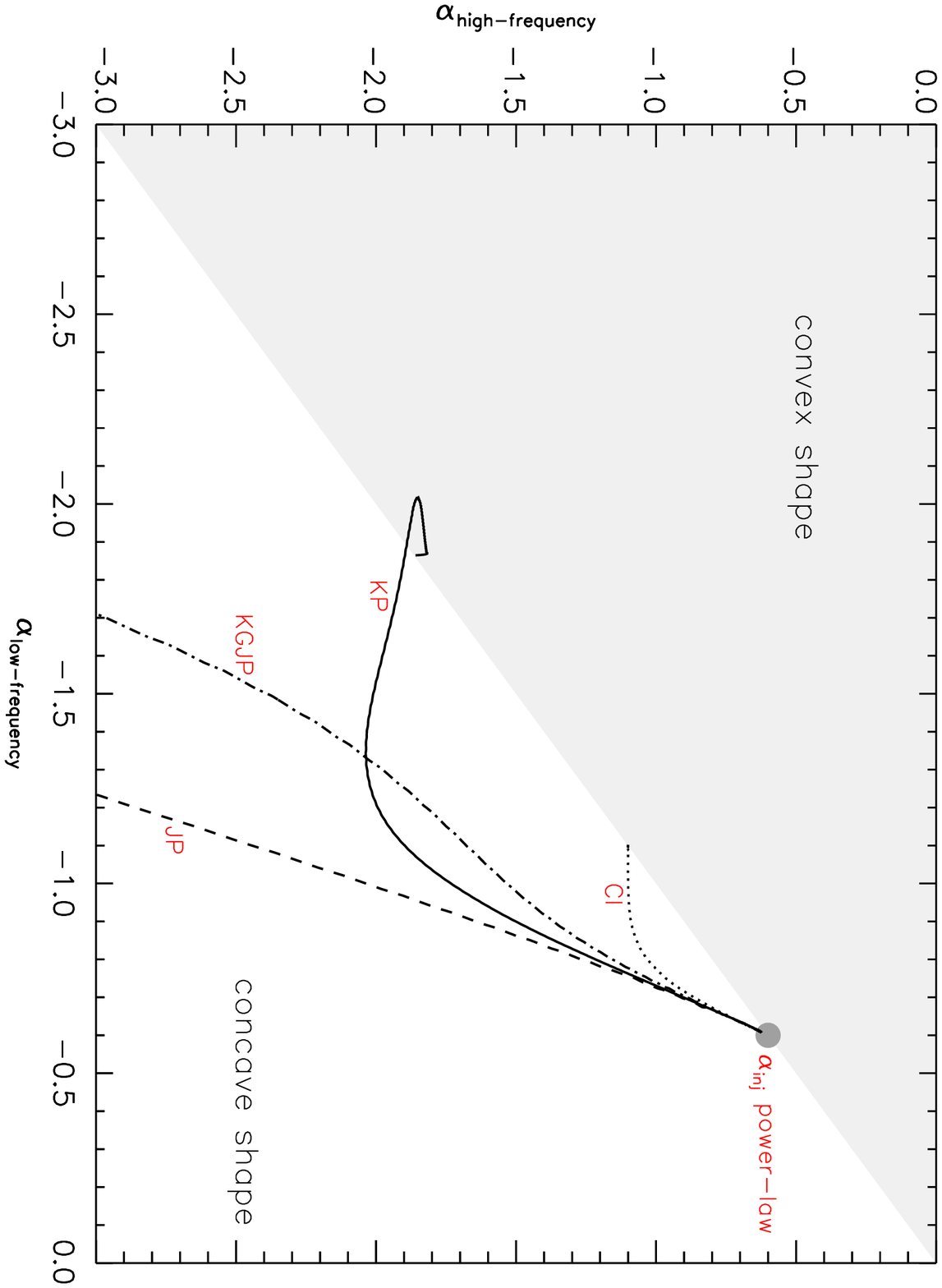}
\end{center}
\caption{{Overview of the spectral models used in this work (left) and corresponding radio color-color corves (right) to highlight differences in the spectral shapes. All spectral models have $\alpha_{\rm{inj}}=0.6$. Solid grey line and  filled grey circle (in the color-color digram): power-law spectrum without any ageing, solid black line: KP, dotted line: CI, dash-dotted line: KGJP, dashed line: JP. Note that a power-law spectrum will translate in a single point on a color-color diagram. The $\alpha_{\rm{low-frequency}} = \alpha_{\rm{high-frequency}}$ line divides concave and convex spectral shapes between the frequencies where the spectral index is computed.}}
\label{fig:rx42specmodels}
\end{figure*}

 \subsection{Effect of resolution on surface brightness}
 \label{sec:rx42effectsurfacebrightness}
If radiative losses are the dominant effect then all  spectra for different parts of the relic should line up at some fiducial low energy/frequency where radiation losses are not important. The spectra displayed in Fig.~\ref{fig:rx42flux_SB} should therefore all line up at low enough frequencies. Part of such a trend is indeed visible. However, the spectra with $\alpha_{\rm{ref}} = -0.65, -0.75$ have a lower surface brightnesses than we naively expect. We attribute this to the fact that the pixels, from which these spectra are determined, are all located on the sharp outer edge of the relic, where the emission drops abruptly to zero within a single beam element. Since the resolution of the spectral index map is only about 20\arcsec~and there are about 4 pixels per beam, the regions with the flattest spectral index are all located on pixels where the surface brightness drops at the front of the relic due to the finite beam size. This can   be seen by looking at the black contours in Fig. \ref{fig:rx42spix_poly}, i.e., the surface brightness does not peak at the region with the flattest spectral index. 
 Comparing these contours to the higher resolution images in Figs.~\ref{fig:rx42gmrt610} and \ref{fig:rx42gmrt241}, we find however that the outer rim of the relic is indeed the brightest part of the relic. The lower surface brightness for the $\alpha_{\rm{ref}} = -0.65, -0.75$ regions is thus caused by the low spatial resolution of the spectral index map. This is also confirmed by the high-resolution 325--610~MHz spectral index map, see Fig.~\ref{fig:rx42spix610-1280}.

It is interesting to note that the spectra for the regions with a flattest spectral index are all relatively straight and show little deviations from a pure power-laws (Fig.~\ref{fig:rx42flux_SB}). Going to the regions with a steeper $\alpha_{\rm{ref}}$, the spectral shapes change and display signs of curvature at the higher frequencies. These changes happens gradually from the regions with a flat  $\alpha_{\rm{ref}}$ to a steeper  $\alpha_{\rm{ref}}$.

\subsection{Color-color diagrams}
To better investigate the spectral shape we use so called three frequency ``color-color'' diagrams for which the spectral indices for different region of the source can all be put in a single diagram \citep{1993ApJ...407..549K}.  On the x-axis we plot the low- frequency spectral index and on the y-axis the high-frequency one. Color-color diagrams are particularly useful to discriminate between theoretical synchrotron spectral models, such as the JP, KP, CI, KGJP, KGKP models described before, {see Fig.~\ref{fig:rx42specmodels} for several examples.} 

Another advantage of color-color diagrams is that the shapes traced out in the diagrams are conserved for changes in the magnetic field, adiabatic expansion or compression, and the radiation losses, for standard spectral models. We observe a different portion of the spectrum for each different frequency or source physical condition (e.g., local $B$ field).  
We can thus use the observed curve to constrain the different models (JP, CI, KP, etc.) and injection spectral indices. Mixing of emission, for example from regions with different $B$ strengths or radiation losses, due the  finite resolution of the observations or projection effects, will lead to different curves in the color-color diagrams.  If a global spectral shape exists, it also allows  for a better effective frequency sampling and provides sampling of a larger range of electrons energies.


\subsubsection{Color-color diagrams for B1 and B2+B3}
The color-color diagrams for B1 and B2+B3 are shown in Fig.~\ref{fig:rx42colorcolor} (left and right panels, respectively). The points in the color-color diagrams seem to trace out single curves. This suggests the existence of a global spectral shape for these regions. The curves are similar for the B1 and B2+B3 regions, but they slightly differ for  $\alpha_{\rm{ref}} \lesssim -1.4$.

Tracing the curves back to the $\alpha_{241}^{610} = \alpha_{1382}^{2272}$ (i.e., power-law) line gives the injection spectral index. In both cases we find $\alpha_{\rm{inj}}$ is about $-0.6$ to $-0.7$ for B1 and B2+B3,  in agreement with the spectral index maps. 
The color-color diagrams highlight the trend of increasing spectral curvature, i.e., distance from the power-law line, with decreasing $\alpha_{\rm{ref}}$ (Fig.~\ref{fig:rx42flux_SB}). The spectra for $\alpha_{\rm{ref}}=-0.65, -0.75$ are power-laws.
We also indicated various spectral models in the diagrams. We compare the observed curves in the color-color diagrams to the some of the standard spectral models. 

The observed curve is clearly different from the continuous injection spectral model. The CI model only steepens to $\alpha = \alpha_{\rm{inj}} - 0.5$ \citep{1970ranp.book.....P}. The KP model (with  $\alpha_{\rm{inj}}= -0.6$) roughly follows the data for flat $\alpha_{\rm{ref}}$, but deviates for $\alpha_{\rm{ref}} \lesssim -1$. In the end the KP curve bends back and returns to the $\alpha_{241}^{610} = \alpha_{1382}^{2272}$ line. We do not see evidence for a turn back to the $\alpha_{241}^{610} = \alpha_{1382}^{2272}$ line in the data, although we do not sample this part of the diagram well because of insufficient SNR at high-frequencies. 
{As  mentioned previously, a KP model is rather unphysical. However, it sometimes provides a good fit to observed radio spectra because it is similar to the shapes caused by losses in a randomly varying magnetic field \citep{1993MNRAS.261...57T}.}

The JP model (plotted for $\alpha_{\rm{inj}} = -0.6$ and $-0.7$) follows the KP model for flat $\alpha_{\rm{ref}}$. Instead of turning back to the  $\alpha_{241}^{610} = \alpha_{1382}^{2272}$ line, the spectral curvature keeps increasing. The data also show this trend of increasing curvature, but less quickly than the JP curve.
A KGJP model with $\alpha_{\rm{inj}} =-0.7$ matches the data for B2+B3 very well. The KGJP curves in Fig.~\ref{fig:rx42colorcolor} are for particles injected continuously for about $0.6 \times 10^{8}$~yr  and $B=9$~$\mu$Gauss (see Sect.~\ref{sec:rx42eqB}). The KGJP model also provides a better match for the B1 region compared to the KP and JP curves. Although, in the regions with the steepest $\alpha_{\rm{ref}}$ values the KGJP model still overestimates the amount of curvature. A KGJP model can be thought of as an integration of JP spectra with a range of spectral ages, up to the oldest population of electrons. At first, the KGJP model follows the CI model until at some point in time the supply of newly injected electrons is shut off. 
Our injection time of $0.6 \times 10^{8}$~yr  leads to a distribution of spectral ages in each $\alpha_{\rm{ref}}$ region that is $0.6 \times 10^{8}$~yr wide at maximum, i.e., it represents the amount of mixing encountered in each $\alpha_{\rm{ref}}$ region. It should not be interpreted as the total injection time for the relic. This value thus depends on the resolution of the spectral maps and unavoidable projection effects.  

\subsubsection{The effect of resolution and mixing}
One  important effect is mixing of emission within the beam. Each beam samples superpositions of regions with potentially physical different conditions. To investigate the effect of resolution in the color-color diagrams, we decreased the resolution of the spectral index maps by convolving them with Gaussians of 60\arcsec~FWHM. The data for the lower resolution maps are also displayed in Fig.~\ref{fig:rx42colorcolor}. The effect of  the lower resolution is that the curve ends up closer to the $\alpha_{241}^{610} = \alpha_{1382}^{2272}$ line.  This is expected because the total integrated spectrum of the relic has a power-law shape (Fig.~\ref{fig:rx42fluxB}).

We also investigated for B1, the effect of a possible underlying flux component from the radio halo C. For this, we determined the average spectrum of the radio halo and subtracted this flux contribution at each frequency. This however did not substantially change the resulting color-color diagram for B1 because the surface brightness of the relic is much higher than that of the radio halo.

Our conclusion is that mixing of emission (from regions with different $B$-fields or different electron energy distributions, {spectral ages}, etc.) pushes the spectra closer to power-law  shapes. The spectral curvature we find can thus be regarded as a lower limit on the actual curvature and it is therefore important that when searching for spectral curvature one retains sufficient spatial resolution (enough to properly resolve the spectral variations).

\subsection{Global spectrum}


Since the data points trace out a well defined curve in the color-color diagrams, we have attempted to map the flux measurements onto a single spectral shape using the ``shift-technique'' described by \cite{1993ApJ...407..549K,  1996cyga.book..158R, 2001ASPC..250..372R}. This allows for a much better sampling in frequency, and maps out the spectrum over a larger range of electron energies. The idea is that the individual radio spectra for different $\alpha_{\rm{ref}}$ regions  each trace some part of the ``global spectrum'' (or electron energy distribution) of the source, depending on the energy losses, and magnetic fields in these regions. The shifts (in $\log{(I)}$-$\log{(\nu)}$ space) remove the effects of possible different local magnetic field strengths, electron densities, radiative energy losses, and adiabatic energy gains/losses.

We shifted the spectra, displayed Fig.~\ref{fig:rx42flux_SB}, in $\log{(I)}$-$\log{(\nu)}$ space and tried to line them all up. This went remarkably well, indicating we indeed have a single electron energy distribution (or global spectral shape) that is consistent with the spectra for each individual region. 
The two resulting spectra (for B1 and B2+B3) are displayed in Fig.~\ref{fig:rx42globalspectra}. As can be seen it allows for an almost continuous sampling of the radio relic spectrum over about 4 orders of magnitude in frequency (two orders of magnitude in energy, see Eq.~\ref{eq:rx42vc}). 

 Similar to the color-color diagrams we find that the low-frequency part of the global spectrum has a power-law shape, with $\alpha=-0.6$ to $-0.7$. At higher frequencies the spectrum steepens. We also compare the global spectra with some of the standard models in Fig.~\ref{fig:rx42globalspectra}.  The results are the same as for the color-color diagrams, i.e., the best match we find for a KGJP model and for B1 the KGJP spectral curvature is somewhat too high for $\nu_{\rm{eff}} \gtrsim 20$~GHz.

\subsubsection{Shift diagrams}
\label{sec:rx42shift}
Interestingly, the shifts made to align up all the individual radio spectra to create Fig.~\ref{fig:rx42globalspectra} provide information about changes in the underlying physical parameters \citep{ 1994ApJ...426..116K}. The shifts made in the frequency direction $(\log{(\nu)})$, to line up the spectra, are related to $\gamma^2 B$. The shifts in $\log{(I)}$ are related to $N_{\rm{T}} B$, where $N_{\rm{T}}$ is the total number of relativistic electrons in the volume determined by the beam size.

\begin{figure}
\begin{center}
\includegraphics[angle =90, trim =0cm 0cm 0cm 0cm,width=0.5\textwidth]{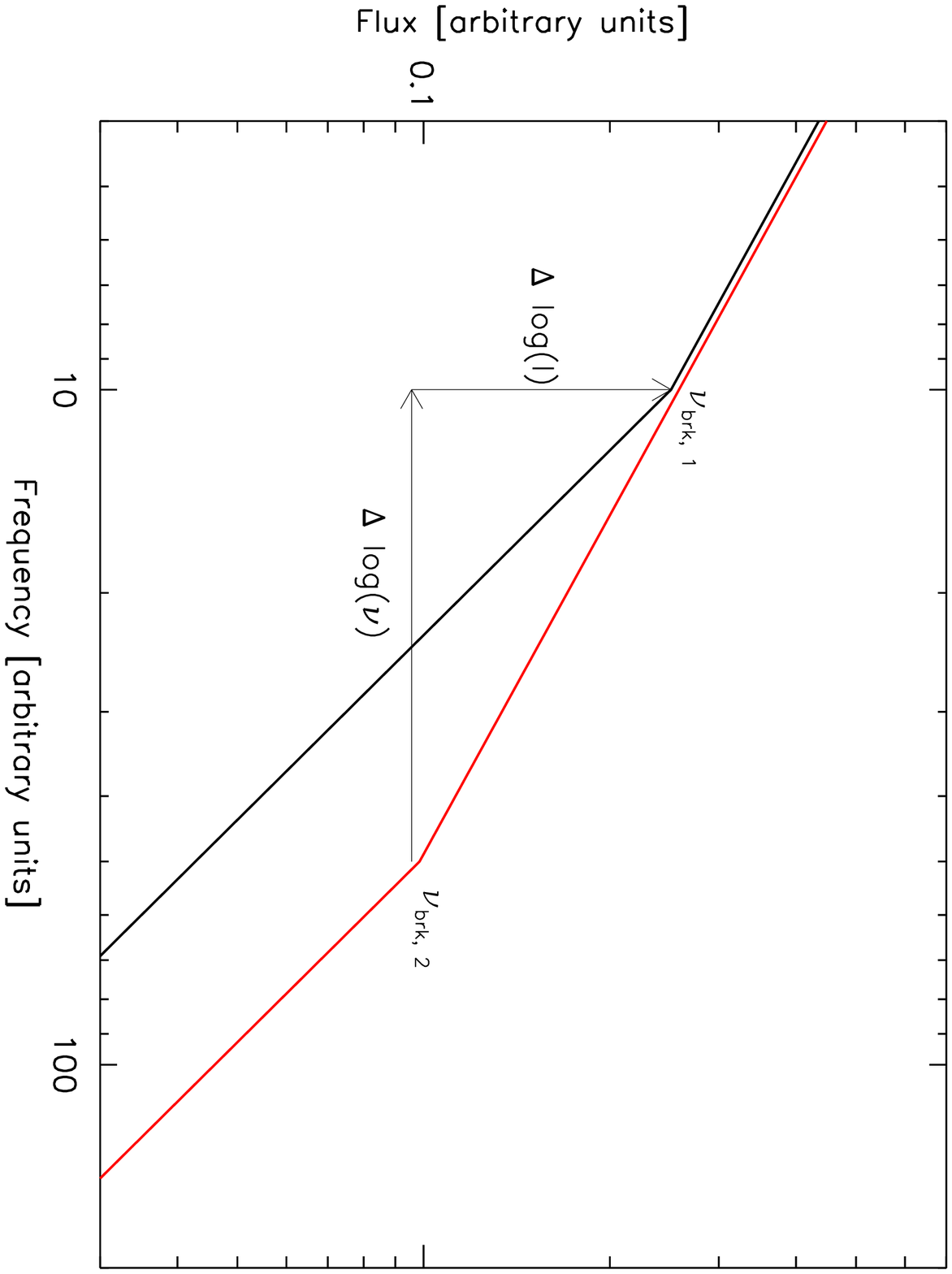}
\end{center}
\caption{{Schematic representation of two radio spectra with different spectral ages (break frequencies). To align the red spectrum to the black one in log-log space, a shift in the frequency direction is required to match up the break frequencies. The red spectrum then falls below the injection power-law spectrum and needs to be shifted upwards. The slope $\Delta\log{(I)}/   \Delta\log{(\nu)} $ equals the injection spectral index in this case (see the main text).} }
\label{fig:refspectra}
\end{figure}

 By plotting the shifts  which we needed to align up the individual spectra ($\log{(\nu)}$ shifts against $\log{(I)}$ shifts),  we can investigate the contributions of the above mentioned quantities, i.e., which quantities cause the spectra to differ from region to region.
 The shift diagrams are shown in Fig.~\ref{fig:rx42shifts}. For example, a slope ($\delta \log{(I)} / \delta \log{(\nu)}$) of $+1$ indicates mainly variations in the magnetic field strength, a slope of infinity indicates density variations, and a slope of zero energy variations. If every line of sight samples the same physical conditions (i.e., the same magnetic field, path length through the source, number of relativistic particles at some fiducial low energy), then ageing alone would give a slope equal to the injection spectral index, see Fig.~\ref{fig:refspectra}. {This can be seen as follows: one needs to shift the red spectrum in Fig.~\ref{fig:refspectra} in $\log{(\nu)}$ space to match up the break frequency of the black spectrum, but then the red spectrum will fall below the initial power-law injection spectrum, so a shift (up) in $\log{(I)}$ is also required. For the shift in $\log{(\nu)}$
  \begin{equation}
 \Delta\log{(\nu)} = \log{\left(\nu_{\rm{brk,2}}\right)} -  \log{\left(\nu_{\rm{brk,1}}\right)} \mbox{ .}
 \end{equation}
  For the shift in $\log{I}$ we have{\tiny   
   \begin{equation}  \Delta\log{(I)} = \log{\left(I_0 \nu_{\rm{brk,2}}^{\alpha_{\rm{inj}}}  \right)} -  \log{\left(I_0 \nu_{\rm{brk,1}}^{\alpha_{\rm{inj}}}  \right)} = \alpha_{\rm{inj}}\left[ \log{\left(\nu_{\rm{brk,2}}\right)} -  \log{\left(\nu_{\rm{brk,1}}\right)}  \right]   
   \end{equation}}
 \noindent and thus the slope in the shift diagram  is $\Delta\log{(I)}/   \Delta\log{(\nu)}  = \alpha_{\rm{inj}}$. }

 We find slopes of $-0.83$ and $-0.67$ in the shifts diagrams for B1 and B2+B3, respectively.  We did not include the first three (B1) and two (B2+B3) points for fitting the slope. This because the $\log{(I)}$-shifts for these points are affected by the reduced surface brightness at the outer edge of the relic, see Sect.~\ref{sec:rx42effectsurfacebrightness}. The slopes are  close to the injection spectral indices of $-0.6$ to $-0.7$ we found from the color-color diagrams. The slope for B1 deviates a little more, but overall the shift diagrams indicate that spectral ageing is likely the dominant factor in explaining the different spectral shapes from region to region. Apparently, the magnetic field and total number of number of relativistic electrons remain more or less constant for the different $\alpha_{\rm{ref}}$ regions.

\begin{figure*}
\begin{center}
\includegraphics[angle =90, trim =0cm 0cm 0cm 0cm,width=0.8\textwidth]{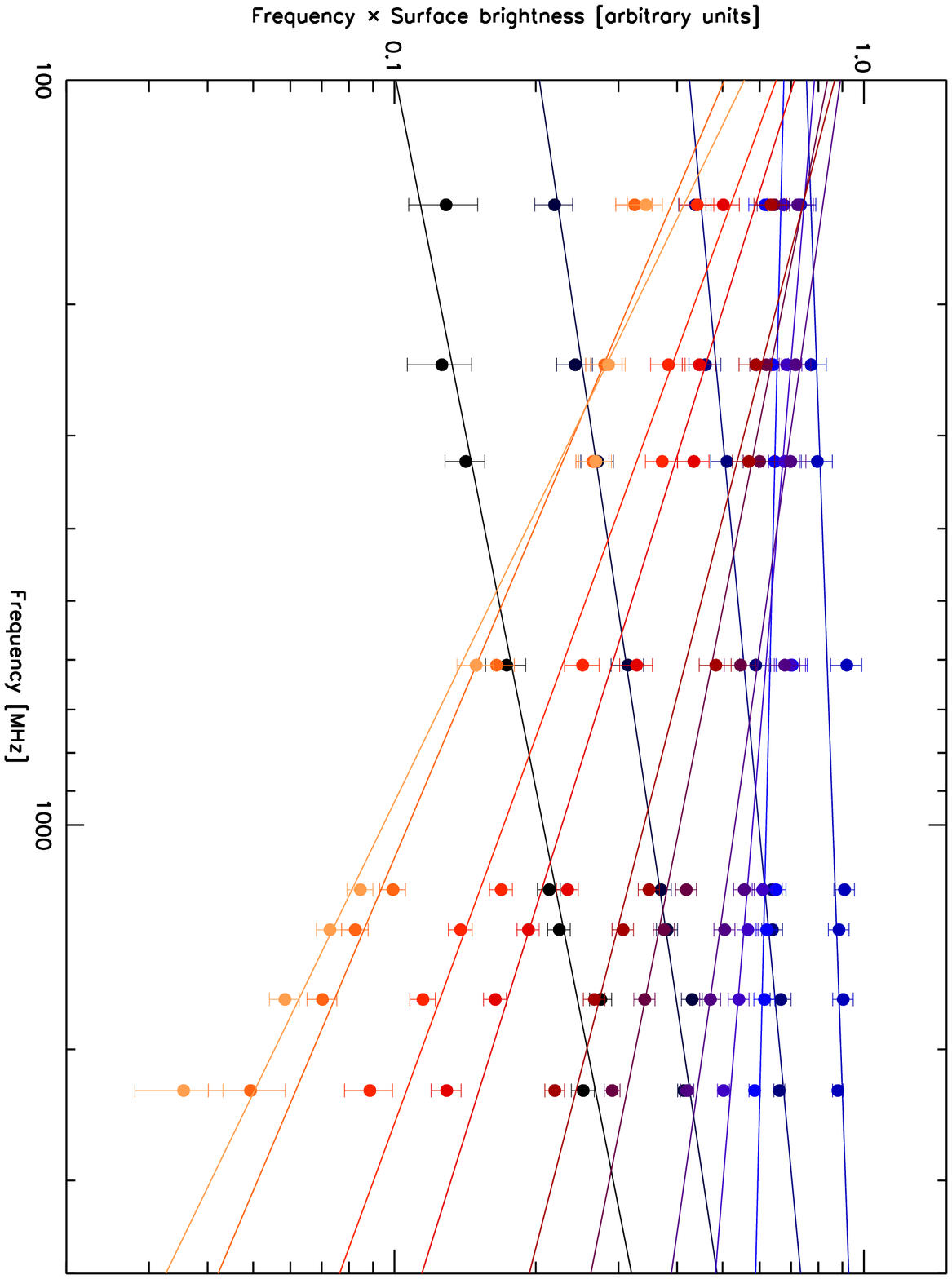}
\includegraphics[angle =90, trim =0cm 0cm 0cm 0cm,width=0.8\textwidth]{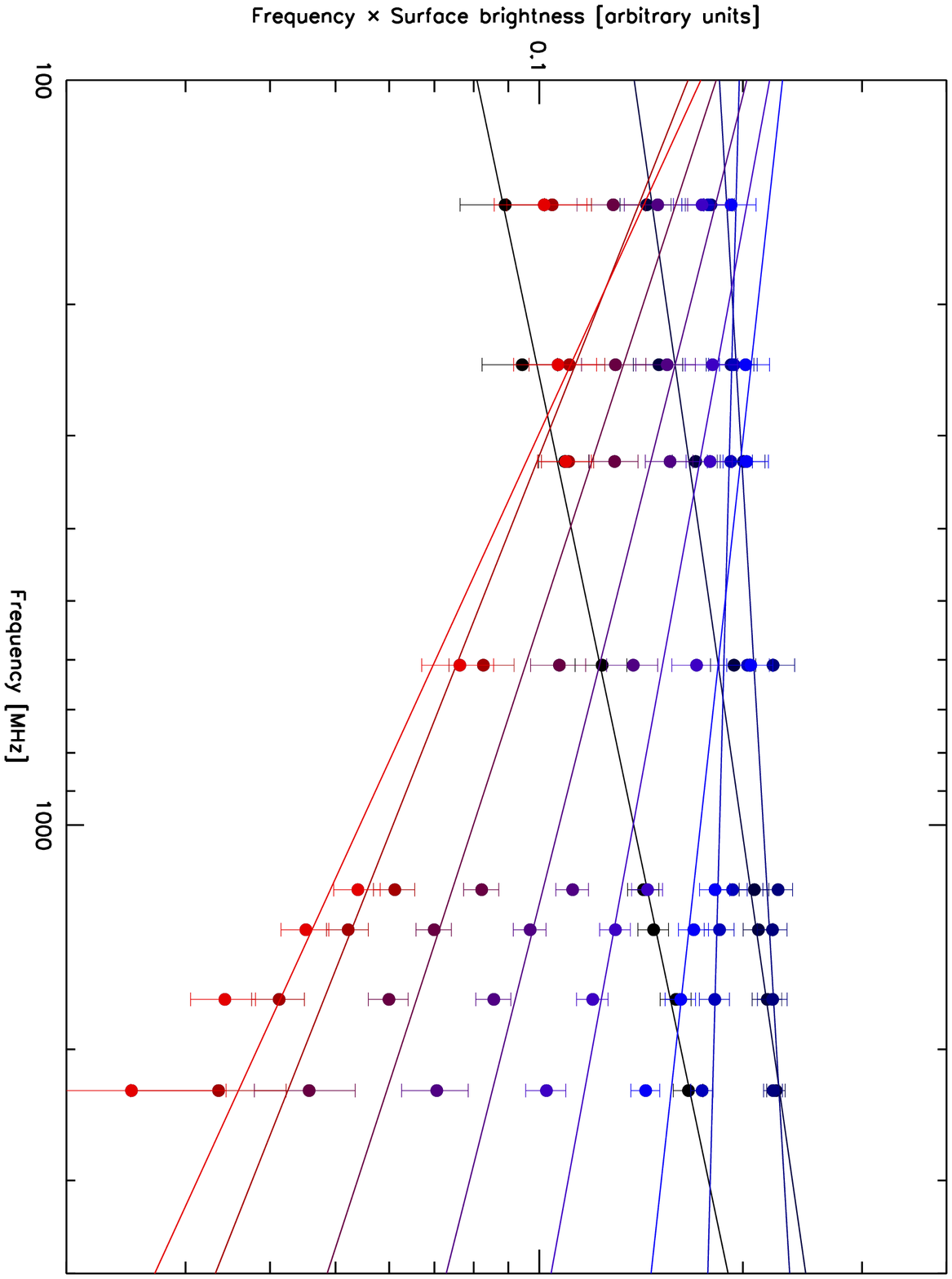}
\end{center}
\caption{Normalized radio spectra for relic B1 (top) and B2+B3 (bottom). Each spectrum corresponds to a region from the spectral index map where  $\alpha_{\rm{ref}} - 0.05 < \alpha_{\rm{ref}}  < \alpha_{\rm{ref}} + 0.05$, with $\alpha_{\rm{ref}} = -0.65, -0.75, \ldots, -1.85$ for B1, and  $\alpha_{\rm{ref}} = -0.75, -0.85, \ldots, -1.65$ for B2+B3. The spectra are plotted in $\log{\nu}-\log{(\nu I)}$ space to emphasize the differences between the spectra.  First order polynomial fits, {done in $\log{(I)}-\log{(\nu)}$ space}, are shown by the solid lines to emphasize possible deviations from power-law radio spectra. Colors correspond to the different  $\alpha_{\rm{ref}}$ and go from black, blue, purple, red, to orange.}
\label{fig:rx42flux_SB}
\end{figure*}

\begin{figure*}
\begin{center}
\includegraphics[angle =90, trim =0cm 0cm 0cm 0cm,width=0.49\textwidth]{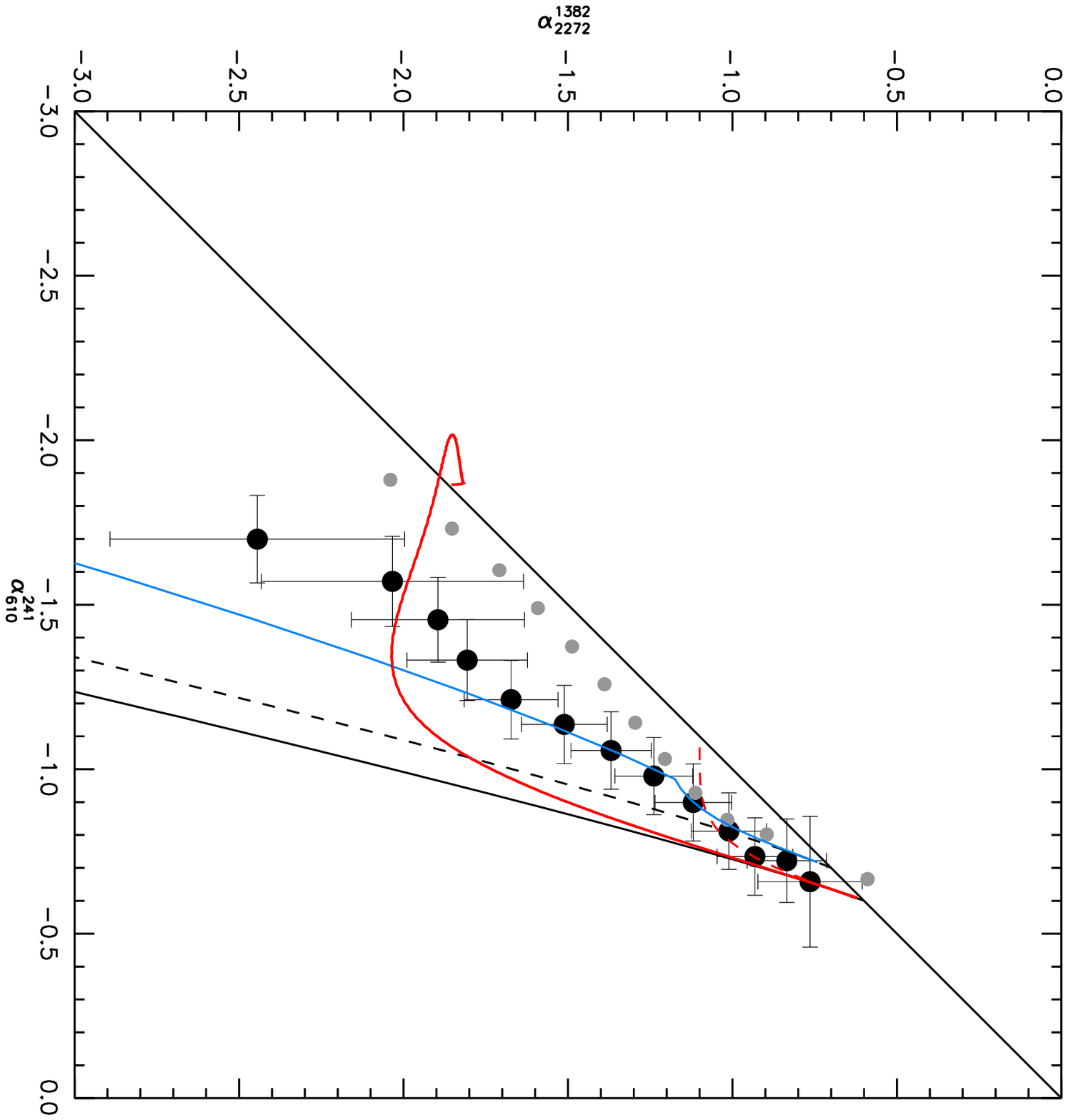}
\includegraphics[angle =90, trim =0cm 0cm 0cm 0cm,width=0.49\textwidth]{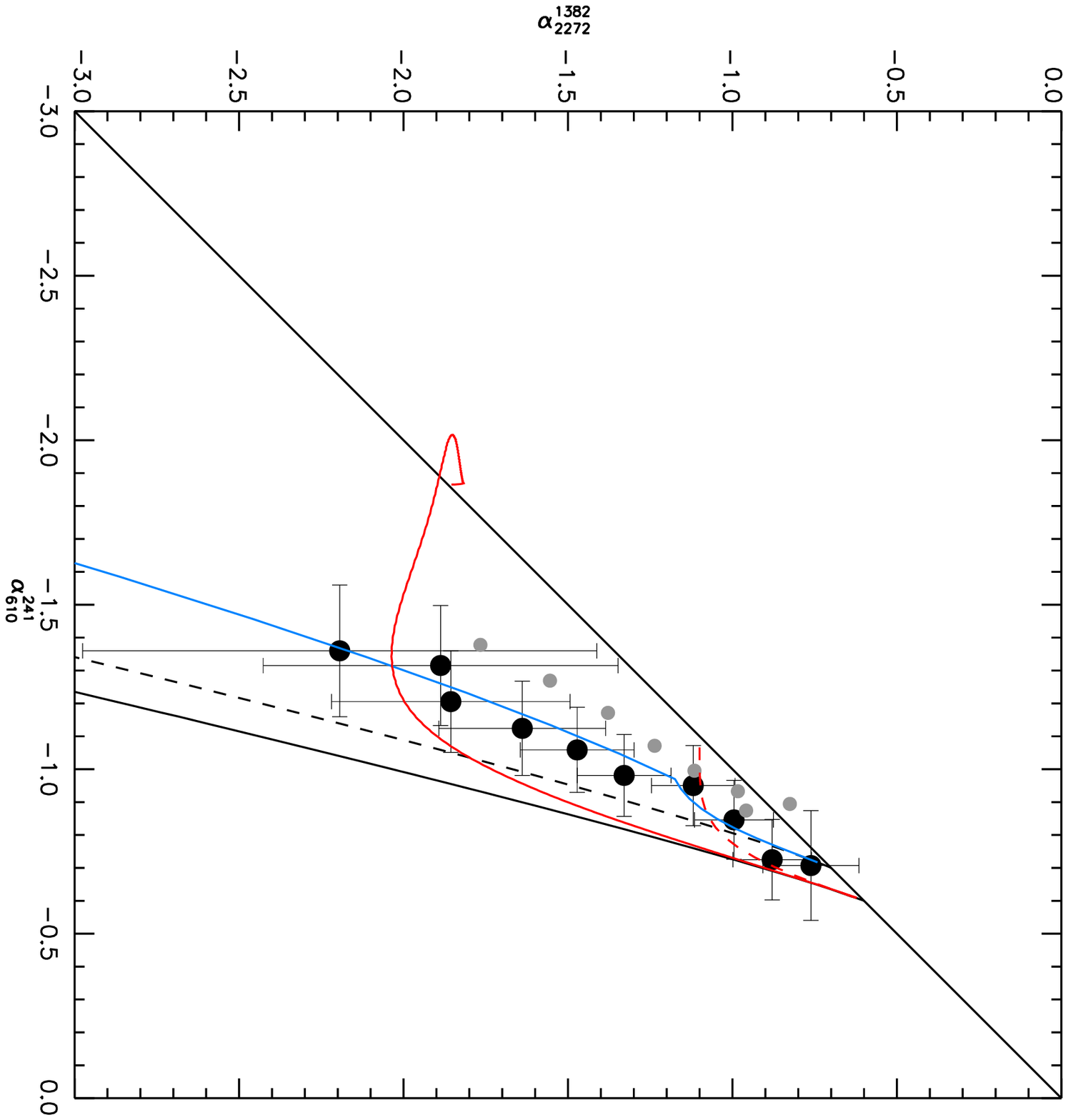}
\end{center}
\caption{Color-color diagrams for B1 (left) and B2+B3 (right). Black data points are from the spectra displayed in Fig.~\ref{fig:rx42flux_SB}. The solid black (slightly curved) line is a JP model with $\alpha_{\rm{inj}}=-0.6$, the dashed black line a JP model with $\alpha_{\rm{inj}}=-0.7$, the solid red line a KP model with $\alpha_{\rm{inj}}=-0.6$, the dashed red line a CI model with $\alpha_{\rm{inj}}=-0.6$, and the blue line a KGJP model with $\alpha_{\rm{inj}}=-0.7$. The grey data points are for the same data as the black points, except the maps were smoothed with a 60\arcsec~FWHM Gaussian. 
The $\alpha_{241}^{610} = \alpha_{1382}^{2272}$ line is also shown for reference.}
\label{fig:rx42colorcolor}
\end{figure*}

\begin{figure*}
\begin{center}
\includegraphics[angle =90, trim =0cm 0cm 0cm 0cm,width=0.49\textwidth]{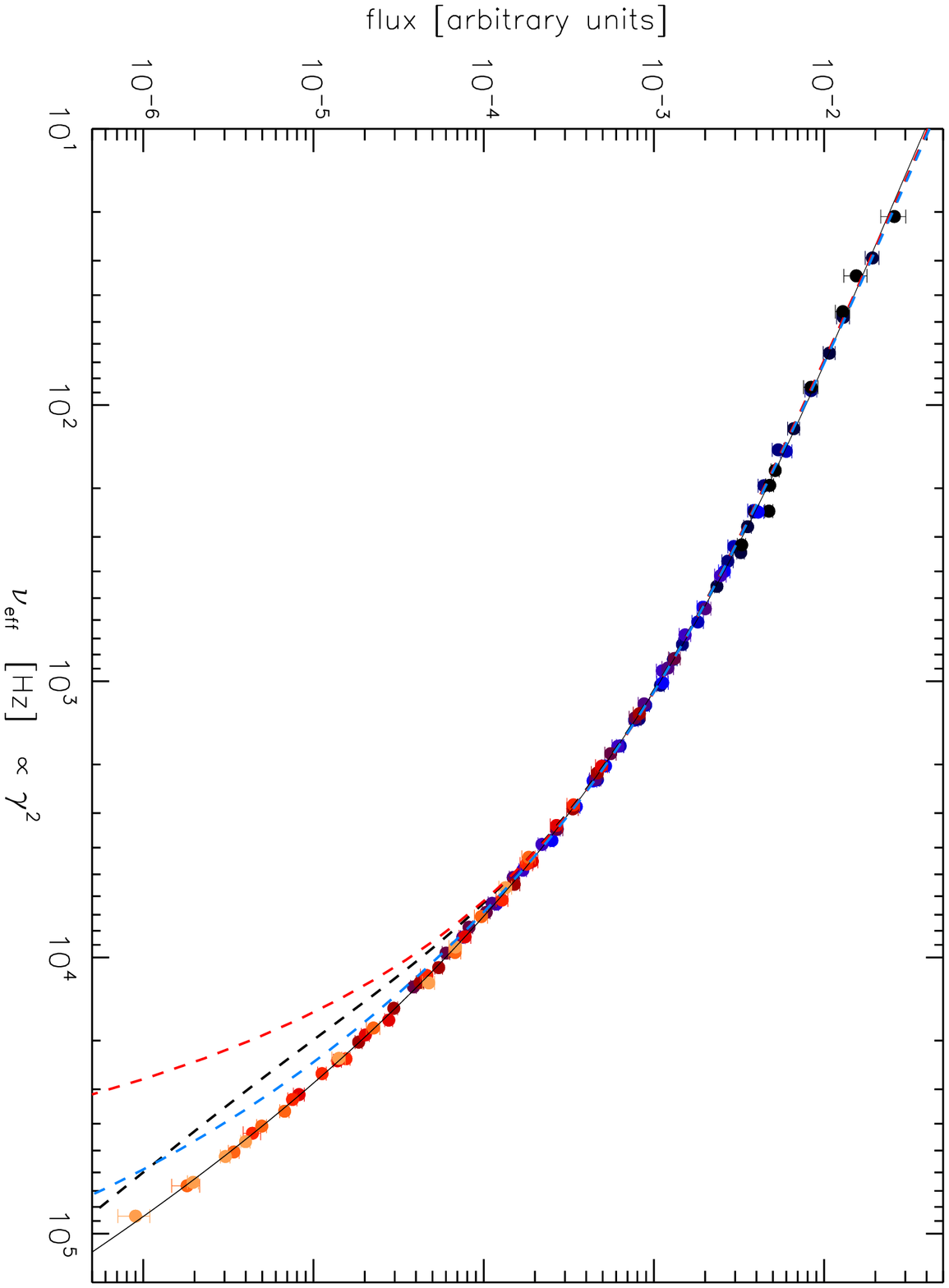}
\includegraphics[angle =90, trim =0cm 0cm 0cm 0cm,width=0.49\textwidth]{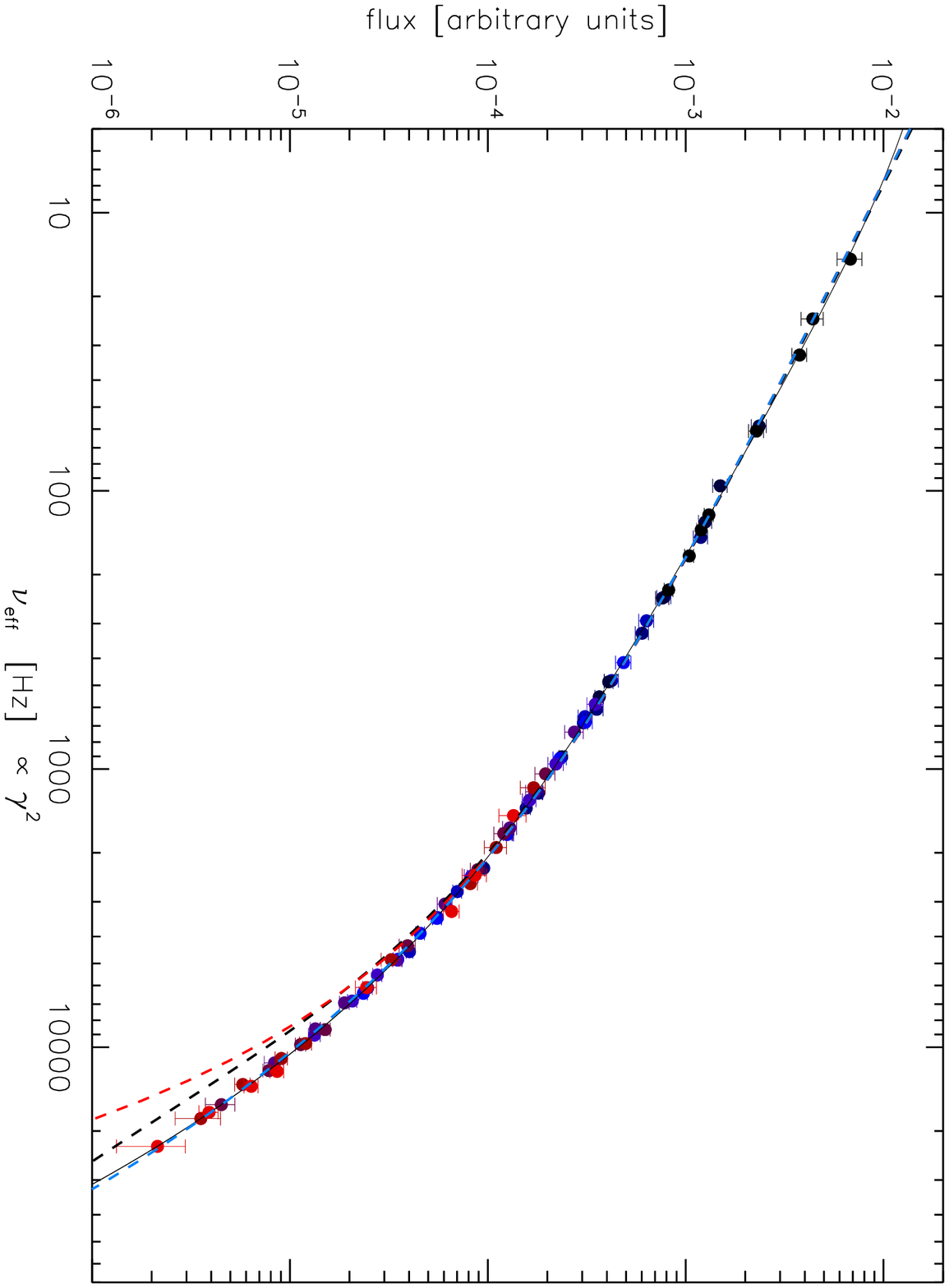}
\end{center}
\caption{``Global'' radio spectra for B1 (left) and B2+B3 (right). The spectrum for each $\alpha_{\rm{ref}}$ spectral region from Fig.~\ref{fig:rx42flux_SB} has been shifted in  $\log{(I)}-\log{(\nu)}$ space to create the ``global'' spectral shape. Fourth order polynomial fits through these data points are shown by solid thin black lines. JP (red), KP (black) and KGJP (blue) models are shown by the dashed lines. The color coding, based on the $\alpha_{\rm{ref}}$ regions, is the same as in Fig.~\ref{fig:rx42flux_SB}.}
\label{fig:rx42globalspectra}
\end{figure*}

\begin{figure*}
\begin{center}
\includegraphics[angle =90, trim =0cm 0cm 0cm 0cm,width=0.49\textwidth]{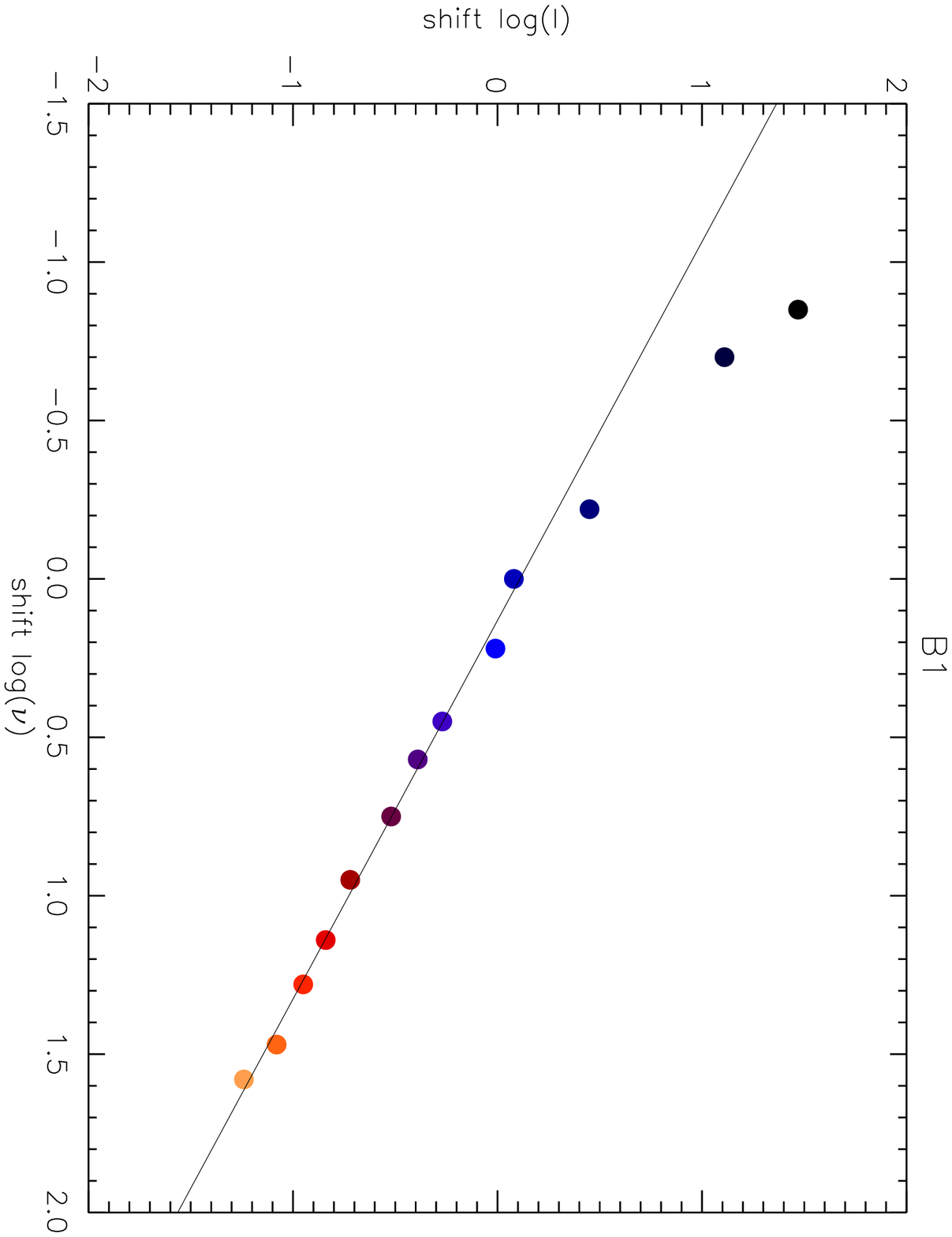}
\includegraphics[angle =90, trim =0cm 0cm 0cm 0cm,width=0.49\textwidth]{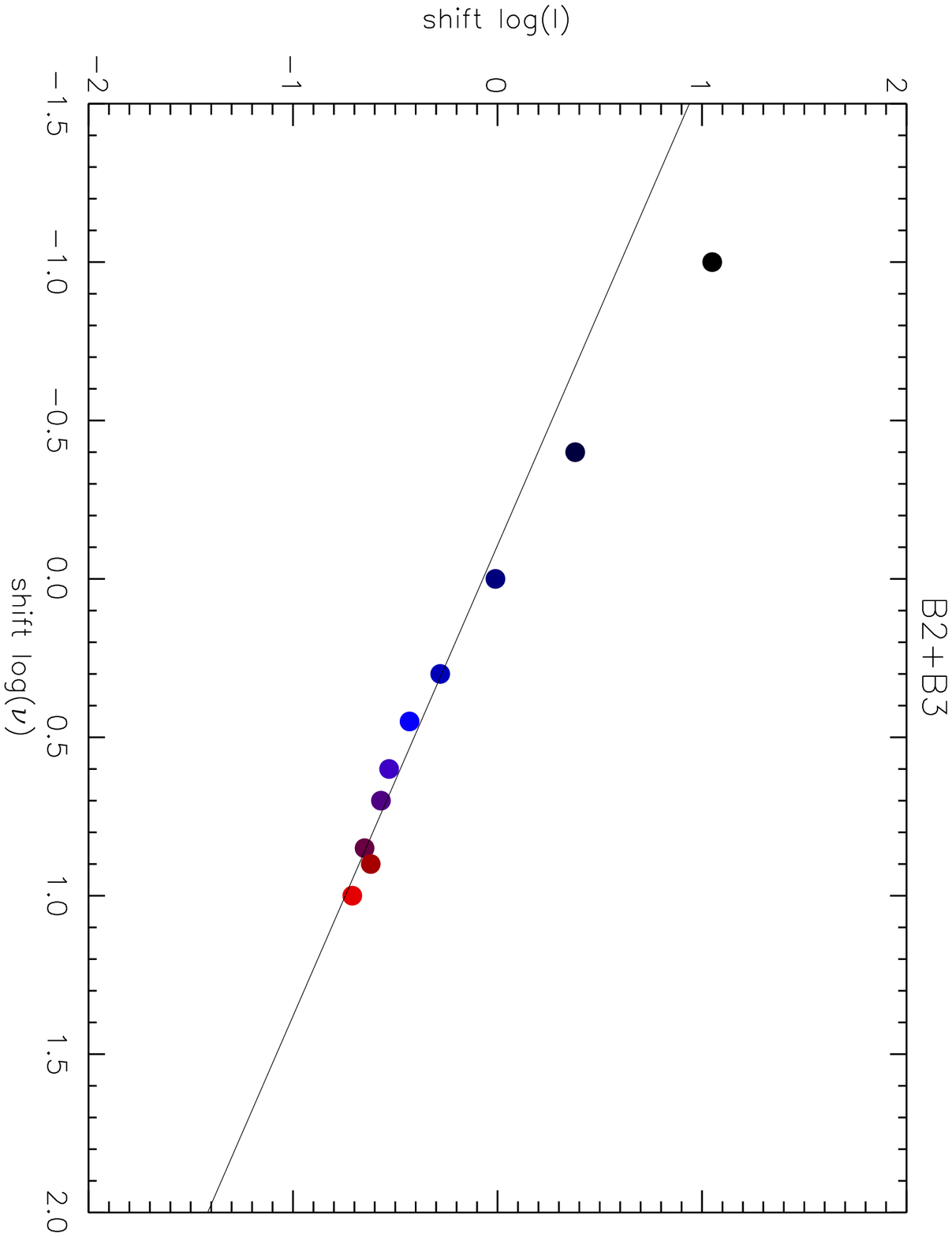}
\end{center}
\caption{Shifts in $\log{(I)}-\log{(\nu)}$ space, for  B1 (left) and B2+B3 (right), needed to line up the radio spectra displayed in Fig.~\ref{fig:rx42globalspectra}. A line is fitted through the data points, excluding the first three (B1) or two (B2+B3) points. Color coding for the  $\alpha_{\rm{ref}}$ regions is the same as in Fig.~\ref{fig:rx42flux_SB}.}
\label{fig:rx42shifts}
\end{figure*}


\section{RM-synthesis \& polarization maps}
\label{sec:rx42pola}
\subsection{Polarization maps}
The distribution of the polarization E-vectors at 4.9 and 1.382~GHz are shown in Fig.~\ref{fig:rx42pol}. In the 4.9~GHz map, we find the relic to be polarized over its entire length (in the region with sufficient SNR). The E-vectors are mainly perpendicular  to the relic's orientation, except at the eastern end of B1 (where the relic bends and is connected to the linear eastward extension B2). The polarization fraction across the relic varies between 10 and 60\%, where it can be measured. At the front of B1, the polarization fraction is mostly between 15 and 30\%, while at the front of B2 the fraction is 50\% or higher. In the 4.9~GHz map we cannot determine the polarization fraction for B3 because the attenuation of the primary beam . It should be noted that the spatially averaged polarization fractions are  lower.  

At 1382~MHz the polarization fraction drops significantly for B1 and B2, compared to at 4.9~GHz. For region B3 we measure a polarization fraction as high as 40\%, while for the brightest part of B2 it is about 5\%, and for B1 it drops below 1\%. 
The average polarization fractions and depolarization properties for relic B are described in Sect.~\ref{sec:rx42depol}.

\begin{figure*}
\begin{center}
\includegraphics[angle =90, trim =0cm 0cm 0cm 0cm,width=0.49\textwidth]{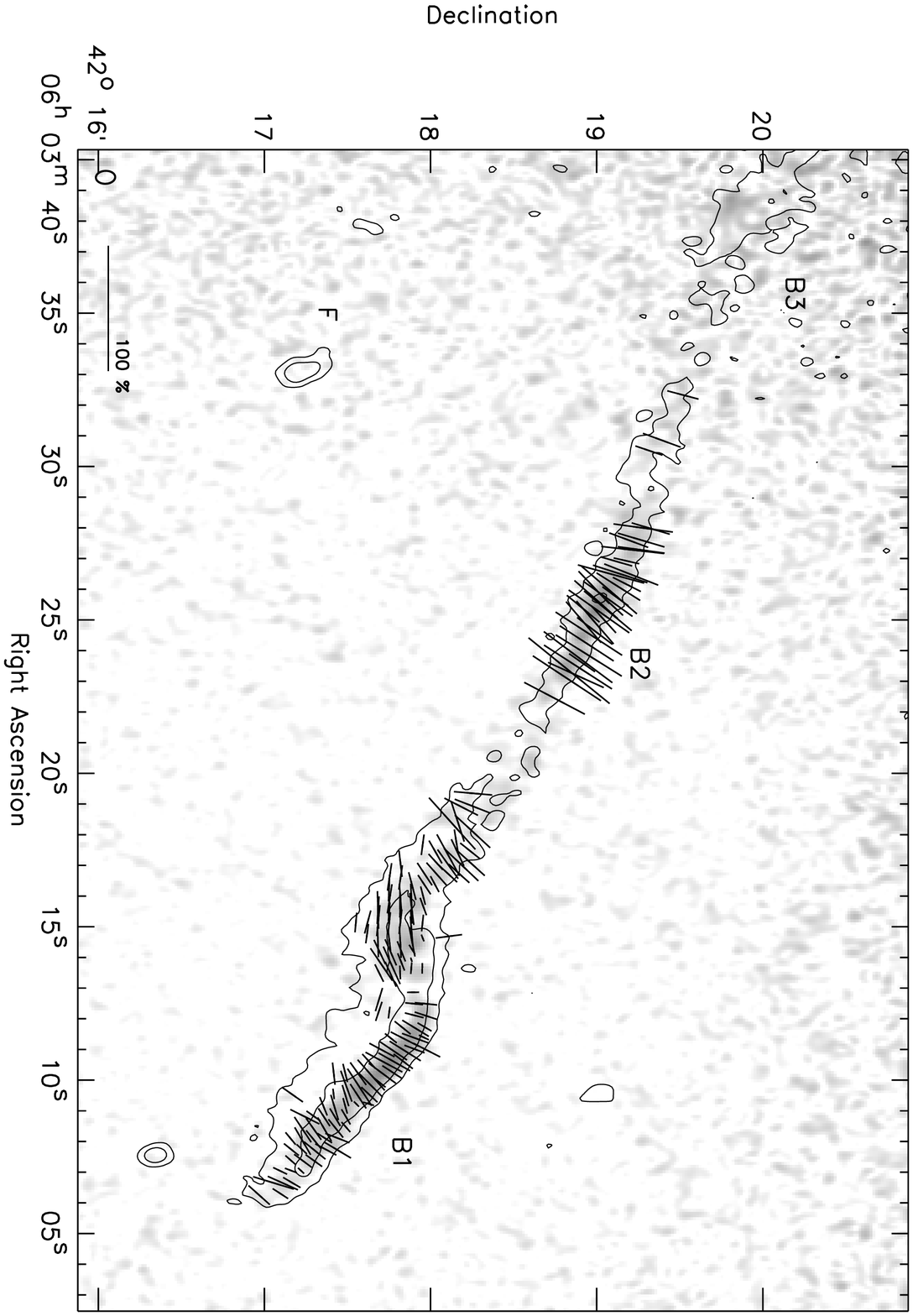}
\includegraphics[angle =90, trim =0cm 0cm 0cm 0cm,width=0.49\textwidth]{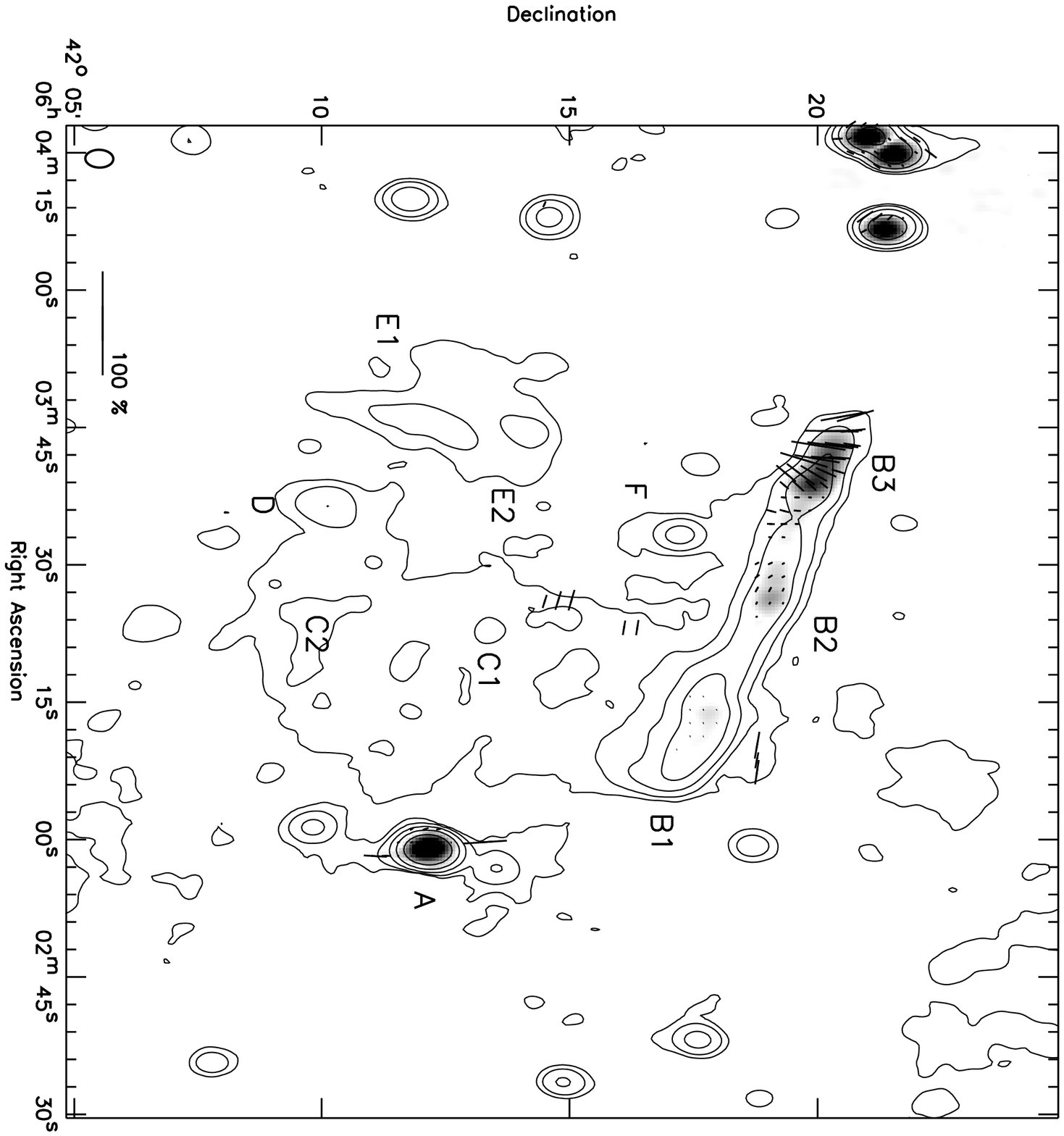}
\end{center}
\caption{Left: WSRT 4.9~GHz polarization E-vector map. Total polarized intensity 
is shown as grayscale image.  Vectors depict the polarization E-vectors, 
their length represents the polarization fraction. The length of the E-vectors are corrected 
for Ricean bias \citep{1974ApJ...194..249W}. 
A reference vector for a polarization fraction of 100\% is shown in 
the bottom left corner. No vectors were drawn for pixels with a 
SNR $< 3$ in the total polarized intensity image. 
Contour levels are drawn at ${[1, 4, 16, 64, \ldots]} \times 3\sigma_{\mathrm{rms}}$ 
and are from the Stokes I image. The beam size 
is $7.0\arcsec \times 4.7\arcsec$ and indicated in the bottom left corner of the image. Right: WSRT 1382~MHz polarization E-vector map made with natural weighting.  Total polarized intensity 
is shown as grayscale image. No vectors were drawn for pixels with a 
SNR $< 4$ in the total polarized intensity image. 
Contour levels are drawn at ${[1, 4, 16, 64, \ldots]} \times 0.18$~mJy~beam$^{-1}$.  
and are from the Stokes I image.}
\label{fig:rx42pol}
\end{figure*}

\subsection{RM-synthesis}
\label{sec:rx42rmsyn}
Faraday rotation changes the intrinsic polarization angle ($\chi_0$) depending on the wavelength ($\lambda$) or frequency of the radiation. The Faraday depth ($\phi$) is related to the properties of the plasma causing the Faraday rotation \citep{1966MNRAS.133...67B, 2005A&A...441.1217B}
\begin{equation}
\phi (\mathbf{r}) = 0.81 \int_{\mathrm{source}}^{\mathrm{telescope}} n_{\mathrm{e}}\mathbf{B} \cdot \, d\mathbf{r} \mbox{  } [\mbox{rad m}^{-2}  ] \mbox{ ,} 
\end{equation}
where $d\mathbf{r}$ is an infinitesimal path length in parsec along the line of sight, $\mathbf{B}$ the magnetic field in $\mu$Gauss and $n_{\mathrm{e}}$ the electron density in cm$^{-3}$. The sign is defined to be positive for a magnetic field pointing towards the observer. The rotation measure (RM) is defined as
\begin{equation}
\mathrm{RM} = \frac{d\chi(\lambda^2)}{d\lambda^2} \mbox{ .}
\end{equation}

If there is only one source along the line of sight (without internal Faraday rotation), the Faraday depth is equal the rotation measure (RM) at all wavelengths. In other words, all polarized emission is observed at a single Faraday depth $\phi$. The observed polarization angle ($\chi$) is then 
\begin{equation}
\chi =   \chi_0 + \phi \lambda^2 \mbox{ .}
\label{eq:chi0}
\end{equation}
In more complicated situations this relation is not valid \citep[e.g.,][]{1980A&A....86..251V, 1998MNRAS.299..189S}. By expressing the polarization as a complex vector $P = p e^{2i\chi} = Q + iU$, with $p$ the intrinsic polarization \cite{1966MNRAS.133...67B} showed that
\begin{equation}
P(\lambda^2) = \int_{-\infty}^{+\infty} F(\phi) e^{2i\phi \lambda^2} \, d\phi \mbox{ ,}
\label{eq:P}
\end{equation}
where $F(\phi)$ is the Faraday dispersion function, i.e., the complex polarized surface brightness per unit Faraday depth. Eq.~\ref{eq:P} is invertible if the intrinsic polarization angle $\chi_0$ is constant as function of $\lambda$, then $F(\phi)$ can be found by measuring $P(\lambda^2)$
\begin{equation}
F(\phi) = \int_{-\infty}^{+\infty} P(\lambda^2) e^{-2i\phi \lambda^2} \, d\lambda^2 \mbox{ .}
\label{eq:F}
\end{equation}
With modern correlator backends, such as the one at the WSRT or EVLA, $P(\lambda^2)$ can measured over a large number of frequency channels.  If $\phi \delta\lambda^2 \ll 1$, with $\delta\lambda^2$ the channel width in wavelength squared, \cite{2005A&A...441.1217B} showed that Eq.~\ref{eq:F} can be approximated as
\begin{equation}
\tilde{F}(\phi)  \approx K \sum_{j=1}^{N} w_{j} P_{j} e^{-2i\phi(\lambda_{j}^{2} - \lambda_{0}^{2} )} \mbox{ ,}
\label{eq:Fapprox}
\end{equation}
with $w_{j}$ being some weights and $K$ a normalization factor
\begin{equation}
K = \frac{1}{\left(    \sum_{j=1}^{N} w_j   \right)} \mbox{ .}
\end{equation}
{As explained by \citeauthor{2005A&A...441.1217B}, adding the term $\lambda_{0}^{2}$ in the exponent of  Eq.~\ref{eq:Fapprox} results in a better sidelobe pattern for the rotation measure spread function (RMSF), with} 
\begin{equation}
\lambda_{0}^{2} = \frac{\sum_{j=1}^{N} w_j \lambda_{j}^{2}    }{\sum_{j=1}^{N} w_j} \mbox{ ,}
\end{equation}
and the RMSF given by
\begin{equation}
\mathrm{RMSF}(\phi) = K \sum_{j=1}^{N} w_j e^{-2i\phi(\lambda_{j}^{2} - \lambda_{0}^{2} )} \mbox{ .}
\end{equation}
Because $P$ is not measured for every possible $\lambda_{j}^{2}$, the true Faraday depth function is related to Eq.~\ref{eq:Fapprox} by a convolution
\begin{equation}
\tilde{F}(\phi) = F(\phi) \ast \mathrm{RMSF}(\phi) \mbox{ .} 
\end{equation}
Depending on the sidelobe structure of the RMSF (which can be adjusted by choosing the weights $w_j$) a brighter component can contaminate the response of fainter components in $\tilde{F}(\phi)$ and therefore deconvolution might be necessary. The one-dimensional deconvolution algorithm which has been used is a simple extension of \cite{1974A&AS...15..417H} {\tt RM-CLEAN} algorithm to the complex domain as described by \cite{brentjens_phd}, see also \cite{2009A&A...503..409H}. It works as follows  
\begin{enumerate}
\item{For a spatial pixel the maximum of $\left|{\tilde{F}(\phi)} \right| $} is found and the location of the peak is stored $(\phi_{\mathrm{peak}})$.  
\item{If this maximum value is higher than a user defined cutoff,  the RMSF is shifted to the location of the peak  $(\phi_{\mathrm{peak}})$ and a scaled version (i.e., using a gain of 0.1) is subtracted from ${\tilde{F}(\phi)}$. Since the RMSF is complex, a multiplication with a phase factor is needed which  depends on the phase of ${\tilde{F}(\phi_{\mathrm{peak}})}$.}
\item{The complex scaling factor, which was used to shift and subtract the  RMSF, is stored as a clean component.}
\item{Steps $2-4$ are repeated until a user defined threshold is reached or the number of iterations has reached a predefined maximum.  }
\item{Optionally, the clean components are restored with Gaussians with a FWHM matched to that of the RMSF. The Gaussians are again shifted to their respective locations ($\phi_{\mathrm{peak}}$), and since this happens in the complex domain, multiplied by a phase factor (i.e., the complex part of the clean component). The restored Gaussians are added to the residual from step 4. If there are multiple RM components within the beam it is however not a good idea to reconvolve the spectra with the RMSF if one wants to investigate the depolarization properties. This is related to the fact that the clean components are complex numbers and can interfere with each other \citep{2011AJ....141..191F}.}
\item{The algorithm then continues with the next spatial pixel.}
\end{enumerate}

The above described RM-synthesis technique works better than linear fitting techniques \citep[e.g.,][]{1979A&A....78....1R, 2005MNRAS.358..726D, 2005MNRAS.358..732V},  which break down in low signal to noise regimes and do not apply to more complex situations when Eq.~\ref{eq:chi0} is not valid.

The RM-synthesis technique (i.e., Eq.~\ref{eq:Fapprox}) and {\tt RM-CLEAN} algorithm were implemented in IDL. 

\subsubsection{Application to the L-band WSRT data}
We performed RM-synthesis on the WSRT 25, 21 and 18 cm data using cubes of Stokes Q and U images with a resolution of $40\arcsec\times28\arcsec$, see also Sect.~\ref{sec:rx42wsrt}. These data give a sensitivity to polarized emission up to a maximum
Faraday depth of $\left| \phi_{\mathrm{max}} \right| \approx \sqrt{3} / \delta \lambda^{2} \approx 8.8 \times 10^{4}$~rad~m$^{-2}$ \citep{2005A&A...441.1217B}. The three bands provide a resolution of $\delta\phi \approx 2\sqrt{3}/\Delta\lambda^2 =  88$~rad~m$^{-2}$, with $\Delta \lambda^2 = \lambda_{\mathrm{max}}^{2} - \lambda_{\mathrm{min}}^{2}$ the total bandwidth in wavelength squared. The maximum scale in $\phi$ space to which the sensitivity has dropped to 50\%  is approximately $ \pi/ \lambda_{\mathrm{min}}^{2} = 111$~rad~m$^{-2}$. The RMSF is shown in Fig~\ref{fig:rx42rmbeam}. Since the first sidelobe of the RMSF is about 65\% of the main lobe, we used the {\tt RM-CLEAN} algorithm, cleaning down to a threshold of 0.12~mJy~beam$^{-1}$~RMSF$^{-1}$. The distribution of the peak of $\left|{\tilde{F}(\phi)} \right|$, the so-called ``rotation measure map'' (or Faraday depth map), is shown in Fig.~\ref{fig:rx42rmmap}.

We extracted the location of the peak in the Faraday depth ($\phi_{\rm{peak}}$) for all compact sources within 0.5\degr~radius of the cluster. In total for 17 sources we find an outlier-resistant mean of $+11.7 \pm 3.0$~rad~m$^{-2}$, clipping values more than two standard deviations away from the median,  with the uncertainty the standard deviation of the mean of the $\phi_{\rm{peak}}$ distribution. This is in agreement with the value of 11~rad~m$^{-2}$ from \cite{2009ApJ...702.1230T} derived from NVSS data measured over scales of 8\degr. The $\left|{\tilde{F}(\phi)} \right|$ spectra for seven compact sources are displayed in Fig.~\ref{fig:rx42rmspec} (left panel). 

For the relic, $\phi_{\rm{peak}}$ values vary spatially from about $-53$ to $+55$~rad~m$^{-2}$. For B3, $\phi_{\rm{peak}}$ is about 10~rad~m$^{-2}$, similar to the compact sources in the field. For B2, we find a strong gradient from $-10$ to $+ 55$~rad~m$^{-2}$ from east to west. For B1, $\phi_{\rm{peak}}$ varies between $-10$ and $-53$~rad~m$^{-2}$. The ``clean component'' spectra are clearly resolved, see Fig.~\ref{fig:rx42rmspec} (right panel). 

Interestingly, the $\phi_{\rm{peak}}$ values for B1 and B2 deviate from the average galactic foreground of $+11.7 \pm 3.0$~rad~m$^{-2}$, indicating that some of the Faraday rotation is probably caused by the ICM. The $\phi_{\rm{peak}}$ value for B3 indicates it is located in the cluster outskirts and the line of sight towards it does not pass deep into the ICM. B2 and B1 seem to be located progressively deeper into the ICM of the cluster (or more behind it) given the larger $\phi_{\rm{peak}}$ deviation from the galactic foreground. Although the number of counts is low in the ROSAT image (Fig.~\ref{fig:rx42wsrtlband}), it is indeed expected that $\int_{\mathrm{source}}^{\mathrm{telescope}} n_{\mathrm{e}} \, dr$ increases from B3, B2, to B1.

\begin{figure}
\begin{center}
\includegraphics[angle =90, trim =0cm 0cm 0cm 0cm,width=0.5\textwidth]{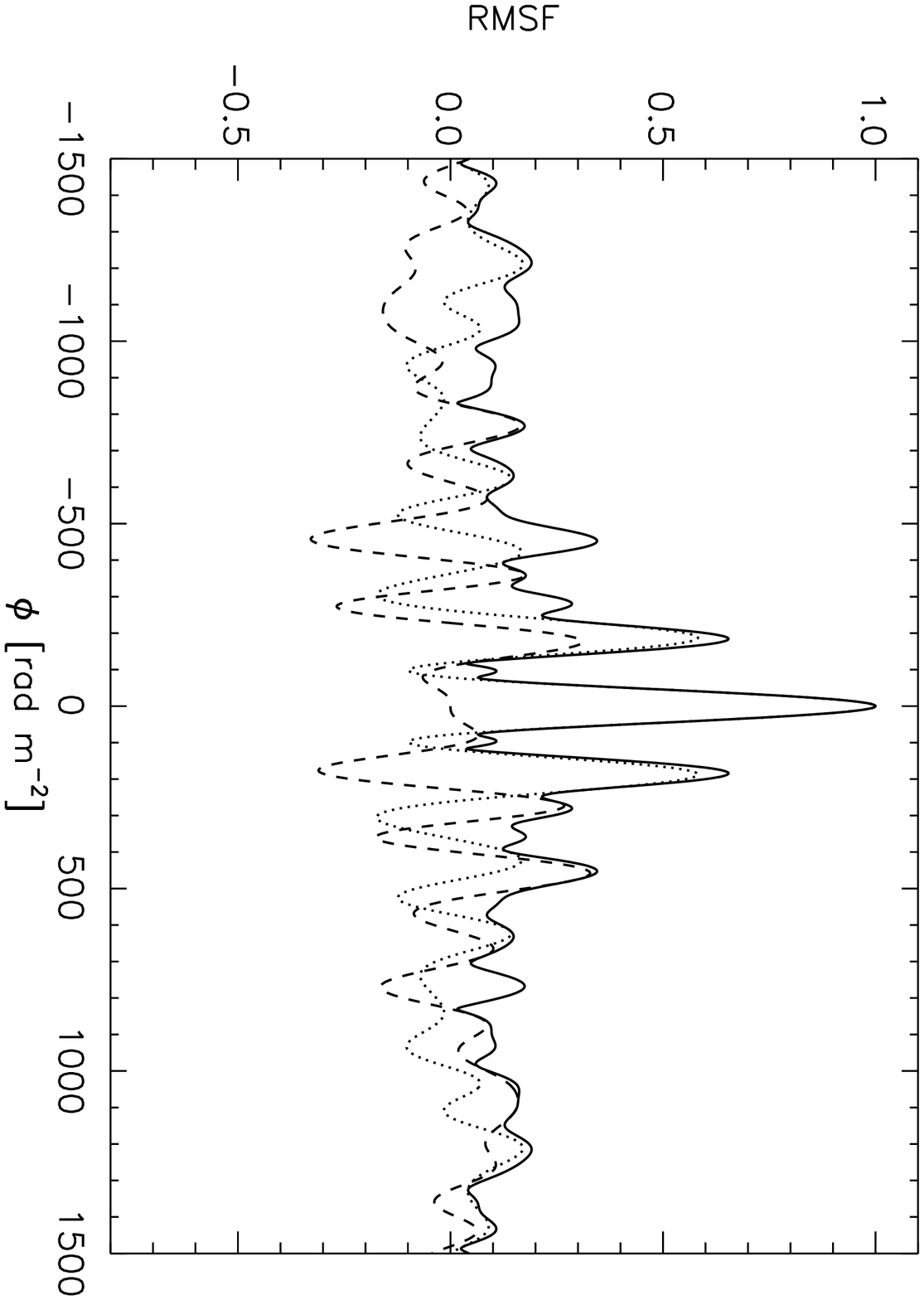}
\end{center}
\caption{RMSF from the 25, 21, 18 cm bands WSRT data. The solid line is the overall amplitude of the complex RMSF. Dotted line depicts the Real part of the RMSF and the dashed line the Imaginary part. The FWHM of the RMSF is 88~rad~m$^{-2}$. }
\label{fig:rx42rmbeam}
\end{figure}

\begin{figure}
\begin{center}
\includegraphics[angle =90, trim =0cm 0cm 0cm 0cm,width=0.49\textwidth]{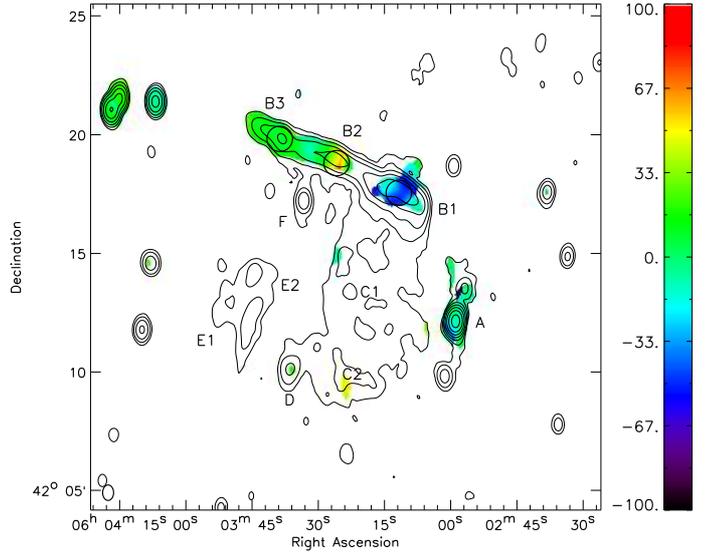}
\end{center}
\caption{Faraday depth value of the peak in $F(\phi)$. Contours are from the WSRT 1382~MHz image and drawn at levels of  ${[1, 3, 9, 27, \ldots]} \times 0.225$~mJy~beam$^{-1}$.}
\label{fig:rx42rmmap}
\end{figure}

\begin{figure*}
\begin{center}
\includegraphics[angle =90, trim =0cm 0cm 0cm 0cm,width=0.47\textwidth]{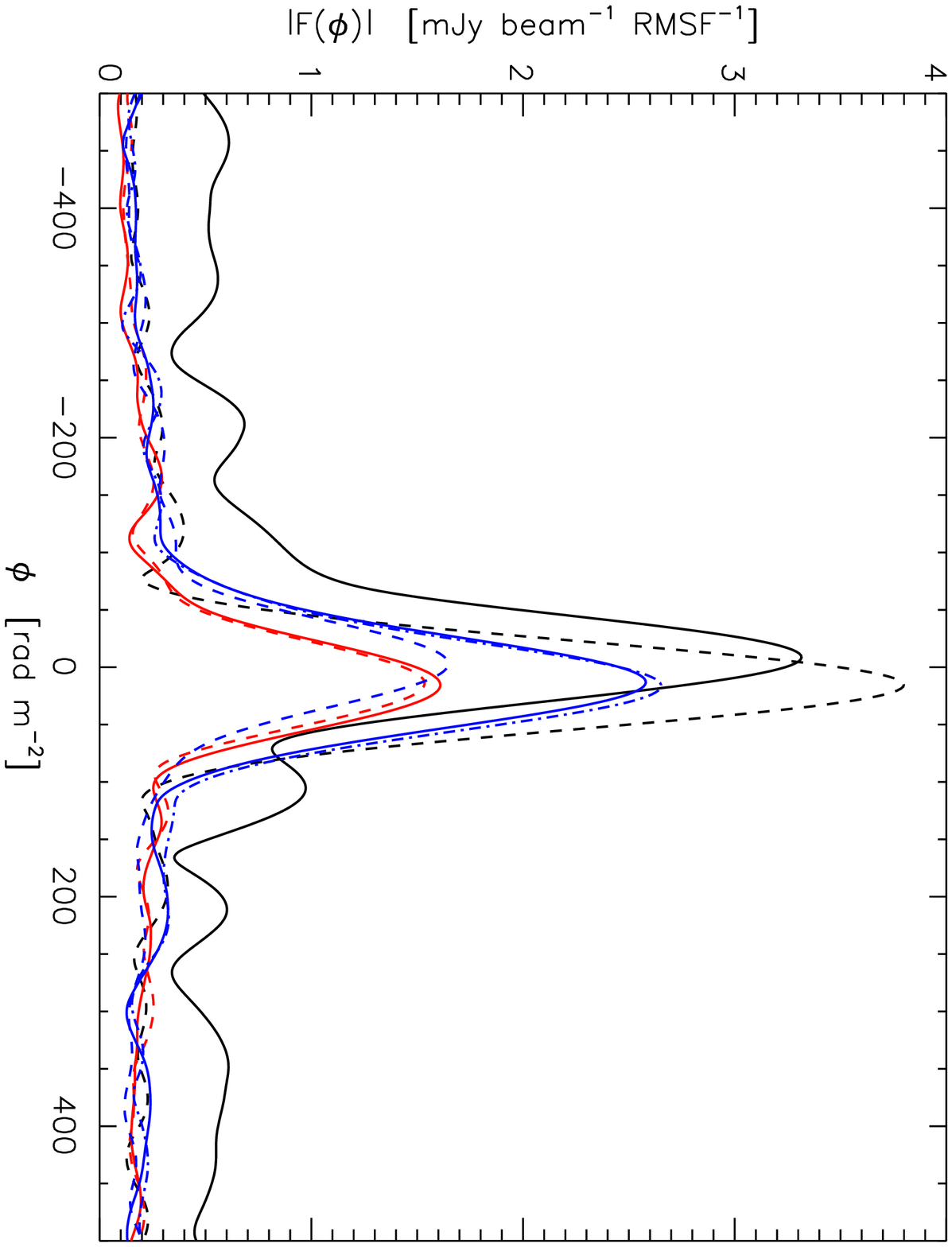}
\includegraphics[angle =90, trim =0cm 0cm 0cm 0cm,width=0.50\textwidth]{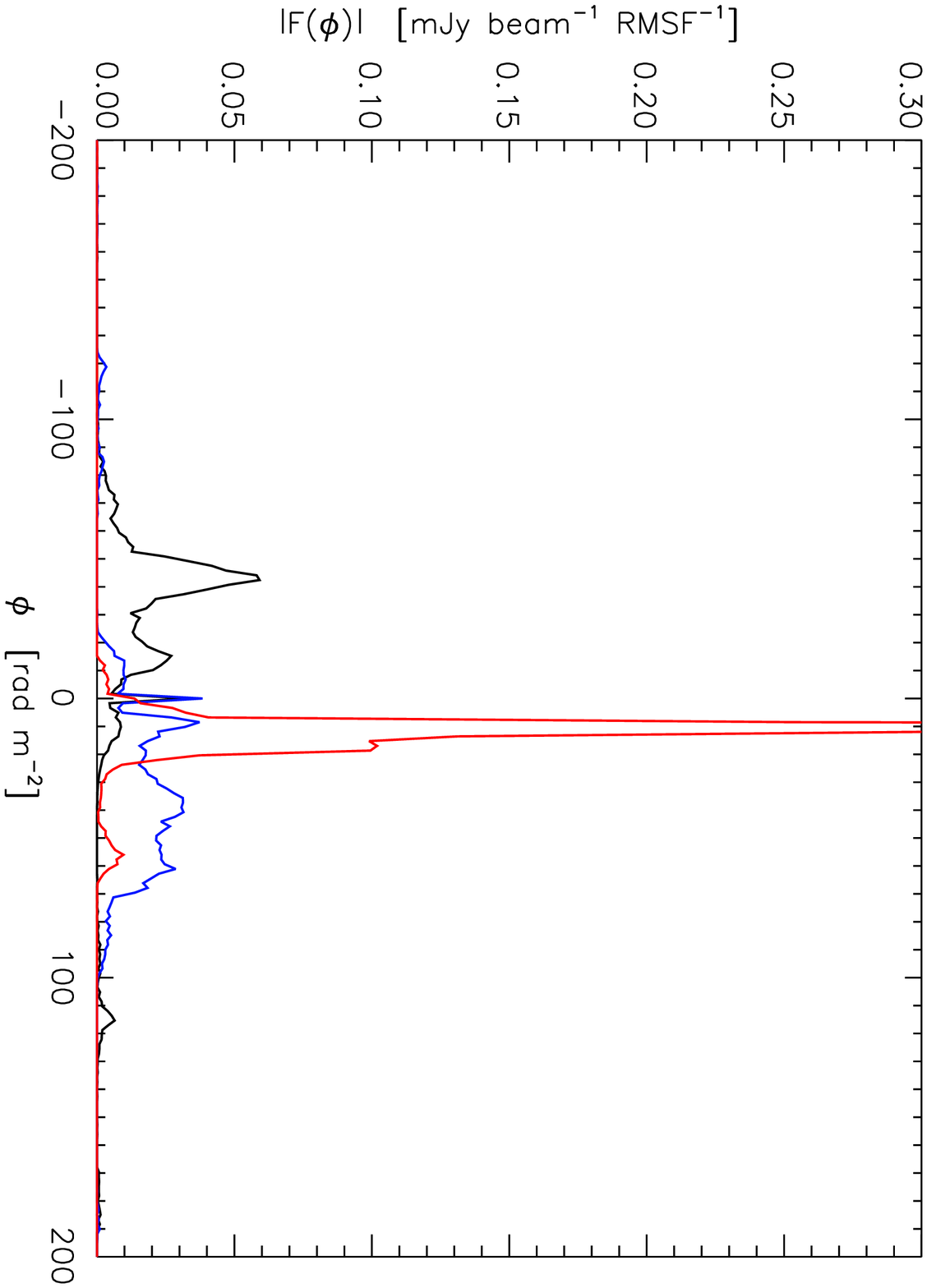}
\end{center}
\caption{Left: Gaussian restored $|F(\phi)|$ spectra of compact sources. B3~0559+422A black solid line, NVSS~J060406+422125 blue dashed line, NVSS~J060416+422110 (south component)  blue solid line, NVSS~J060416+422110 (north component) dashed-dotted blue line, NVSS~J060231+422811 (west component) red dashed line, NVSS ~J060231+422811 (east component) red solid line, and NVSS~J060206+422045 black dashed line.  The high ``residuals'' for B3~0559+422A are an artifact due to remaining calibration errors. Right: $|F(\phi)|$ spectra, consisting of clean components, extracted at three different positions from relic B. The positions are indicated by the circles drawn in Fig.~\ref{fig:rx42rmmap}. Red line: position B3,  blue line: position B2, and black line: position B1 (see also Fig.~\ref{fig:rx42wsrtlband}).}
\label{fig:rx42rmspec}
\end{figure*}

\subsection{The depolarization properties of the radio relic}
\label{sec:rx42depol}
Our observations show a significant decrease in the polarization fraction towards lower frequencies. There are several factors that can cause such an effect. We rule out the effect of bandwidth depolarization, as the frequency resolution of the observations is sufficient to detected polarized emission with a Faraday depth $>10^{4}$~rad~m$^{-2}$ using RM-synthesis. A second possibility is beam depolarization due to multiple RM components within the beam (spatially varying). These different RM components can be caused by  variations of the magnetic field and/or electron density.  The spatially averaged polarization fraction over B1 and B2 is about 13\% at 4.9~GHz, 1.6\% at 1714~MHz cm and $< 1\%$ at 1382 and 1221~MHz. We excluded B3 from the analysis because the SNR is too low at 4.9~GHz due to the primary beam attenuation.

 Two types of depolarization, \emph{internal} and \emph{external}, have been identified \citep{1966MNRAS.133...67B,1991MNRAS.250..726T}. Internal depolarization takes place within the radio source itself, while external depolarization takes place in a medium between the source and the observer, e.g., the ICM or magnetized plasma from our own galaxy. A simple case of internal depolarization is for a uniform magnetic field and electron density within the source. In this slab model \citep{1966MNRAS.133...67B}, the fractional polarization as function of $\lambda^2$ is given by a sinc-function. From this, the extent in Faraday depth of the source can be estimated. Observations indicate that the uniform slab model does not fit the data for the radio relics in A2256 and the Coma cluster \citep{2008A&A...489...69B, pizzo_phd}. 
 
For B1 and B2 there is a clear trend of depolarization towards higher $\lambda^2$.  The only way to get depolarization as a function of wavelength is to have multiple RM components within the beam, either across it or along it -- external --, or mixed with the synchrotron emitting plasma -- internal. Therefore, the Faraday spectrum must be resolved wherever there is depolarization. In practice, because of the finite width of the RMSF, it may not always be possible to see these multiple RM components, even though they are depolarizing. In the case of B1 and B2, the spectra are indeed clearly resolved, consistent with the high amount of depolarization found for these regions. It cannot be directly determined whether this is due to beam depolarization (spatially varying RM components) or due to multiple RM components along the line of sight. In the first case, the depolarization should vanish when going to high enough spatial resolution. The fact that the RM varies spatially from beam to beam across B1 and B2 makes it very likely that at least a significant part of the depolarization is indeed caused beam depolarization.
 
The trend from east to west over the relic of (i) a larger $\phi_{\rm{peak}}$ deviation from the galactic foreground, and (ii) an increasing depolarization towards higher $\lambda^2$, suggest that most of the depolarization is probably due to the ICM of the cluster itself and not due to internal depolarization.

\section{Discussion}
\label{sec:rx42discussion}
The complex diffuse radio emission and extended X-ray emission give strong support for the fact that the cluster is currently undergoing a merger event.  There are however a few puzzling aspects to the diffuse emission in this cluster. The linear extension of relic B to the east is quite peculiar. Usually shock surfaces are curved and not very linear over distances of more than a Mpc \citep[e.g.,][]{2008MNRAS.391.1511H, 2009MNRAS.393.1073B, 2009MNRAS.395.1333V,2011ApJ...735...96S}. The ongoing merger event for  1RXS~J0603.3+4214 is probably more complex than the ``simple'' binary merger events that are thought to give rise to symmetric double radio relics \citep[e.g.,][]{1999ApJ...518..603R, 2011MNRAS.418..230V}. Deep X-ray observations will be needed to investigate whether the relic is indeed associated with such a linear shock front. 

Relic E  could also trace a shock wave with DSA, given its large extent and relatively flat integrated spectral index. The nature of relic D remains unclear, its small size and morphology would suggest a radio phoenix scenario, but the integrated spectral index of $-1.10 \pm 0.05$ is not consistent with this interpretation. Higher-resolution observations ($\sim 1 \arcsec$) would be helpful to better delineate the morphology.

An interesting point is the relation between relic B and the radio halo C. The radio halo is located directly south of B1, the brightest part of relic B. The halo and relic are connected by a region with $\alpha \lesssim -2$, but then the spectral index flattens to $\alpha \sim -1.2$ and increases a bit towards the center of the halo. 
We speculate that relativistic particles, previously accelerated at the shock front, are re-accelerated due to merger induced turbulence and then form the radio halo. Another interesting feature of the halo is the increased surface brightness at its southern end. Why does the surface brightness increase here?
For a binary cluster merger event one expects two shock waves to form traveling diametrically outwards \cite[e.g.,][]{1999ApJ...518..603R}. If relic B traces the main shock wave one could also expect a second shock waves at the south side of the cluster center. This could be the region of increased surface brightness at the southern end of C. Although, this remains somewhat speculative, also because the merger event seems to be complex given the peculiar linear extension of relic B. 
 X-ray observations could reveal whether the south part of radio halo C is associated with a shocked region with a higher ICM temperature. 

There is also a possibility that source C is not a radio halo, but another relic seen close to face-on, because the surface brightness does not clearly peak at the location of the ROSAT X-ray peak. However, the radio power of source C is in agreement with the $L_{\rm{X}}$--$P_{\rm{1.4GHz}}$ correlation. Also, the brightness distribution over C is regular and quite smooth, which argues against a relic interpretation.  Relics are often composed of smaller filamentary structures  \citep[e.g.,][]{2006AJ....131.2900C}. We note that source C is somewhat similar to the diffuse emission seen between the relics in CIZA~J2242.8+5301 \citep{2010Sci...330..347V}.

\subsection{Relic spectra}
The analysis of the radio spectra for radio relic B are consistent with the existence a two global spectral shapes, one that describes the spectrum for the east part and one for the west part. This because it is possible to line up all radio spectra onto a single shape with shifts along $\log{(I)}$ and $\log{(\nu)}$.

The spectral shapes of the west (B1) and east (B2+B3) regions only differ in regions in the spectral index map with  $\alpha \lesssim -1.4$ (e.g., see Fig.~\ref{fig:rx42colorcolor}). 
At the front part of the relic we find power-law radio spectra, consistent with an injection spectral index of $-0.6$ to $-0.7$. The injection spectral index is also consistent with the total integrated spectral index of $-1.1$, i.e., the integrated spectral index should be $-0.5$ units steeper than the injection spectral index \citep[e.g.,][]{2002MNRAS.337..199M, 2002NewA....7..249B} for a simple shock model (with the properties of the shock not changing over time). We also find that the shape of the radio spectra are mostly governed by the effects of spectral ageing, by looking at the shifts made to align up all individual spectra to the global spectral shape (see Sect.~\ref{sec:rx42shift}). Changes in the magnetic field, total electron content, or electron energy (except energy changes due to ageing) do not seem to play a major role. 

The observed radio spectra are not consistent with a JP model (which consists of a single burst of injection and spectral ageing). Instead, we find that  the KGJP model, which consist of a sum of individual JP spectra with different amounts of ageing, provides a good match to the spectrum of the east part of the relic. This can be thought of as mixing of emission, most likely because of projection effects. For the west part of the relic (in the regions with the steepest spectral index), we find deviations from the KGJP model. We speculate that additional mixing of emission takes place here, or the spectra are contaminated by the emission from radio halo C (although we attempted to subtract this emission). 

The KGJP spectra indicate that (1) at the front of the relic we see freshly injected radio plasma, giving a power-law spectrum. Little mixing of emission takes place here. (2) For the middle part of the relic (the region $-0.9 \lesssim \alpha \lesssim -1.1$ from the spectral index map), we observe a mix of radio aged radio plasma and freshly injected radio plasma. For the back part of the relic  (the region with $\alpha \lesssim -1.1$), we see a mix of aged radio plasma, but no freshly injected plasma. 

Mixing of emission thus seems to be an important factor determining the spectral shape. This suppresses spectral curvature and pushes back the radio spectra to power-law shapes. It is therefore crucial that this effect is considered. The lack of spectral curvature does therefore not directly imply spectral ageing is not important. 

An injection spectral index of {$-0.6$ in combination with DSA, implicates a Mach number between $4.6$, quite high. The minimum allowed injection spectral index of $-0.7$, gives a Mach number of $3.3$}. Instead of DSA, another possibility is shock re-acceleration of fossil electrons. In practice, these radio spectra will probably be indistinguishable from DSA \citep{2005ApJ...627..733M}. {The $\mathcal{M} > 3$ we find is interesting. Simulations indicate that merger shocks with these Mach numbers are rare \citep[e.g.,][]{2011MNRAS.418..960V,2012MNRAS.tmp.3005A}. Deep all-sky radio surveys are therefore very important to pinpoint these extreme cluster merger shocks.}

\subsection{Alternative models to explain the relativistic electrons from the radio relic}

An alternative explanation for radio relics, based on a secondary cosmic ray electron model, has been proposed by \cite{2010arXiv1011.0729K}. This model predicts a spectral index of $-1$ at the front of the relic. We do not observe this, although there is a hint of a steeper spectral index at the front of B1 at some places in the 325--610~MHz spectral index map. However, this could be an artifact, as edge pixels is the spectral index map are not reliable. Also these steeper regions are smaller than the beam size.
We do not find indications of magnetic field changes across the relic from the shifts needed to align up the spectra from individual regions.
On the other hand, in the images we see a morphological connection between the radio relic and halo, these halo-relic ``bridges'' are included in the model from \cite{2010arXiv1011.0729K}.  

\cite{2001A&A...366...26E} and \cite{2002MNRAS.331.1011E} proposed that (some) radio relics could be the result of adiabatically compressed fossil radio plasma due to a shock wave. This could boost the radio luminosity by about two orders of magnitude. The question is whether the large relic in 1RXS~J0603.3+4214 can be explained in this way. 
The model of \cite{2001A&A...366...26E} consist of five phases: (1) injection by a radio galaxy, (2) expansion of the radio cocoon, during this phase the cocoon becomes undetectable, (3) the ``lurking''-phase, when pressure equilibrium is reached and the volume of the cocoon remains constant. Due to significant adiabatic losses most of the electrons reside at low energies which reduces the radiation losses. The losses can be further reduced by a low-magnetic field. (4) Adiabatic compression of the cocoon, the radio emission is boosted so it becomes visible again and the break frequency moves upwards. (5) The radio emission fades away due to  synchrotron and IC losses.

In the case of 1RXS~J0603.3+4214, the fossil plasma should be quite old ($\gtrsim 1$~Gyr) as it takes a considerable amount of time for a radio galaxy to move over a distance of 2~Mpc and dump the radio plasma.  
The change in the break frequency from the adiabatic compression is \citep{2008A&A...486..347G}
\begin{equation}
\frac{\nu^{\rm{after}}_{\rm{brk}}}{\nu^{\rm{before}}_{\rm{brk}}} \approx \frac{5\mathcal{M}^2 -1}{4} \mbox{ .}
\end{equation}
Adopting a reasonable value of $\nu^{\rm{before}}_{\rm{brk}} = 50$~MHz for 1~Gyr old fossil radio plasma \citep[e.g.,][]{2001A&A...366...26E}, we would need a $\mathcal{M} > 8$ shock to push $\nu^{\rm{after}}_{\rm{brk}}$ above 4~GHz. For more realistic Mach numbers of $\mathcal{M} \lesssim 4$ we should thus observe steep and curved radio spectra. This contradicts the flat spectral indices of $\alpha=-0.6$ we find and the straight power-law integrated radio spectrum.

Only in the unlikely scenario that the plasma is  about 0.1~Gyr old, before it is compressed by the shock, the break frequency could be pushed up to a few GHz. We also note that to produce a spectral index gradient a high mass load is required, which would allow the shock wave to successively compress the radio plasma. In the case of a high mass load due to undetectable cool gas, a shock forms and the adiabatic gains are restricted due to the limited compression factor of such a shock wave \citep{2001A&A...366...26E}.

\subsection{Magnetic field}
\label{sec:rx42eqB}
The RM for the west part of the main relic (B1) deviates from the foreground RM, determined from compact sources in the field around the cluster. This suggests that the west part of the relic is located deeper into the ICM,  also consistent with the higher amount of depolarization found for that part of the relic. A constant magnetic field of 1~$\mu$Gauss,  $n_{\rm{e}} = 1 \times 10^{-4}$~cm$^{-3}$ and a path length 500~kpc, will give a Faraday depth of about 40~rad~m$^{-2}$. The observed difference of about 60~rad~m$^{-2}$, compared to the galactic foreground value, suggests a relatively high magnetic field ($B \gtrsim 1$~$\mu$Gauss). Also because the magnetic field topology is likely more complex, i.e., the magnetic field could have reversals along the line of sight. 

We estimate the magnetic field at the front of relic B1 by assuming 
minimum energy densities in the relics \citep[e.g.,][]{2004IJMPD..13.1549G}} and using the same procedure 
as described in \cite{2009A&A...506.1083V}. We take $k=100$, i.e, the 
ratio of energy in relativistic protons to that in electrons,  250~kpc for the depth ($d$) along the line of sight, a spectral index of $-0.6$, and a surface brightness of 0.205~mJy~arcsec$^{-2}$ at 610~MHz. We use low and high 
energy cutoffs ($\gamma_{\mathrm{min}}$ and $\gamma_{\mathrm{max}}$, with $\gamma_{\mathrm{min}} \ll  \gamma_{\mathrm{max}}$) instead of low and high frequency cutoffs \citep{2005AN....326..414B, 1997A&A...325..898B}, giving the so-called  ``revised'' equipartition magnetic strength ($B^{\prime}_{\mathrm{eq}}$). For $\gamma_{\mathrm{min}}=100$  this gives $B^{\prime}_{\mathrm{eq}}=9.2$~$\mu$Gauss.  If we take 
$\gamma_{\mathrm{min}}=5000$ we get $B^{\prime}_{\mathrm{eq}} = 7.4$~$\mu$Gauss. The revised equipartition 
magnetic field strength scales with $(1+k)^{1/(3-\alpha)}$. 

Both the polarization and equipartition results indicate  a relatively high magnetic field strength. These relatively high values have also been found for other relics by \cite{2010ApJ...715.1143F,2010Sci...330..347V} using limits on the IC X-ray emission or modeling the relic's brightness profile.

\section{ Conclusions}
\label{sec:rx42conclusion}
We presented detailed GMRT and WSRT radio observations of a newly discovered cluster of galaxies at $z = 0.225$. We find the cluster to host three radio relics and also a giant $\sim 2$~Mpc radio halo. The radio power follows the $L_{\rm{X}}$--$P_{\rm{1.4GHz}}$ correlation for giant radio halos. 
The extended X-ray emission and complex diffuse radio emission indicate we are witnessing an ongoing cluster merger event. 

The WSRT observations reveal that the front of the relic is highly polarized, with a polarization fraction of up to $\sim 60\%$ at 4.9~GHz. 
At lower frequencies (in the L-band), the polarization fraction drops considerably, as has been seen for other well studies relics, e.g., in A2256 \citep{2008A&A...489...69B} and A2255 \citep{2011A&A...525A.104P}. The observed depolarization and Faraday depth suggest that the west part of the main relic is located deeper into the ICM of the cluster. Using equipartition argument we find a high magnetic field strength of about 7--9~$\mu$Gauss for the bright relic.

For the radio halo we find a spectral index of $\alpha = -1.15 \pm 0.06$. The  spectrum for the bright radio relic is a power-law between 74~MHz and 4.9~GHz with $\alpha = -1.10\pm 0.02$. If this relic traces a shock, where particles are accelerated by the DSA mechanism, {then the injection spectral index of $-0.7$ to $-0.6$ would imply a Mach number of $3.3$ to $4.6$}.  The spectral index steepens systematically across the width of the relic from about $-0.6$ to $\sim-2$. For the bright western part of the relic we can trace the spectral steepening to $\alpha<-2.5$.   The radio spectrum at the front of the relic has a power-law shape with $\alpha = -0.6$ to $-0.7$, while the amounts of spectral curvature increases gradually towards the back of the relic. 

We  analyzed the radio spectra making use of color-color and shift diagrams \citep{1993ApJ...407..549K,1994ApJ...426..116K} and divided the relic into an eastern and western part. Both parts were further divided into smaller regions based on the spectral index map in Fig.~\ref{fig:rx42spix_poly} (left panel). We found that the individual radio spectra in each region could all be lined up. This implies the existence of a single global electron energy distribution. The shifts needed to align up the individual spectra, indicate that spectral ageing is the dominant factor explaining the changes in the spectra from one region to another.  Changes in the magnetic field, total number of electrons, or adiabatic expansion/compression do not seem to be important. 

In addition, we find evidence for mixing of radio emission with different amounts of spectral ageing within the beam. The amount of mixing increases away from the front of the relic. A so-called KGJP model (see Sect.~\ref{sec:rx42spectralmodels}), which can be thought of as a mix of spectra with different ages,  describes the global spectral shape for the east part of the relic. For the west part of the relic, in particular  the region with the steepest spectral index, we find a small deviation from this shape, possibly because the amount of mixing increases here.

The effects of spectral curvature are suppressed by mixing and this pushes the radio spectra closer towards power-law shapes. This means that if a spectral index gradient is seen, but no spectral curvature (or only very little), this does not necessarily imply that the spectral index gradient is not related to spectral ageing.

We conclude that the color-color and shift diagrams provide a  powerful tool to constrain physical conditions that shape the radio spectra. In the future we plan (i) to expand the color-color analyses to smaller regions, by obtaining higher resolution maps, (ii) increase the frequency range to $< 150$~MHz or $> 3$~GHz, and (iii) apply the technique to other radio relics and halos.

 For the origin of the relativistic  electrons (in the bright relic), the results favor a scenario where particles are accelerated or re-accelerated in a shock. {Interesting is the high Mach number we found, simulations indicate that merger shocks with these Mach numbers are rare.} A crucial test would be the to measure the Mach number directly from the jump conditions and compare this with $\alpha_{\rm{inj}}$. Deep X-ray observations of this cluster and simulations are also needed to constrain the merger scenario and determine the dynamical state of the cluster.

\begin{acknowledgements}
We would like to thank the anonymous referee for useful comments. 
We thank the staff of the GMRT who have made these 
observations possible. The GMRT is run by the National 
Centre for Radio Astrophysics of the Tata Institute of 
Fundamental Research. The Westerbork Synthesis Radio 
Telescope is operated by ASTRON (Netherlands 
Institute for Radio Astronomy) with support from the 
Netherlands Foundation for Scientific Research (NWO). 
The William Herschel Telescope are operated on the island of 
La Palma by the Isaac Newton Group in the Spanish 
Observatorio del Roque de los Muchachos of the 
Instituto de Astrof\'{\i}sica de Canarias. 

This publication makes use of data products from the Two Micron All Sky Survey, which is a joint project of the University of Massachusetts and the Infrared Processing and Analysis Center/California Institute of Technology, funded by the National Aeronautics and Space Administration and the National Science Foundation.

Support for this work was provided by NASA through Einstein Postdoctoral
grant number PF2-130104 awarded by the Chandra X-ray Center, which is
operated by the Smithsonian Astrophysical Observatory for NASA under
contract NAS8-03060.
RJvW acknowledges funding from the Royal 
Netherlands Academy of Arts and Sciences. MB and MH acknowledges support by the research group FOR 1254 funded by the Deutsche Forschungsgemeinschaft. LR acknowledges support from the U.S. National Science Foundation, under grant AST0908668 to the University of Minnesota. HTI is Jansky Fellow of the National Radio Astronomy Observatory. 
RJvW would like to thank  C.~Pfrommer for discussions and G.~Heald for his explanation of the {\tt RM-CLEAN} algorithm. 

\end{acknowledgements}
\bibliographystyle{aa}
\bibliography{ref_4x42}

\begin{appendix}

\section{Spectral index error maps} 

\label{sec:spixerror} 

\begin{figure*}
\begin{center}
\includegraphics[angle =90, trim =0cm 0cm 0cm 0cm,width=1.0\textwidth]{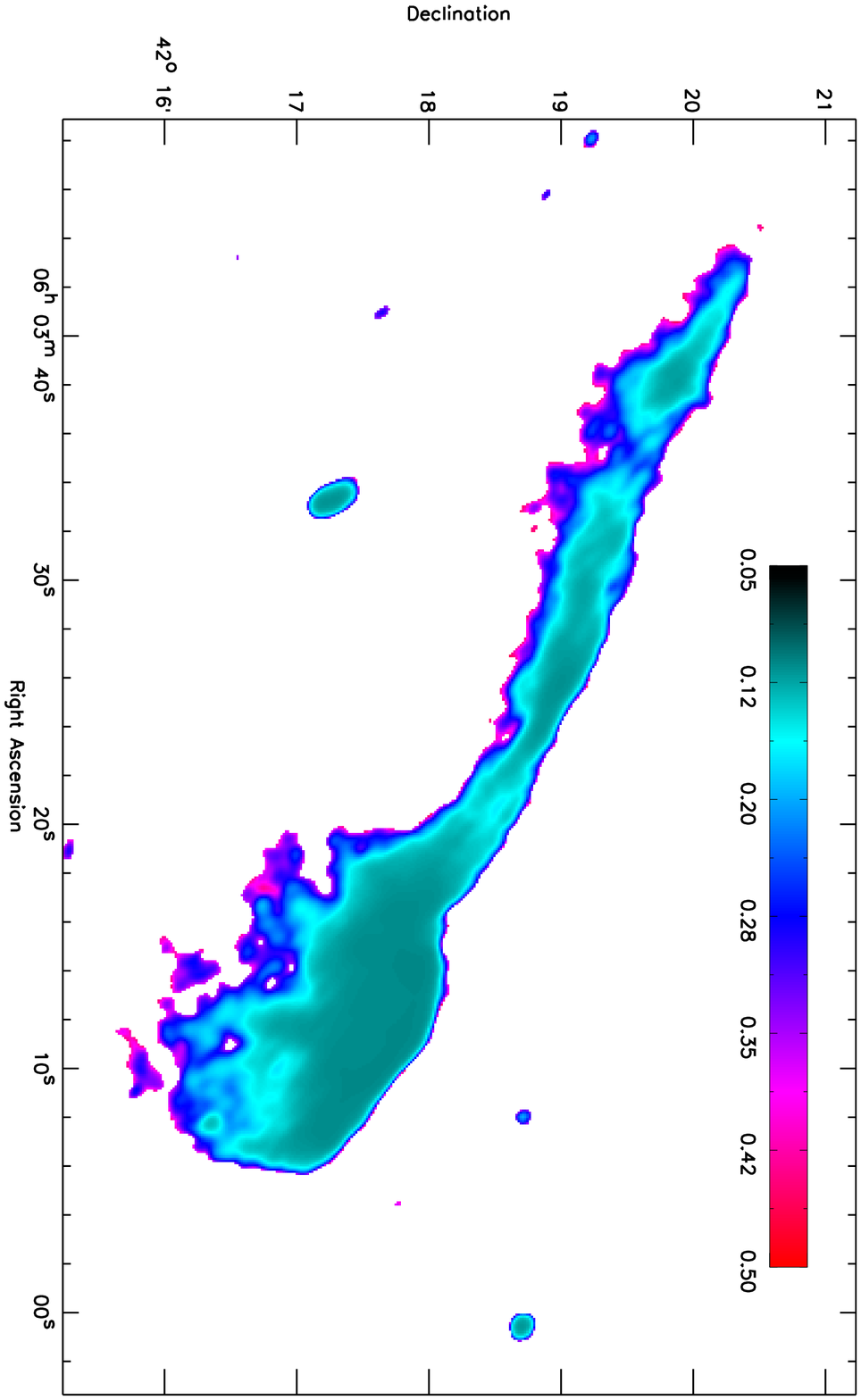}
\end{center}
\caption{Spectral index error map corresponding to Fig.~\ref{fig:rx42gmrt610}. The map is computed on the basis of $\sigma_{\mathrm{rms}}$ values for the individual maps and the uncertainties in the absolute flux scale.}
\label{fig:rx42spix610-1280_error}
\end{figure*}

\begin{figure*}
\begin{center}
\includegraphics[angle =90, trim =0cm 0cm 0cm 0cm,width=0.49\textwidth]{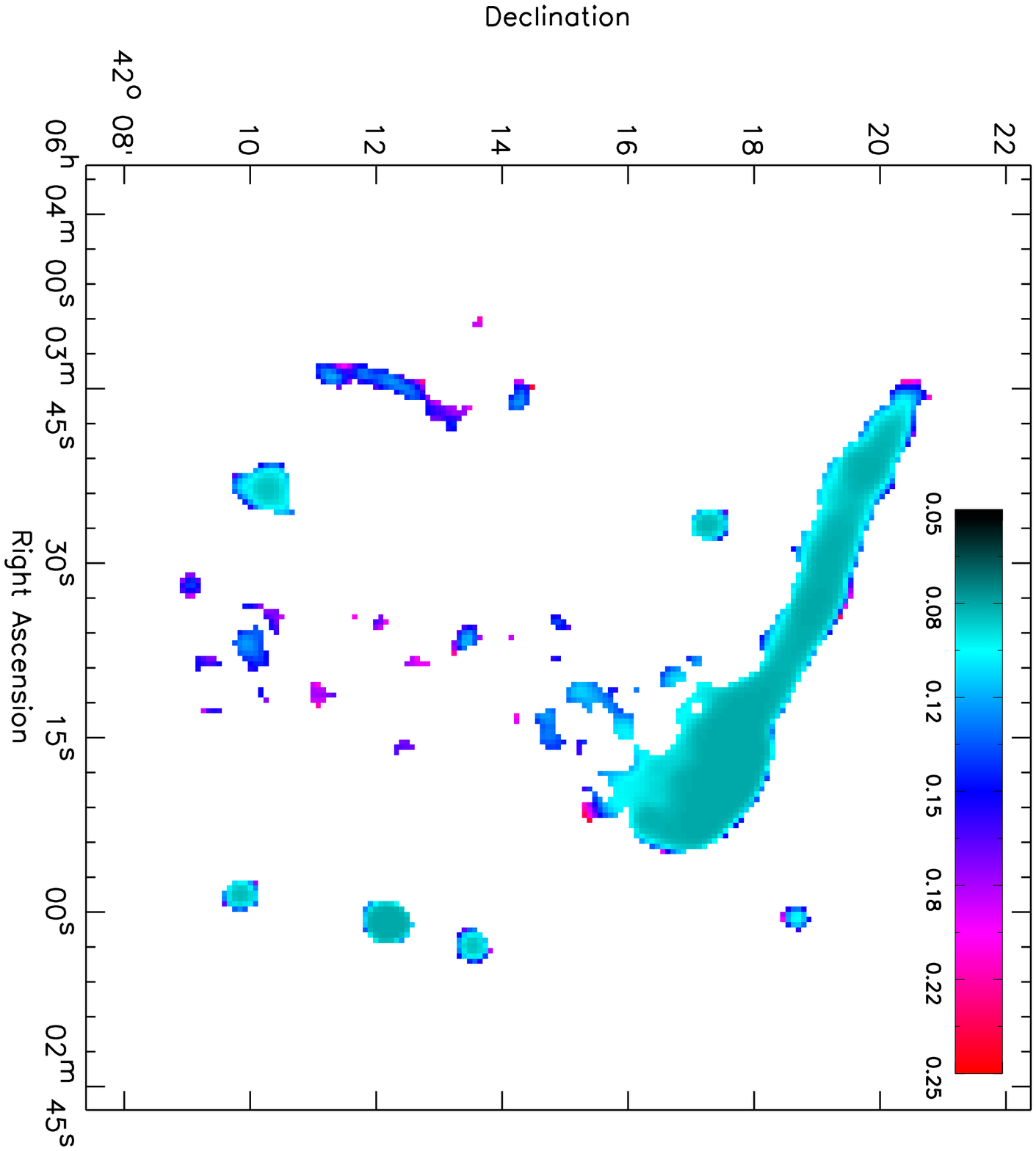}
\includegraphics[angle =90, trim =0cm 0cm 0cm 0cm,width=0.49\textwidth]{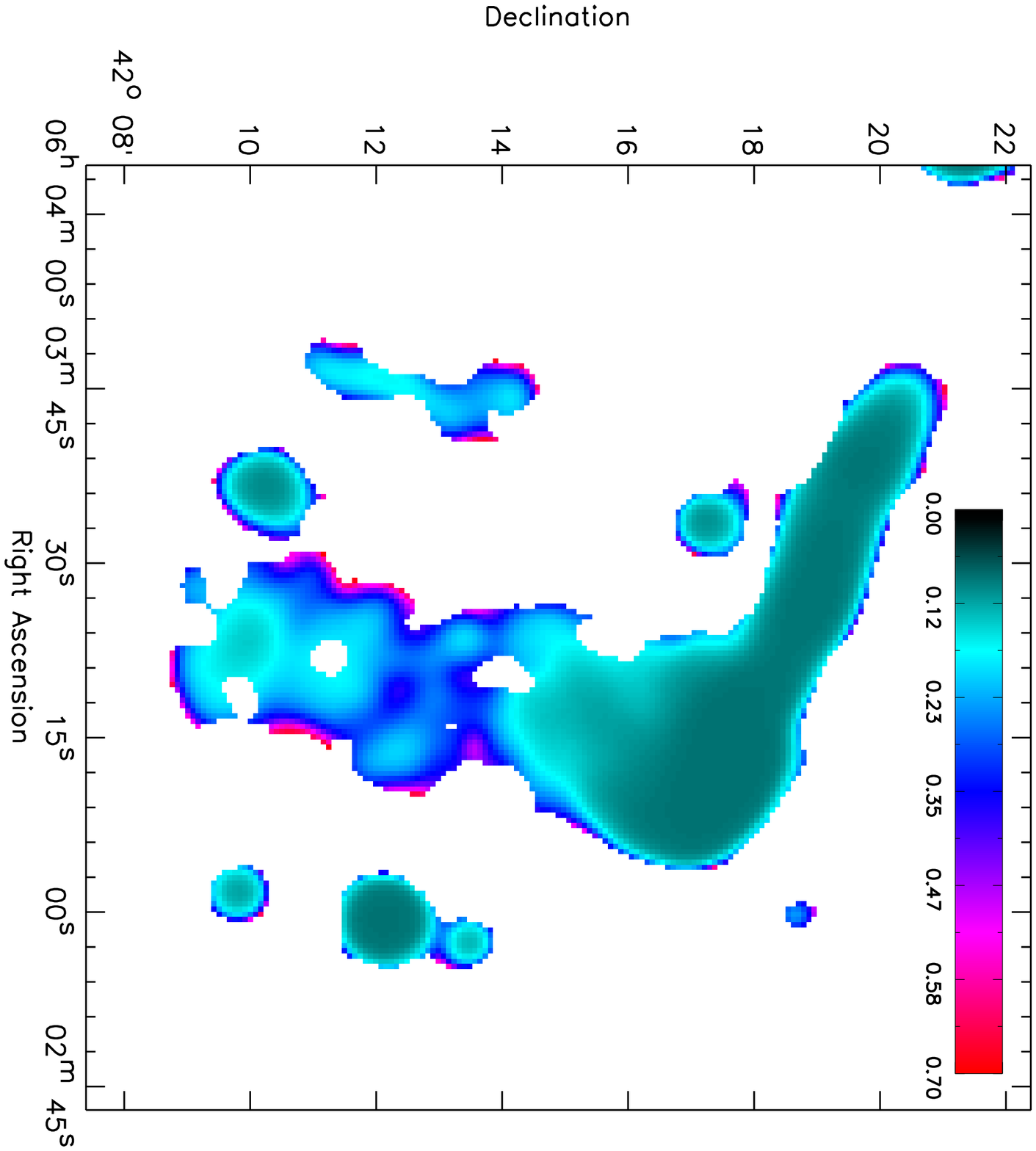}
\end{center}
\caption{Left: Spectral index error map corresponding to Fig.~\ref{fig:rx42spix_poly} left panel. Right:  Spectral index error map corresponding to Fig.~\ref{fig:rx42spix_poly} right panel.}
\label{fig:rx42spix_poly_error}
\end{figure*}

\end{appendix}
\end{document}